\documentclass[conference]{IEEEtran}
\IEEEoverridecommandlockouts

\def\BibTeX{{\rm B\kern-.05em{\sc i\kern-.025em b}\kern-.08em
    T\kern-.1667em\lower.7ex\hbox{E}\kern-.125emX}}

\usepackage{booktabs} 
\usepackage{xpatch,amsthm}
\usepackage[normalem]{ulem}
\usepackage{amsmath}
\usepackage{amsfonts}
\usepackage{amssymb}
\usepackage{algorithmic}
\usepackage{graphicx}
\usepackage{tikz}
\usetikzlibrary{decorations.pathreplacing}
\usepackage{textcomp}
\usepackage{epsfig}
\usepackage{xspace}
\usepackage{balance}
\usepackage{caption}
\usepackage{comment}
\usepackage{multirow}
\usepackage{url}
\usepackage{enumitem}
\usepackage{threeparttable}
\usepackage{xcolor}
\usepackage[ruled,noend,linesnumbered]{algorithm2e}
\usepackage{comment}
\usepackage{makecell}
\usepackage{subcaption}

\usepackage{xurl}
\usepackage{hyperref}

\usepackage{etoolbox}
\newtoggle{journalversion}
\togglefalse{journalversion} 

\SetKw{Break}{break}
\SetKw{KwBy}{by}

\expandafter\def\expandafter\normalsize\expandafter{%
    \normalsize%
    \setlength\abovedisplayskip{0pt}%
    \setlength\belowdisplayskip{8pt}%
    \setlength\abovedisplayshortskip{-8pt}%
    \setlength\belowdisplayshortskip{2pt}%
}

\usepackage{hyperref}     
\usepackage{cite}         

\hypersetup{
  colorlinks=true,        
  linkcolor=blue,         
  citecolor=blue,         
  filecolor=magenta,      
  urlcolor=blue           
}

\newcommand{\eatTR}[1]{} 
\newcommand{\eatNTR}[1]{}  

\newcommand{\eatOld}[1]{}
\makeatletter
\DeclareRobustCommand*\cal{\@fontswitch\relax\mathcal}
\makeatother

\xpatchcmd{\@thm}{\fontseries\mddefault\upshape}{}{}{}

\newtheorem{example}{Example}

\newcommand{\kw}[1]{{\ensuremath {\mathsf{#1}}}\xspace}

\newcommand{\stitle}[1]{\noindent{\bf #1}}

\newcommand{\fei}[1]{{\noindent \color{red}{Comment(Fei): #1}}}

\newcommand{\mostafa}[1]{{\noindent \color{teal}{Comment(Mostafa): #1}}}

\newcommand{\jj}[1]{{\noindent \color{magenta}{Comment(JP): #1}}}

\newcommand{\blue}[1]{{\noindent \color{blue}{#1}}}

\newcommand{\mc}[1]{\mathcal{#1}}
\renewcommand{\paragraph}[1]{\vspace{2mm}\noindent {\bf #1:}}
\newcommand{\trainP}{\textsf{LearnNormalModel}\xspace}
\newcommand{\detectP}{\textsf{FindDriftAnomalies}\xspace}

\newcommand{\activeM}{M_A}
\newcommand{\pattern}{N}
\newcommand{\latency}{\kappa}
\newcommand{\pmember}{\phi_i}
\newcommand{\threshold}{\tau}
\newcommand{\freq}{\nu}
\newcommand{\pweight}{w}
\newcommand{\cwindow}{\ensuremath{W}}

\newcommand{\trainTS}{T}
\newcommand{\cutoffTh}{\textit{cf}}

\newcommand{\TClustering}{\mc{C}}

\newcommand{\TCluster}{C}
\newcommand{\avgOf}{\textsf{avg}}
\newcommand{\minOf}{\textsf{min}}
\newcommand{\stdOf}{\textsf{std}}

\newcommand{\ascore}[1]{\textit{score}(#1)}

\newcommand{\AHC}{\kw{AHC}}
\newcommand{\soln}{\kw{AnDri}}
\newcommand{\subseq}{$T_{j, j+l}$}
\newcommand{\activep}{$M_{A}$}

\newcommand{\ldist}{d}
\newcommand{\dist}{d}
\newcommand{\membership}{${\phi_i}$}

\newcommand{\seq}{\ensuremath{T_j}}
\newcommand{\sizeC}{$R_{min}$}
\newcommand{\cwindowMax}{\ensuremath{W_{max}}}

\definecolor{cbDarkOrange}{RGB}{213, 94, 0}
\definecolor{cbBlue}{RGB}{0,114,178}
\definecolor{cbGreen}{RGB}{0,158,115}
\definecolor{cbLightGray}{RGB}{100,100,100}

\newcommand{\ecg}{\kw{ECG}}
\newcommand{\iops}{\kw{IOPS}}
\newcommand{\elec}{\kw{Elec}}

\newcommand{\sensor}{\kw{Sensor}}
\newcommand{\smd}{\kw{SMD}}

\newcommand{\swat}{\kw{SWaT}}

\newcommand{\climate}{\kw{Climate}}

\newcommand{\traffic}{\kw{Traffic}}

\newcommand{\nab}{\kw{NAB}}
\newcommand{\ucr}{\kw{Environ}}
\newcommand{\cat}{\kw{CATS}}

\newcommand{\andri}{\kw{AnDri}}

\newcommand{\tranad}{\kw{TranAD}}
\newcommand{\norma}{\kw{NormA}}
\newcommand{\sand}{\kw{SAND}}
\newcommand{\damp}{\kw{DAMP}}
\newcommand{\andrioff}{\kw{AnDri}-\kw{Offline}}

\newcommand{\cabd}{\kw{CABD}}
\newcommand{\arcus}{\kw{ARCUS}}
\newcommand{\divad}{\kw{DIVAD}}
\newcommand{\meter}{\kw{METER}}
\newcommand{\candi}{\kw{CANDI}}

\newcommand{\eat}[1]{}

\makeatletter
\DeclareRobustCommand*\cal{\@fontswitch\relax\mathcal}
\makeatother

\newcommand{\sstab}{\rule{0pt}{8pt}\\[-1.8ex]}

\newcommand{\bi}{\begin{itemize}}
	\newcommand{\ei}{\end{itemize}}
	{\end{itemize}} 

\newcommand{\ignore}[1]{}

\newcommand{\be}{\begin{enumerate}}
	\newcommand{\ee}{\end{enumerate}}
\newcommand{\beqn}{\begin{eqnarray*}}
	\newcommand{\eeqn}{\end{eqnarray*}}

\newcommand{\eop}{\hspace*{\fill}\mbox{$\Box$}}

\newcommand{\nthesection}{\arabic{section}}

 \newcounter{lemma}
 
 \newcounter{axiom}
 \renewcommand{\theaxiom}{\arabic{axiom}}

\ignore{\newcounter{theorem}
 \renewcommand{\thetheorem}{\arabic{theorem}}
  }

\newenvironment{ctheorem}[1]{\begin{em}
        \refstepcounter{theorem}
        {\vspace{1ex}{\noindent \bf Theorem  \thetheorem~[#1]}: }}{
        \end{em}\eop\vspace{1.5ex}}

\newcounter{cor}
\renewcommand{\thecor}{\arabic{cor}}

\newcounter{prop}
\renewcommand{\theprop}{\arabic{theorem}}

 \newcounter{definition}[section]
 \renewcommand{\thedefinition}{\nthesection.\arabic{definition}}

\newcounter{alg}[section]
\renewcommand{\thealg}{\nthesection.\arabic{alg}}

\newcounter{arule}
\renewcommand{\thearule}{\arabic{arule}}

\definecolor{gray}{rgb}{0.5,0.5,0.5}

\newcounter{ccc}

\newenvironment{ab}
{\mathactivatecomma
	\mathcode`\,=\string"8000
	\ignorespaces}
{\ignorespacesafterend}

\newcommand{\mathactivatecomma}{%
	\begingroup\lccode`~=`\,
	\lowercase{\endgroup\edef~}{\mathchar\the\mathcode`\,\penalty0 }}

\newcommand{\bab}{\begin{ab}}
	\newcommand{\eab}{\end{ab}\xspace}

\begin{document}

\title{Adaptive Anomaly Detection in the Presence of Concept Drift}
%

\author{
\IEEEauthorblockN{Jongjun Park}
\IEEEauthorblockA{\textit{McMaster University} \\
Hamilton, Ontario, Canada \\
{parkj182@mcmaster.ca}}
\and
\IEEEauthorblockN{Fei Chiang}
\IEEEauthorblockA{\textit{McMaster University} \\
Hamilton, Ontario, Canada \\
{fchiang@mcmaster.ca}}
\and
\IEEEauthorblockN{Mostafa Milani}
\IEEEauthorblockA{\textit{Western University} \\
London, Ontario, Canada \\
{mostafa.milani@uwo.ca}}
}




\maketitle

\begin{abstract}
The presence of concept drift poses challenges for anomaly detection in time series.  While anomalies are caused by undesirable changes in the data, differentiating abnormal changes from varying normal behaviours is difficult due to differing frequencies of occurrence,  varying time intervals when normal patterns occur, and identifying similarity thresholds to separate the boundary between normal vs. abnormal sequences.   Differentiating between concept drift and anomalies is critical for accurate analysis as studies have shown that the compounding effects of error propagation in downstream tasks lead to lower detection accuracy and increased overhead due to unnecessary model updates.  Unfortunately, existing work has largely explored anomaly detection and concept drift detection in isolation.   We introduce \soln, a framework for \uline{An}omaly detection in the presence of \uline{Dri}ft.  \soln\ introduces the notion of a \emph{dynamic}  normal model where normal patterns are activated, deactivated or newly added, providing flexibility to adapt to concept drift and anomalies over time.  We introduce  a new clustering method, \emph{Adjacent Hierarchical Clustering (\AHC)}, for learning normal patterns that respect their temporal locality; critical for detecting short-lived, but recurring patterns that are overlooked by existing methods.  Our evaluation on real datasets with varying types, proportions, and distributions of concept drift and anomalies shows that AnDri outperforms or matches existing baselines, with the largest gains on univariate data with recurring, short-lived normal patterns. 
\end{abstract}


%
%



\section{Introduction}
\label{sec:intro}

Real-world time series evolve due to changes in user behaviour, system dynamics, and environmental conditions. Identifying the inherent patterns underlying this evolution is a central task in time-series analysis. When the underlying data distribution changes, the phenomenon is known as \emph{concept drift}~\cite{widenanddeep-16, ARF-17, Review-19, DDG-DA}. Traditional anomaly detection methods often assume stationary distributions, while drift detection assume anomalies are rare—assumptions that rarely hold in practice.  In this work, we shift the focus from detecting drift and anomalies independently to understanding the challenges that arise when they occur simultaneously. Their interaction produces behaviours that cannot be captured when drift and anomalies are treated in isolation.  

\begin{figure}[t]
\center
    \begin{subfigure}[t]{\linewidth}
        \vspace{0pt}
        \centering
        \includegraphics[width=\linewidth]{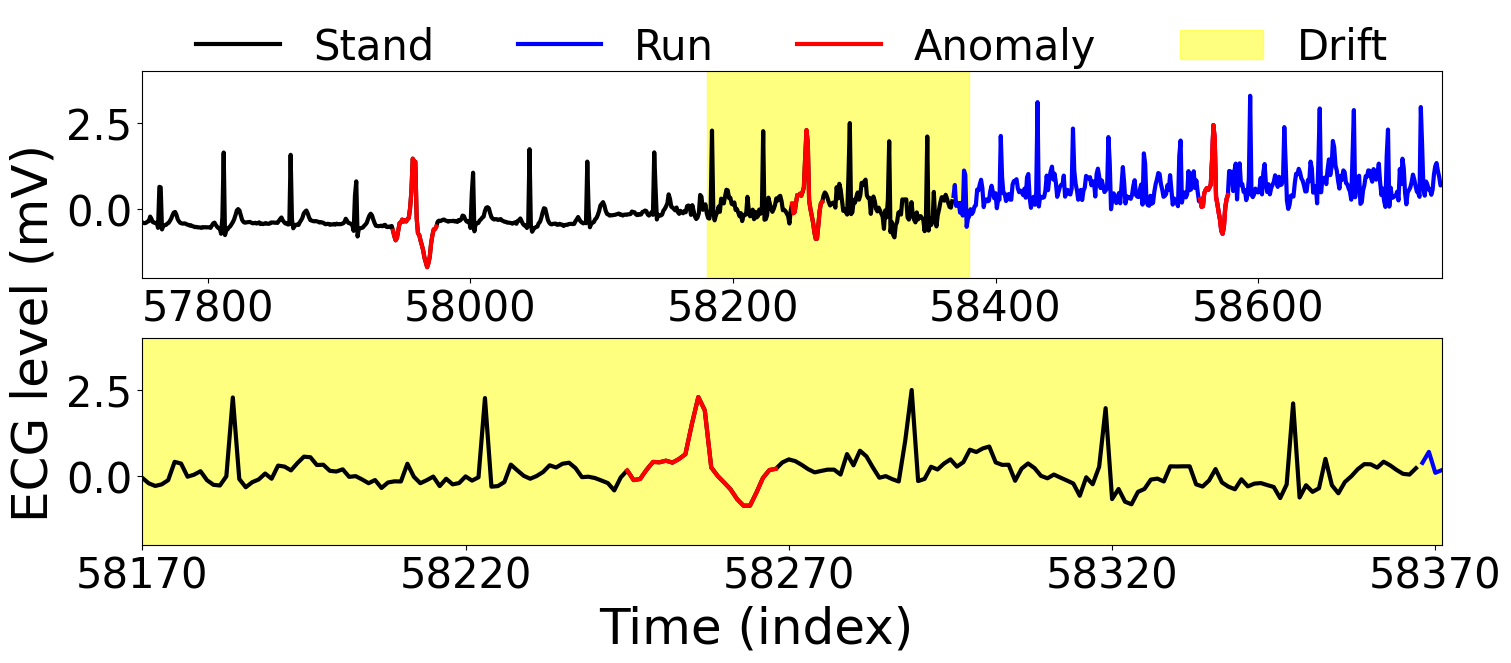}
        \vspace{-5mm}
        \caption{ECG recordings from standing to running
        \eat{\fei{Example is good, but picture needs revision to be more clear, the black box is not readable, and doesn't say much. Show the actual similar patterns from Summer'23 to the heavy rain period during drift. }}}
        \label{fig:ecg_ex}
    \end{subfigure}
    \begin{subfigure}[t]{\linewidth}
        \vspace{0pt}
        \centering
        \includegraphics[width=\linewidth]{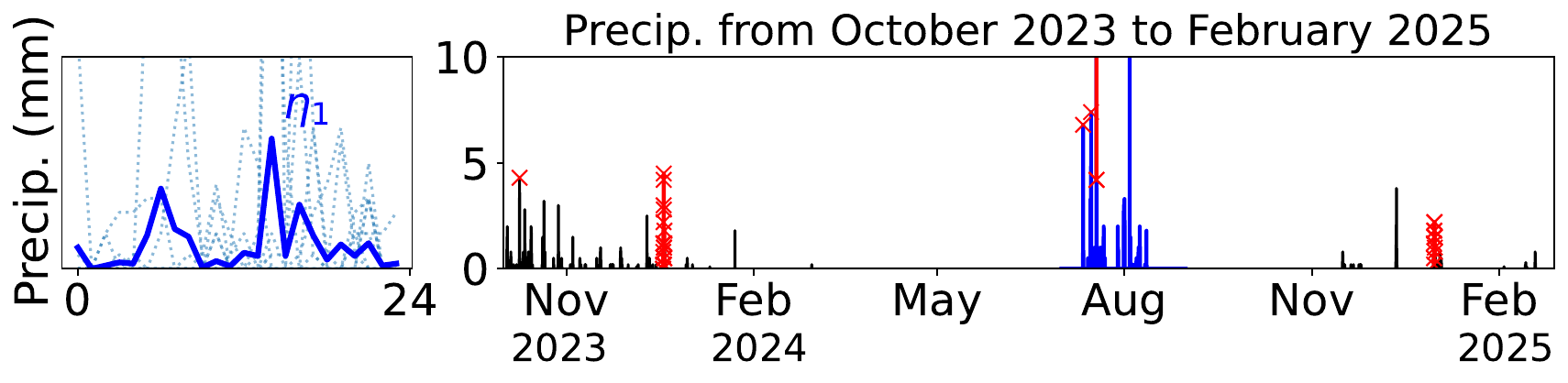}
        \caption{Seasonal precipitation patterns.  \eat{Update figure 1b (right) to show missed normal patterns from baselines, and distinctive locally frequent pattern.} \eat{The $\eta_1$ is the pattern what baselines missed, but \andri\ finds.}
        \eat{\fei{the top-2 figures are not readable, cannot see the difference in normality that is short-lived vs. more frequent ones. Please make this more clear in the pictures.  Too small, and not clear what is short-lived vs. not.} \eat{, and similar but temporally distant anomalies (red).}}}
        \label{fig:precip_ex_b}
\end{subfigure}
    \caption{Anomalies and drift interactions. 
    \eat{\fei{update your x-axis in (b) to Nov'24, and the last one Nov'25. Your title on (b) is not needed and wrong dates anyway. "Precip from Oct..."}}}
    \label{fig:anom_ex}
    \vspace{-5mm}
\end{figure}

\begin{example}
Electrocardiogram (ECG) patterns fluctuate over different normal baseline patterns depending on physical activity.  A person with heart rhythm problems, known as arrhythmias, may experience a sudden rapid, uncoordinated heartbeat while running, \eat{walking} changing the rate and regularity of their heartbeat.  Such anomalies should be identified relative to a person's physical activity, and differentiated from normal fluctuations such as transitioning from standing to running. \eat{walking to jogging.}  
Figure~\ref{fig:ecg_ex} shows sample ECG recordings of a person \emph{standing} (denoted in black) and \emph{running} (blue), the transition between the two activities is highlighted in yellow, and anomalies, representing arrhythmias, are denoted in red~\cite{mhealth_2014}.   \eat{We synthetically injected sequential anomalies of arrhythmia patterns from a patient while projecting the person's ECG period (in red). As you seen, (normal) heartbeat pattern changes corresponding to a change in physical activity.} We observe a change in the baseline normal pattern from a low frequency, long period (standing) pattern toward a higher frequency, shorter period, with large ECG variations, while running. Existing baselines identify normal patterns that are sufficiently frequent and distinct, without considering such transitions between normal patterns (drift periods)~\cite{Norma-2021, boniol2021sand}.  These transitions, and anomalies that occur during these periods are often missed, leading to an increased number of false positives. 

Figure~\ref{fig:precip_ex_b}  shows precipitation levels in Wabush, Newfoundland, Canada, where labeled anomalies are defined according to hourly precipitation levels exceeding 4mm~\cite{heavyRain}, or daily historical averages~\cite{canada-climate-normal}.  The figure shows a rainfall pattern $\eta_1$ occurring with high-frequency in August 2024. Although such rainfall amounts are not pervasive throughout the entire time series, short-lived but locally frequent patterns such as $\eta_1$ occur with expected regularity every summer.  Existing methods mis-classify patterns such as $\eta_1$ as anomalies due to their dissimilarity and global infrequency, resulting in higher false positive rates~\cite{MatrixProfile1-2016, lu2022matrix,Norma-2021, boniol2021sand}.  However, we argue that such behaviours, which are frequent and similar over short time intervals, variations in the period, or having varying baseline normal behaviours, occur in real phenomena, and are normal.
\end{example}

As exemplified in Figure~\ref{fig:anom_ex}, the presence of concept drift poses challenges for anomaly detection in time series.  While anomalies are caused by undesirable changes in the data, differentiating abnormal changes from varying normal behaviours is difficult due to differing frequencies of occurrence,  varying time intervals when normal patterns occur, and identifying similarity thresholds to separate the boundary between normal vs. abnormal sequences.   Differentiating between concept drift and anomalies is critical for accurate analysis as studies have shown that the compounding effects of error propagation in downstream data analysis tasks lead to lower detection accuracy and increased overhead due to unnecessary model updates~\cite{audio-ADCD-21, Review-anomaly-2022, Norma-2021}.  Unfortunately, existing work has largely explored anomaly detection and concept drift detection in isolation~\cite{ARF-17, DeepGBM-2019, DDG-DA}.   

\noindent \textbf{State-of-the-Art.}
The problem of subsequence anomaly detection has been extensively studied. Pattern-based anomaly detection methods identify normal behaviour by searching for frequent motifs~\cite{MatrixProfile1-2016, lu2022matrix} or by comparing subsequences to learned normal models~\cite{MatrixProfile-2017, Norma-2021}. These approaches implicitly assume that normal behaviour is frequent, persistent, and representative across the entire dataset. Such assumptions break down when normal patterns change over time or appear only in short intervals. Neural-network–based methods~\cite{TranAD-2022, LSTM-AD-2015, MAD-GAN-2019} predict future behaviour from historical trends, but their predictive accuracy degrades under drift and they can unintentionally adapt to recurring anomalies. Concept drift detection techniques, in turn, typically assume that anomalies are negligible or absent~\cite{Norma-2021, ADWIN-07, Review-19, ARF-17}.

\noindent \textbf{Challenges.} We study time series anomaly detection in the presence of drift, and address the following challenges.

\noindent \uline{Drift--Anomaly Ambiguity:} During drift transitions, subsequences temporarily exhibit mixed characteristics between the old and new normal patterns. 
Similarly, recurring anomalies may appear deceptively regular and mimic the onset of a new normal pattern. This \emph{drift--anomaly ambiguity} makes it difficult to determine whether a subsequence reflects an evolving normal behaviour or a genuine anomaly. This ambiguity is inherent whenever concept drift and anomalies coexist.\\
\noindent \uline{Short-Lived Normality:} Normal patterns do not necessarily persist across the entire dataset. In many real-world settings—such as seasonal weather patterns, periodic machine states, or activity-specific ECG behaviour—normal patterns may be \emph{locally frequent but globally rare}. Frequency-based methods that aggregate over the full series tend to misclassify such short-lived normality as anomalies. \\
\noindent \uline{Recognizing temporal context:} Existing methods define normal models independent of time~\cite{Norma-2021, boniol2021sand}.   We argue that local context is critical for accurate anomaly detection where baseline behaviour that was normal in the past, but is no longer currently observed, should be augmented with recent events.

\paragraph{Contributions}
To address these challenges, we rely on two core principles: (1) explicitly model the \emph{temporal locality} of normal patterns; and  (2) dynamically update the normal model so normal patterns become active, inactive, or revived as drift evolves. 

\sstab
(1) We introduce \soln (\uline{An}omaly detection in the presence of \uline{Dri}ft), an adaptive subsequence anomaly detection method designed to co-detect anomalies and drift.

\sstab 
(2) We propose a \emph{dynamic normal model} where normal patterns can be added, activated, and deactivated over time. Activation and deactivation allow the model to capture short-lived normality and prevent intermittent recurring anomalies from being absorbed as normal behaviour. The ability to add new normal patterns enables the system to distinguish anomalies from genuine drift by examining local frequencies when  unseen patterns emerge.

\sstab 
(3) We develop \emph{Adjacent Hierarchical Clustering (AHC)}, a clustering method that restricts merges to $k$-adjacent temporal neighborhoods.  
This locality constraint is essential for identifying drift phases, capturing short-lived normality, and preserving temporal structure.  
We further introduce a rollback optimization that recovers merges missed under small $k$, allowing clustering to use small adjacency widths without fragmenting recurring patterns.

\sstab 
(4) We perform extensive experiments against eight (univariate, multivariate) baselines, and six real datasets.  
\soln\ shows superior and comparable performance against existing baselines, and adapts to evolving normal patterns, in cases where existing methods fail. To support reproducibility, we publicly release our implementation, and datasets~\cite{datasite}.


\section{Preliminaries}
\label{sec:preliminaries}


We review time series concepts, notation, and the anomaly detection problem in Section~\ref{sec:problem}. Section~\ref{sec:normA} covers normal models and patterns from the literature, and Section~\ref{sec:distance} defines the distance functions we use in this work.

\subsection{Background and Problem Definition} \label{sec:problem}

A \emph{time series} $T \in \mathbb{R}^n$ is a sequence of real-valued observations $[x_1, \ldots, x_n]$, where $|T|=n$ denotes the length of the series. Traditional anomaly detection methods typically focus on \emph{point anomalies}, aiming to produce a binary label sequence
$Y = [y_1, \ldots, y_n] \in \{0,1\}^n$, where $y_i = 1$ indicates that $x_i$ is detected as an anomalous (noisy) point~\cite{SWP22}. The corresponding ground-truth labels are denoted by
$Y^* = [y^*_1, \ldots, y^*_n]$, where $y^*_i = 1$ means that $x_i$ is truly anomalous.

In contrast to point anomalies, \emph{subsequence anomalies} refer to unusual patterns that occur within contiguous segments of a time series. A subsequence of length $\ell$ starting at position $j$ is denoted by $T_{j,\ell} = [x_j, x_{j+1}, \ldots, x_{j+\ell-1}]$. Given a time series $T=[x_1,\ldots,x_n]$, subsequences are extracted using a sliding window of length $\ell$. Two common extraction strategies are widely used in the literature. With \emph{maximally overlapping} extraction ($h=1$), a subsequence is extracted at every time step, yielding $n-\ell+1$ subsequences
$T_{1,\ell}, T_{2,\ell}, \ldots, T_{n-\ell+1,\ell}$.
Alternatively, with \emph{non-overlapping} extraction ($h=\ell$), subsequences are extracted back-to-back, yielding $\lfloor n/\ell \rfloor$ subsequences.
Unless stated otherwise, we assume maximally overlapping extraction ($h=1$). 

\emph{To simplify notation, we let $N$ denote the number of subsequences extracted from $T$, and use $T_j$ to denote the $j$-th extracted subsequence, independent of the hop size or extraction strategy; unless stated otherwise, we assume $h=1$, in which case $T_j = T_{j,\ell}$.}

Subsequence anomaly detection aims to identify such segments that deviate significantly from expected or ``normal'' patterns. Assuming a unit hop size ($h=1$), the goal is to produce a label sequence
$Y = [y_1, \ldots, y_{N}] \in \{0,1\}^{N}$, where $y_j = 1$ indicates that the subsequence $T_j$ is considered anomalous. The corresponding ground truth labels are denoted by
$Y^* = [y^*_1, \ldots, y^*_{N}].$  

Subsequence anomaly detection methods often assign labels $y_j$ by first computing a sequence of anomaly scores
$S = [s_1, \ldots, s_{N}] \in \mathbb{R}^{N}$,
where each score $s_j$ corresponds to the subsequence $T_j$. A common approach is to apply a threshold to these scores to derive the binary label sequence $Y$. Alternatively, some methods identify anomalies by selecting the top-ranked subsequences with the highest anomaly scores. In this case, $y_j = 1$ if $T_j$ is among the selected subsequences, and $y_j = 0$ otherwise.

In an \emph{online} setting, the anomaly label $y_j$ for subsequence $T_j$ can be produced only after all points in $T_j$ have been observed, that is, at time $t = j + \ell - 1$ (for $h=1$). At this time, only subsequences $T_1, \ldots, T_j$ are available for detection. In contrast, \emph{offline} detection assumes access to the entire time series $T$ in advance to compute the full label sequence $Y$.

\eat{
More generally, some online settings allow a reporting \emph{latency} $\latency \geq 0$, in which case the label $y_j$ may be produced at time $t = j + \ell - 1 + \latency$. Allowing latency provides access to future observations and can improve detection accuracy. When $\latency = 0$, labels are produced immediately after observing $\seq$, corresponding to real-time detection; as $\latency$ increases, the online setting approaches the offline setting. 
}

\subsection{Normal Models and Normal Patterns} \label{sec:normA} 

Identifying normal patterns in a time series is essential for detecting anomalies, which are defined as deviations from these patterns. We build on the notion of normality introduced in NormA~\cite{Norma-2021}, where normal patterns are derived from their recurrence in the time series. While our approach shares this core idea, we adopt a modified formulation of the normal model to support more flexible and adaptive behavior. We review the definition of normal patterns~\cite{Norma-2021}.

Given a time series $T$, a \emph{normal model} $M = \{(\pattern^M_1, \pweight^M_1), \ldots, (\pattern^M_m, \pweight^M_m)\}$ is a set of $m$ pairs, where each $\pattern^M_i$ is a subsequence of size $\ell^M \geq \ell$, and $\pweight^M_i > 0$ is a real-valued weight representing the frequency of $\pattern^M_i$ in $T$. Each $\pattern^M_i$ is referred to as a \emph{normal pattern}. 

The anomaly score of a subsequence $T_{j} \subseteq T$ with respect to a model $M$ is defined as a weighted sum of its distances to the normal patterns $\pattern^M_i$, where $\dist$ is a distance function between subsequences, $\pattern^M_i$ and $T_{j}$:
\begin{align}
\ascore{T_{j}, M} = \sum_{i=1}^{m} \pweight^M_i \cdot \dist(\pattern^M_i, T_{j}) \label{eq:score}
\end{align}

\noindent where $\dist$ is a distance function between two subsequences. 

\subsection{Distance Measures for Subsequences}\label{sec:distance}
A widely used distance function $\dist$ for comparing subsequences is the \emph{Z-normalized Euclidean distance}:
\begin{align} \label{eq:d-znorm}
\dist(A, B) = \sqrt{\sum_{i=1}^{\ell} \left( \frac{A_{i,1} - \mu(A)}{\sigma(A)} - \frac{B_{i,1} - \mu(B)}{\sigma(B)} \right)^2}
\end{align}
where $A$ and $B$ are subsequences of length $\ell$, and $\mu(\cdot)$ and $\sigma(\cdot)$ denote the mean and standard deviation, respectively. This normalization removes both offset and scale, enabling shape-based comparison. A related alternative is the \emph{zero-mean Euclidean distance}:%
\begin{align} \label{eq:d-zm}
\dist(A,B)
= \sqrt{\sum_{i=1}^{\ell}
\bigl((A_{i,1}-\mu(A))-(B_{i,1}-\mu(B))\bigr)^2}.
\end{align}
Zero-mean normalization removes only the offset, preserving amplitude information.

When $\ell \neq \ell^M$, the distance $\dist(\pattern^M_i,T_j)$ is computed as the minimum distance between $T_j$ and any contiguous subsequence of length $\ell$ within $\pattern^M_i$, under a chosen normalization. In our experiments, we consider both \emph{zero-mean} and \emph{z-normalized} variants of the distance $d$. Zero-mean normalization preserves relative scale and is effective when amplitude carries semantic meaning, whereas z-normalization removes offset and scale and is preferable when anomalies are primarily shape-based. The normalization is selected per dataset based on validation performance. For clarity, we use $d$ to denote the chosen distance metric (zero-mean or z-normalized) throughout the paper, as the subsequent discussion applies to either variant.

\noindent The distance $\dist(A,B)$ therefore captures \emph{value-based similarity} between two subsequences: it measures how closely their normalized shapes and amplitudes align, independent of scale or offset. In addition to this content-based distance, we also consider a simple \emph{temporal distance} function that reflects how far apart two subsequences occur within the same time series. \noindent For two subsequences $T_{i,\ell}$ and $T_{j,\ell'}$ starting at time indices $i$ and $j$, we define their temporal distance as the absolute difference between their starting positions $\delta(T_{i,\ell},\, T_{j,\ell'}) = |\, i - j \,|$. This temporal distance quantifies how subsequences are ordered in time and is used to reason about their adjacency and local continuity.  Table~\ref{tbl:symbol} summarizes our  notation.
 
\begin{table} [t]
\vspace{-3mm}
\begin{small}
\begin{center}
\renewcommand{\arraystretch}{1.1}
\caption{Notation summary. \label{tbl:symbol}}
\vspace{-1mm}
\begin{tabular}{  l | l  }
      \hline
      \textbf{Symbol} & \textbf{Description}  \\
      \hline \hline
      $M$ & Normal model: set of normal patterns\\
      \hline
      \activep & Active normal patterns \\
      \hline
      $\pattern^M_{i}$ & Normal pattern $i$ in model $M$ \\
      \hline
      \seq\ & Subsequence at time $j$  \\
      \hline
      $\threshold^M_i$ & Distributional threshold of normal pattern $i$ \\
      \hline
      $\freq^M_i$ & Frequency threshold of normal pattern $i$ \\
      \hline
      \membership\ & Membership function of normal pattern $i$ \\
      \hline
      \cwindow & Currency window size  \\
      \hline
      $\TCluster_j$ & Cluster $j$  \\
      \hline
      \sizeC & Minimum cluster size in \AHC \\
      \hline
\end{tabular}
\end{center}
\end{small}
\vspace{-5mm}
\end{table}

\section{\soln Overview}
\label{sec:overview}

Our anomaly detection method, \soln (\uline{An}omaly detection in the presence of \uline{Dri}ft), has several key technical features that set it apart from existing approaches such as \norma and \sand~\cite{Norma-2021,boniol2021sand}.  First, \soln redefines the notion of a normal model (a set of normal patterns) to be {\em dynamic}: normal patterns are not fixed after training but can be activated, deactivated, or added over time. This flexibility enables the model to adapt as the time series evolves. Second, unlike \norma, which computes anomaly scores using a weighted sum of distances to all normal patterns (Eq.~\ref{eq:score}), \soln assigns scores based only on the distance to a {\em matching pattern}, which is the most similar active pattern. This avoids the need to aggregate over all patterns and improves precision when multiple normal behaviours coexist. Third, \soln introduces a new clustering method, \emph{Adjacent Hierarchical Clustering (\AHC)}, for learning normal patterns. \AHC groups similar subsequences while respecting their temporal locality; crucial for detecting short-lived or recurring patterns that would be overlooked by traditional clustering methods.

These key features directly address the core challenges posed by concept drift. By allowing normal models to change, \soln can respond to both gradual and abrupt changes in data distribution. The use of active pattern management prevents stale or outdated patterns from misleading the anomaly detection process, while \AHC enables the discovery of emerging local behaviours. Comparing subsequences only to the closest active patterns reduces the chance of misclassifying new normal patterns as anomalies, especially during transitions. Together, these features allow \soln to maintain accuracy and robustness in non-stationary environments where both anomalies and concept drift occur.

\begin{figure}[t]
 \centering
    {    
    \includegraphics[width=\linewidth]{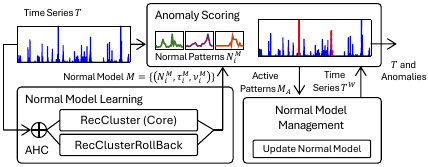}
    }
    \vspace{-4mm}
 \caption{\soln (online) \ Architecture. }
  \label{fig:overview}
  
\end{figure}

Figure~\ref{fig:overview} presents an overview of \soln, which consists of three main modules: (1) learning a (dynamic) normal model, (2) managing the patterns in a normal model over time, and (3) computing anomaly scores. Given a time series $T$, \soln first extracts normal patterns from an initial portion of $T$ (or a separate training sequence) and uses them to initialize a model of normal behaviour. These normal models serve as baselines for detecting anomalies in the remainder of the time series. We introduce each component of the solution in the remainder of this section, and provide technical details in Sections~\ref{sec:detail} and~\ref{sec:normals}.

\eat{
\begin{figure}[t]
 \centering
    {    
    \includegraphics[width=\linewidth]{figures/architecture_detail.png}
    }
    \vspace{-4mm}
 \caption{\soln (online) \ Architecture.\jj{I edited more detailed version. Based on the paper length, we can choose one of them.}}
  \label{fig:overview2}
\end{figure}
}

\subsection{Learning the Normal Model} \label{sec:learning-normal-models}

To detect anomalies, we rely on a model of normal behaviour built from recurring subsequences, referred to as \emph{normal patterns}, as introduced in Section~\ref{sec:normA} in the context of \norma. 
While the high-level intuition is similar—using previously observed patterns to characterize normality—our formulation differs substantially in both structure and dynamics. We define a \emph{dynamic normal model} as a set of triples $M = \{(\pattern^M_i, \threshold^M_i, \freq^M_i)\}_{i=1}^m,$
where $\pattern^M_i$ is a normal pattern, $\threshold^M_i$ is a \emph{distributional threshold}, and $\freq^M_i$ is a \emph{frequency threshold}. 
Unlike the static model used in \norma, our model evolves over time: patterns may be added, deactivated, or reactivated in response to newly observed data. The two thresholds jointly determine which patterns are considered active and replace the single weight parameter $\pweight^M_i$ used in \norma. The distributional threshold $\threshold^M_i$ controls how similar a new subsequence must be to match a pattern, while the frequency threshold $\freq^M_i$ captures how often the pattern must occur within a recent time window to remain active. This two-parameter design provides finer-grained control and enables better adaptation to shifting and recurring behaviours in non-stationary time series.

To construct the dynamic normal model, we extract subsequences of a fixed length $\ell$ from a training time series $\trainTS$, which may be a prefix of the full time series or a separate dataset. We use clustering to extract normal patterns from the training data by grouping similar subsequences into clusters $\TClustering = \{\TCluster_1, \ldots, \TCluster_{|\TClustering|}\}$. Each normal pattern $\pattern^M_i$ is then defined as the centroid of the subsequences in cluster $\TCluster_i$.

Traditional time-series clustering methods typically ignore the temporal locations of subsequences and cluster them solely based on similarity in shape or value. This overlooks important time-dependent correlations and fails to account for non-stationarity in the data. For example, a subsequence that is considered normal at time $t$ may be anomalous at time $t'$, due to a shift in the underlying distribution. In our setting, we assume that normality is consistent only within a limited time window and may change over the course of the series; a key aspect of concept drift that prior work fails to address~\cite{boniol2021sand,Norma-2021}.

To handle this, \soln introduces \AHC, a temporal-aware clustering algorithm for learning normal patterns. We give an intuitive overview of \AHC here, while full details and pseudocode are provided in Section~\ref{sec:ahc}. \AHC captures time-domain correlations by prioritizing the clustering of subsequences that are not only similar in shape but also close in time. It begins with singleton clusters and iteratively merges the closest pairs drawn from a neighborhood of \emph{$k$-adjacent} clusters, where adjacency refers to temporal proximity, ensuring that merges are considered only among clusters within $k$ consecutive positions. Distances above the cumulative average are used as cut points to form new clusters, and merging continues until all subsequences are grouped or a stopping condition is met. By incorporating temporal locality, \AHC extracts clusters that reflect time-dependent normal patterns. This allows the normal model to better capture normality as it evolves, adapting to changes over time and maintaining detection performance in the presence of concept drift.

Each normal pattern $\pattern^M_i$ learned through \AHC is paired with a distributional threshold $\threshold^M_i$, which defines the maximum allowable distance for a subsequence to be considered an instance of the pattern. It is also assigned a frequency threshold $\freq^M_i$, which captures how often the pattern must appear within a recent window to remain active. Together, these thresholds allow the system to track intra-cluster similarity and ensure that only recently and frequently observed patterns are used during detection. These parameters play a key role in distinguishing true anomalies from natural shifts in normal behaviour. Both $\freq^M_i$ and $\threshold^M_i$ are computed from the clusters and the time series used during clustering, with details provided in Section~\ref{sec:detail}.

\subsection{Normal Model Management} \label{sec:normal_model_management}

As the time series $T$ evolves, changes in the observed subsequences may indicate a shift in normality; this is the essence of \emph{concept drift}. In \soln, such a drift is interpreted as a transition between normal patterns. This may occur either between the existing patterns in the model—requiring a change in which patterns are currently active—or between existing patterns in $M$ and a previously unseen pattern not in $M$. In practice, distinguishing a genuinely new normal pattern from a temporary anomalous deviation can be challenging, as both involve subsequences that differ from all currently active patterns. A robust solution must therefore decide whether a subsequence reflects a new concept or a transient anomaly.

To monitor and manage these changes, \soln maintains a set of \emph{active normal patterns} that represent current normality. For a given \emph{currency window size} $\cwindow$, shared across all patterns in $M$, \soln checks whether the subsequences observed within $\cwindow$ are similar to any existing normal patterns in $M$. If they are, the corresponding patterns are activated (if they are not already active), or remain remain active  (if they are already active). If the subsequences are not similar to any known pattern, \soln triggers further analysis to determine whether the subsequences represent a genuine anomaly or the emergence of a new pattern.

To assess similarity between incoming subsequences and the active patterns, we define a \emph{membership function} $\phi(\seq, N^M_i)$, which measures how well a subsequence $\seq$ matches a normal pattern $N^M_i$. The core idea is that the more similar $\seq$ is to $N^M_i$, the higher the membership value. This function is based on the distance between $\seq$ and $N^M_i$ relative to that pattern’s distributional threshold, and is formally defined in Equation~\ref{eq:membership} (Section~\ref{sec:detail}). When anomalies appear sporadically, their membership values tend to fluctuate and remain low. In contrast, gradual or recurring drift produces a sustained shift in membership values, allowing \soln to differentiate stable changes in normality from isolated deviations.  \soln maintains a dynamic normal model $M$, and tracks the \emph{status} of each pattern as either active or inactive at any given moment. This status is determined using the membership function evaluated over the currency window $\cwindow$, which defines the range of most recent subsequences considered. \eatTR{\blue{The choice of $\cwindow$ is important: a small $\cwindow$ makes the system more sensitive to short-term variations, potentially detecting drift more quickly but increasing false positives. A larger $\cwindow$ smooths short-term noise but may delay drift detection. We study the impact of $\cwindow$ in our experiments (see Section~\ref{sec:param}). If a pattern $N^M_i$ maintains sufficiently high membership across $\cwindow$, it is marked as active; otherwise, it becomes inactive.}}

When the set of active patterns becomes empty, \soln applies \AHC to the subsequences in $\cwindow$ to determine whether a new coherent cluster can be formed. \AHC returns no cluster if none of the groups meet the minimum size requirement, or the criteria for distinctiveness. If a valid cluster is found, i.e., one that exceeds a size threshold $R_\textsf{min}$, it is considered as a new normal pattern $N^M_{i'}$, and added to $M$. This update is interpreted as concept drift. If no such cluster is formed, the subsequence $\seq$ is temporarily labeled as anomalous. 

\subsection{Anomaly Scoring}

Given a normal model $M$, current window \cwindow, and subsequence \seq, \soln\ computes an anomaly score for \seq\ by measuring its similarity to the set of \emph{active normal patterns}, which we denote by \activep.  In contrast,  ($M \setminus$ \activep) denotes the set of \emph{inactive} patterns that have appeared in $T$, but beyond the current window \cwindow, not recently.

\soln computes the anomaly score $s_j$ as the minimum distance between \seq\ and the active patterns in \activep, or the maximum score $s_\textsf{max}$ if no pattern is active. \soln\ is amenable to any distance metric.  We use the zero-mean distance to overcome the sensitivity of the Z-norm distance to sequential changes, as it can overestimate the distance when comparing subsequences with small variations~\cite{z-norm-2020}.  The zero-mean distance excludes the normalization by standard deviation, and is robust to small variations.  In contrast to existing methods that compute anomaly scores using the weighted sum of all normal pattern frequencies (independent of temporal locality)~\cite{Norma-2021, boniol2021sand}, \soln directly uses the distance values in the anomaly score to measure the magnitude of deviation from the norm, going beyond just using frequency. If an observed subsequence \seq\ is not an active pattern, \soln\ still computes its distance to non-active patterns ($M\setminus$ \activep) to track the occurrence of historical patterns. \eat{We compute the distance of \subseq\ to each pattern in $M$. }  To determine the presence of each normal pattern $\pattern^M_i$, we compute the average of their membership within \cwindow, and compare this with $\freq^M_i$.

\vspace{-1ex}
\section{Spatio-Temporal Normality Learning} \label{sec:detail}
\vspace{-1ex}

\soln\ learns a dynamic normal model by identifying normal patterns that are \emph{locally frequent} and \emph{temporally recent}, even if they do not occur frequently or cover many subsequences across the entire time series $T$. Existing methods typically focus on patterns with broad coverage and high global frequency, and therefore miss such short-term or emerging patterns. Capturing these locally relevant patterns is essential for adapting to concept drift and for distinguishing between true anomalies and newly emerging normal behavior; challenges that are not adequately addressed by existing methods~\cite{Norma-2021,boniol2021sand,TSB-UAD-2022}.

\setlength{\textfloatsep}{6pt}
\begin{algorithm}[t]
\SetKwInOut{Input}{Input}
\SetKwInOut{Output}{Output}
\caption{\trainP}\label{alg:training}
\SetAlgoLined
\SetNoFillComment
\Input{Time series $\trainTS = [x_1, \ldots, x_r]$}
\Output{Normal model $M$ \eat{and currency window $\cwindow$}}

$M \gets \emptyset;$ $\TClustering\gets\AHC(\trainTS);$ \label{ln:learn-cluster}\\

\ForEach{$\TCluster_i \in \TClustering$}{
    $\pattern^M_i \gets \avgOf(\TCluster_i.\textsf{sq})$; \label{ln:learn-pattern} 
   
    $D \gets \{ \dist(\seq, \pattern^M_i) \;|\; \seq \in \TCluster_i.\textsf{sq} \}$; \label{ln:learn-d}
    
    $\threshold^M_i \gets \avgOf(D) + 3 \times \stdOf(D)$; \label{ln:learn-threshold}
    
    $G \gets \emptyset$; \label{ln:learn-g}

    \ForEach{\cwindow\ {\bf overlapping with} $\TCluster_i$}{
        $G \gets G \cup \{\avgOf(\{ \pmember(\seq, \pattern^M_i) \;|\; \seq \in \cwindow \})\}$;
    }

    $\freq^M_i \gets \minOf(G)$; \label{ln:learn-freq} 
    
    $M\!\gets\!M\!\cup (\pattern^M_i, \threshold^M_i, \freq^M_i)$; \label{ln:learn-addtoM} 
}

\KwRet $M;$ \label{ln:learn-return}
\end{algorithm}

Algorithm~\ref{alg:training} learns a normal model from a training time series $\trainTS$. \AHC\ first clusters similar subsequences into
$\TClustering=\{\TCluster_1,\ldots,\TCluster_{|\TClustering|}\}$
(Line~\ref{ln:learn-cluster}). For each cluster $\TCluster_i$, it computes a normal pattern $\pattern^M_i$ as the cluster centroid (Line~\ref{ln:learn-pattern}) and derives its distributional and frequency thresholds, $\threshold^M_i$ and $\freq^M_i$ (Lines~\ref{ln:learn-threshold}--\ref{ln:learn-freq}). The resulting triples $(\pattern^M_i,\threshold^M_i,\freq^M_i)$ are added to the normal model $M$ (Line~\ref{ln:learn-addtoM}).

For each normal pattern $\pattern^M_i$, \soln\ computes two parameters: a distributional threshold $\threshold^M_i$ and a frequency threshold $\freq^M_i$ (Lines~\ref{ln:learn-d}--\ref{ln:learn-freq}). The distributional threshold defines how close a subsequence must be to match $\pattern^M_i$. It is computed from the distances between $\pattern^M_i$ and the subsequences in $\TCluster_i$ (Lines~\ref{ln:learn-d}--\ref{ln:learn-threshold}).

The frequency threshold $\freq^M_i$ captures how consistently $\pattern^M_i$ appears within a recent window. To compute it, \soln\ slides the currency window $\cwindow$ over the subsequences in $\TCluster_i$. For each window, it computes the membership value $\pmember(\seq,\pattern^M_i)$ for each subsequence $\seq \in \cwindow$:
\begin{equation}
\pmember(\seq,\pattern^M_i) =
\begin{cases}
1, & \hspace{-5mm}\text{if } \dist(\seq,\pattern^M_i) \le \threshold^M_i \\
e^{-\eta (\dist(\seq,\pattern^M_i) - \threshold^M_i)}, & \text{otherwise}
\end{cases}
\label{eq:membership}
\end{equation}

\noindent which extends a boolean membership to a fuzzy membership. If the distance between $\seq$ and $\pattern^M_i$ ($\dist(\seq,\pattern^M_i)$) is less than the distributional threshold ($\threshold^M_i$), then the membership is $1$, indicating that the subsequence $\seq$ follows the pattern $\pattern^M_i$.  Otherwise, the membership function returns a value in $[0,1)$ indicating the similarity between  $\seq$ and the normal pattern, since the membership value is negatively correlated with the gap between $\seq$'s distance with $\pattern^M_i$, and the distributional threshold $\threshold^M_i$. For each moving window, we compute the average membership of the subsequences in \cwindow, and save this average in $G$, representing the likelihood that $\pattern^M_i$ is the current pattern in \cwindow. We set $\freq^M_i$  to be the minimum average over all moving windows (Line~\ref{ln:learn-freq}).  We add each normal pattern and its parameters,  $(\pattern^M_i, \threshold^M_i, \freq^M_i)$, to $M$ (Line~\ref{ln:learn-addtoM}).  After processing all clusters, we return the final model $M$ and window size \cwindow\ (Line~\ref{ln:learn-return}).

\subsection{Adjacent Hierarchical Clustering}
\label{sec:ahc}

We introduce the \emph{Adjacent Hierarchical Clustering} (\AHC) algorithm, which performs hierarchical clustering over temporally adjacent subsequences in a time series. \AHC identifies normal temporal patterns occurring close in time while preserving their local structure and continuity. 
Unlike conventional hierarchical clustering, \AHC\ maintains and dynamically updates explicit \emph{neighborhoods} between clusters, ensuring that merging decisions are restricted to nearby temporal regions. This adjacency constraint reduces the number of pairwise comparisons during clustering—thereby improving scalability.

We describe the \emph{core} version of \AHC, which applies standard hierarchical clustering but restricts distance computations to $k$-adjacent clusters defined by temporal neighborhoods. This introduces a natural tradeoff in choosing $k$. Very small $k$ (e.g., $k=1$) minimize comparisons and maximize efficiency, but may prevent merging subsequences that share the same underlying pattern, yet are separated by even one intervening subsequence. In such cases, the algorithm cannot form the larger coherent clusters needed to recognize normal patterns, and incorrectly flags them as anomalies. Larger $k$ values alleviate this by allowing merges between temporally close, similar pattern subsequences, but at the cost of more comparisons and higher runtime per iteration. Thus, there exists a tradeoff where $k$ must be large enough to capture local, temporal structure while remaining small enough to retain the computational benefits of adjacency-based pruning. In the following subsection (Section~\ref{sec:ahc-rollback}), we extend this core algorithm with a rollback mechanism that allows smaller $k$ values while  recovering missed merges between non-immediate neighbors, improving efficiency without sacrificing clustering quality.  We evaluate this tradeoff for varying $k$, and show that while the number of rollbacks increase with small $k$,  improved and comparable accuracy is achieved (Section~\ref{sec:param}).


\noindent \textbf{Linkage and Temporal Distance.}  We first introduce necessary notation and core concepts.  A cluster $C$ contains a set of subsequences, written as $C.\textsf{sq}$.   We use $C.\textsf{mb}$ to denote its member (child) clusters in the hierarchy: this set is empty for leaf clusters and contains exactly two clusters for internal nodes. Because \AHC\ only merges temporally adjacent regions, each cluster $C$ also maintains an explicit set of neighboring clusters, denoted by $C.\textsf{nb}$.


In the algorithm, we use cluster-level distances that extend the subsequence distance $d$ (Eq.~\ref{eq:d-zm}) to clusters. Following standard hierarchical clustering practice~\cite{mullner2011modern,Norma-2021,z-norm-2020}, several \emph{linkage distances} can be defined between clusters $C$ and $C'$ based on pairwise distances between their subsequences—most commonly single (minimum), complete (maximum), average, and Ward linkage. Since our method relies primarily on local temporal adjacency, single linkage (Eq.~\ref{eq:min}) provides clear illustrative examples.  In our experiments, we use Ward linkage (Eq.~\ref{eq:ward}) due to its robust empirical performance and tendency to produce compact, well-separated clusters~\cite{ward14}.  

\vspace{2mm}
\begin{align}
d_{\min}(C,C') &= \min_{A\in C.\textsf{sq},\, B\in C'.\textsf{sq}} d(A,B) \label{eq:min} \\[3pt] 
d_{\mathrm{ward}}(C,C') &= \frac{|C.\textsf{sq}|\,|C'.\textsf{sq}|}{|C.\textsf{sq}|+|C'.\textsf{sq}|}\,(\bar{d}(C,C'))^{2}\label{eq:ward}
\end{align}

In Eq.~\ref{eq:cdist}, $\bar{d}(C,C')$ is the distance between the mean subsequences of the two clusters:
\vspace{2mm}
\begin{align}
\bar{d}(C,C') = d\!\left(\frac{1}{|C.\textsf{sq}|}\!\sum_{A\in C.\textsf{sq}}\!A,\; \frac{1}{|C'.\textsf{sq}|}\!\sum_{B\in C'.\textsf{sq}}\!B\right). \label{eq:cdist}
\end{align}


\begin{algorithm}[t]
\SetKwInOut{Input}{Input}
\SetKwInOut{Output}{Output}
\caption{\AHC}
\label{alg:ahc-core}
\SetAlgoLined
\Input{Training time series $\trainTS$ and adjacency $k$}
\Output{Set of clusters $\TClustering$}

$\mathcal{C} \leftarrow \emptyset;$\\
\ForEach{$j \in [1,N]$}{
    $C_j.\textsf{sq} \gets \{T_j\};\;\; C_j.\textsf{mb} \gets \emptyset;\;\; C_j.\textsf{nb} \gets \emptyset;$ \label{ln:ahc-singleton}\\
    \ForEach{$i \in [1,k]$}{
        \If{$j-i \ge 1$}{
            $C_j.\textsf{nb} \gets C_j.\textsf{nb} \cup \{C_{j-i}\};$ \label{ln:ahc-neighstart}\\
            $C_{j-i}.\textsf{nb} \gets C_{j-i}.\textsf{nb} \cup \{C_j\};$ \label{ln:ahc-neighend}
        }
    }
    $\mathcal{C} \gets \mathcal{C} \cup \{C_j\};$\\
}

$\mathcal{C} \gets \textsf{RecCluster}(\mathcal{C});$ \label{ln:ahc-recurse}\\
$\mathcal{C} \gets \textsf{Refine}(\textsf{Cutoff}(\mathcal{C}));$ \label{ln:ahc-cutoff-refine}\\

\KwRet $\mathcal{C};$
\end{algorithm}

\noindent \textbf{Overview.}
Algorithm~\ref{alg:ahc-core} outlines the core procedure of \AHC. 
Given a training time series $\trainTS$, \AHC outputs a final clustering $\TClustering$ capturing sets of temporally adjacent subsequences that represent recurring normal patterns. 
More generally, \AHC can be applied to any sequence of subsequences extracted from $\trainTS$ using a sliding window of length $\ell^M$ and an arbitrary hop size $h$. 

Each extracted subsequence $T_j$ initializes a leaf cluster $C_j$ whose member set $C_j.\textsf{mb}$ is empty and whose subsequence set is $C_j.\textsf{sq}=\{T_j\}$. 
The temporal neighborhood of each cluster, denoted by $C_j.\textsf{nb}$, stores its $k$ nearest temporal neighbors in the extracted subsequence stream. 
During initialization, every cluster $C_j$ is linked to the $k$ clusters immediately preceding it in this sequence (Lines~\ref{ln:ahc-neighstart}–\ref{ln:ahc-neighend}), establishing adjacency constraints that restrict subsequent merges to nearby temporal regions. 
As clusters merge, their neighborhoods are updated to reflect newly formed temporal boundaries, ensuring that adjacency-aware merging is maintained throughout the hierarchy.

The recursive routine \textsf{RecCluster} (Line~\ref{ln:ahc-recurse}) performs agglomerative clustering while restricting candidate merges to pairs of clusters appearing in each other's neighborhoods. 
The procedure produces a binary hierarchy whose leaves correspond to the initial subsequences extracted from $\trainTS$, and whose internal nodes store their two merged components in the \textsf{mb} field. 
This hierarchy forms the adjacency-constrained dendrogram used in the cutoff step that extracts the final flat clusters.

A key difference from standard hierarchical clustering is that the linkage distances $\ldist_i$ in \AHC\ are not guaranteed to be non-decreasing. 
Because merges are restricted to $k$-adjacent clusters, new neighborhoods formed after each merge may introduce previously non-adjacent pairs with smaller linkage distances than earlier merges. 
This can produce dendrograms with non-monotonic link heights—a direct consequence of the locality constraint.  
To compute the final clustering, we apply a dendrogram cutoff based on \emph{inconsistency coefficients}~\cite{hierarchical-2016,mullner2011modern} (\textsf{Cutoff}, Line~\ref{ln:ahc-cutoff-refine}). 
For the $i$-th merge with linkage distance $\ldist_i$, its inconsistency coefficient is
\vspace{2mm}
\begin{align}
\gamma_i \;=\; \frac{\ldist_i - \mu_i}{\sigma_i},    
\end{align}
where $\mu_i$ and $\sigma_i$ are the mean and standard deviation of linkage distances among the merges within a fixed depth preceding the $i$-th merge. 
Intuitively, large $\gamma_i$ values highlight merges that connect internally coherent subclusters that are relatively dissimilar from one another.

Since \AHC\ may generate non-monotonic linkage values, we impose a monotonic non-decreasing ordering of $\{\ldist_i\}$ before computing $\gamma_i$, ensuring a consistent hierarchical scale for the cutoff while preserving relative differences in linkage magnitude. 
(Alternatively, a local temporally ordered window may be used, though we adopt the monotonic variant in our experiments.) 
A threshold $\cutoffTh$ is then applied: all links with $\gamma_i > \cutoffTh$ are treated as boundaries, yielding a clustering in which each cluster corresponds to a coherent group of temporally adjacent subsequences representing a normal pattern.  
Lastly, the \textsf{Refine} step filters and deduplicates the clusters obtained from the cutoff. 
Clusters with cardinality $|C_i.\textsf{sq}| < R_\textsf{min}$ are removed, as such small regions typically reflect anomalous behavior rather than stable patterns. 
Among the remaining clusters, near-duplicates—identified as having highly similar average subsequences—are merged by retaining only the cluster with larger membership. 
The refined set $\TClustering$ contains distinct, representative clusters capturing the dominant normal temporal patterns in the training series.

\begin{algorithm}[t]
\SetKwInOut{Input}{Input}
\SetKwInOut{Output}{Output}
\caption{\textsf{RecCluster}}
\label{alg:ahc-rec-core}
\SetAlgoLined
\Input{Current set of clusters $\mathcal{C}$}
\Output{Updated clustering $\mathcal{C}$ of a complete hierarchy}
\lIf{$|\mathcal{C}| = 1$}{ \KwRet $\mathcal{C}$}
$\ldist_\textsf{min} \leftarrow \infty;$\quad 
$C^1_\textsf{min} \leftarrow \emptyset;$\quad 
$C^2_\textsf{min} \leftarrow \emptyset;$\\[2pt]
\ForEach{$C_i \in \mathcal{C}$ \textbf{\em and} $C_j \in C_i.\textsf{\em nb}$}{
    \If{$\ldist(C_i, C_j) < \ldist_\textsf{\em min}$}{
        $\ldist_\textsf{min} \leftarrow \ldist(C_i, C_j);$\\
        $C^1_\textsf{min} \leftarrow C_i;$\quad $C^2_\textsf{min} \leftarrow C_j;$ \label{ln:ahc-core-deltamin}
    }
}
$C \leftarrow \textsf{Merge}(\mathcal{C}, C^1_\textsf{min}, C^2_\textsf{min});$ \label{ln:ahc-core-
mergecall}\\
$\mathcal{C} \leftarrow (\mathcal{C} \setminus \{C^1_\textsf{min}, C^2_\textsf{min}\}) \cup \{C\};$\\
\KwRet $\textsf{RecCluster}(\mathcal{C});$ \label{ln:ahc-core-recursivecall}
\end{algorithm}

\begin{algorithm}[t]
\SetKwInOut{Input}{Input}
\SetKwInOut{Output}{Output}
\caption{\textsf{Merge}}
\label{alg:merge-core}
\SetAlgoLined
\Input{Clustering $\mathcal{C}$ and $C^1_\textsf{min}, C^2_\textsf{min} \in \mathcal{C}$ to merge}
\Output{Merged result $C$}
$C \leftarrow \textsf{NewCluster}()$;\\
$C.\textsf{mb} \leftarrow \{C^1_\textsf{min}, C^2_\textsf{min}\}; \;C.\textsf{sq} \leftarrow C^1_\textsf{min}.\textsf{sq} \cup C^2_\textsf{min}.\textsf{sq};$\\
$C.\textsf{nb} \leftarrow (C^1_\textsf{min}.\textsf{nb} \cup C^2_\textsf{min}.\textsf{nb}) 
       \setminus \{C^1_\textsf{min}, C^2_\textsf{min}\};$ \label{ln:merge-nbstart}\\
\ForEach{$C' \in (C^1_\textsf{min}.\textsf{\em nb} \cup C^2_\textsf{min}.\textsf{\em nb})$}{
    $C'.\textsf{nb} \leftarrow (C'.\textsf{nb} \setminus \{C^1_\textsf{min}, C^2_\textsf{min}\}) \cup \{C\};$ \label{ln:merge-nbend}
}
\KwRet $C;$
\end{algorithm}

\eat{
Algorithm~\ref{alg:ahc-rec-core} defines the recursive procedure that drives the core version of \AHC. 
The algorithm operates on the current set of clusters $\mathcal{C}$ and recursively merges the closest pair of clusters that are within each other's temporal neighborhoods. 
At each step, the pair $(C^1_\textsf{min}, C^2_\textsf{min})$ with the smallest linkage distance $\ldist(C^1_\textsf{min}, C^2_\textsf{min})$ among all adjacent pairs is selected for merging. 
The merge operation (Algorithm~\ref{alg:merge-core}) creates a new cluster $C$ that stores its children in $C.\textsf{mb}$ and inherits the union of their neighborhoods (excluding the merged pair itself). 
Any cluster that previously listed either child now lists $C$ as its neighbor. 
The process continues recursively until all clusters have been merged into a single root cluster, producing a complete binary hierarchy constrained by temporal adjacency.
}

Algorithm~\ref{alg:ahc-rec-core} recursively merges the pair of neighboring clusters with minimum linkage distance; clusters are visited in temporal order, so ties are broken in favor of the earliest pair. Given clusters $C^1_\textsf{min}$ and $C^2_\textsf{min}$, the merge operation creates a new cluster $C_\textsf{new}$ whose subsequences are $C^1_\textsf{min}.\textsf{sq}\cup C^2_\textsf{min}.\textsf{sq}$, whose children are stored in $C_\textsf{new}.\textsf{mb}$, and which records its linkage $\ldist_\textsf{min}$ as the height of the merge used by the cutoff. The neighborhood of $C_\textsf{new}$ is initialized as the union of the neighborhoods of the merged clusters, excluding the clusters themselves, and every neighbor $X$ of $C_\textsf{new}$ replaces $C^1_\textsf{min}$ and $C^2_\textsf{min}$ in $X.\textsf{nb}$ by $C_\textsf{new}$. The process repeats recursively until a single root cluster remains, producing an adjacency-constrained binary hierarchy.

\begin{figure}[t]
    \centering
    \includegraphics[width=0.9\linewidth]{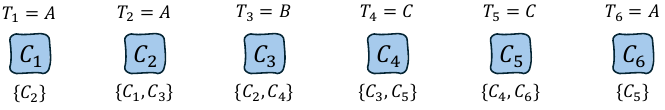}
    \caption{\AHC's initialization}
    \label{fig:AHC_ex1}
    \vspace{3mm}
    \includegraphics[width=0.9\linewidth]{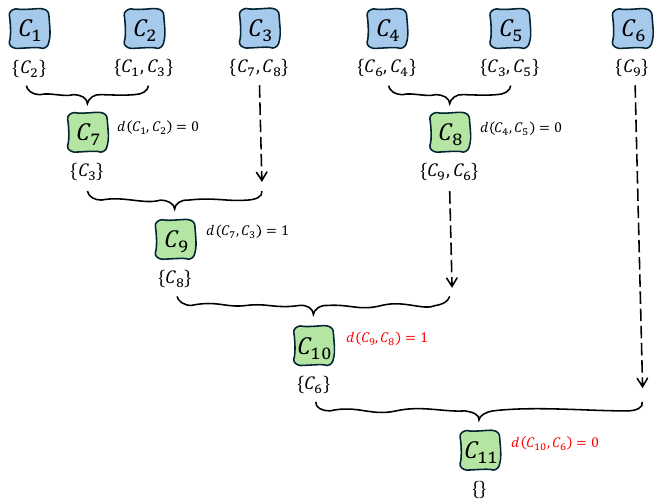}
    \caption{\AHC's resulting dendrogram}
    \label{fig:placeholder}
\end{figure}

\begin{example} \label{ex:ahc}
Figure~\ref{fig:AHC_ex1} shows a short time series composed of six adjacent subsequences extracted using a sliding window of length $\ell^M$ with hop size $h=\ell^M$, resulting in non-overlapping segments:
$T_1=T_{1,\ell}=A$, $T_2=T_{\ell+1,\ell}=A$, $T_3=T_{2\times \ell+1,\ell}=B$, $T_4=T_{3\times \ell+1,\ell}=C$, $T_5=T_{4\times \ell+1,\ell}=C$, and $T_6=T_{5\times \ell+1,\ell}=A$. We chose $h=\ell$ to simplify our running example; other hop sizes can be used without affecting the core algorithm. We explore other hop sizes in our experiments. We assume that the extracted subsequences are \emph{instances} of a small set of underlying patterns, denoted by $A$, $B$, and $C$. Instances of the same pattern are highly similar to one another, whereas instances of different patterns have larger pairwise distances.

We assume $T_1,T_2,$ and $T_6$ are instances of a pattern $A$, while $T_3$ follows a another distinct pattern $B$, and $T_4$ and $T_5$ are of a different pattern $C$ where the distinct patterns $A$, $B$, and $C$ have equal pairwise distances
$d(A,B)=d(A,C)=d(B,C)=1$ (subsequence distance $d$ as defined in Eq.~\ref{eq:d-zm}).
For simplicity, let the linkage function be the minimal linkage distance and $k=1$. Initially, $\mathcal{C}=\{C_1,\dots,C_6\}$, where each cluster $C_j$ contains a single subsequence ($C_j.\textsf{mb}=\{T_j\}$), and maintains adjacency through its neighborhood field (e.g., $C_1.\textsf{nb}=\{C_2\}$, $C_2.\textsf{nb}=\{C_1,C_3\}$, etc.).

For $k=1$, \AHC\ iteratively merges the most similar adjacent clusters. 
The first merges are $(C_1,C_2)\!\rightarrow\!C_7$ and $(C_4,C_5)\!\rightarrow\!C_8$, 
followed by $(C_7,C_3)\!\rightarrow\!C_9$, then $(C_9,C_8)\!\rightarrow\!C_{10}$, 
and finally $(C_{10},C_6)\!\rightarrow\!C_{11}$. Because new temporal neighborhoods are created after each merge, the linkage distance can \emph{decrease} in later merges. For instance, the merge $(C_{10},C_6)\!\rightarrow\!C_{11}$ occurs with a smaller linkage distance than the preceding merge $(C_9,C_8)\!\rightarrow\!C_{10}$, illustrating the non-monotonic behavior of linkage values under adjacency constraints. 

To handle such cases, \AHC\ applies a \emph{monotonic reordering} step before computing inconsistency coefficients. This step ignores temporary decreases in linkage distance and restores a consistent hierarchical order. In the reordered sequence, the main jumps in linkage distance occur at merges $(C_7,C_3)$ and $(C_9,C_8)$, which define the effective cluster boundaries. Assuming $R_\textsf{min}=2$ (clusters must contain at least two subsequences), the resulting flat clustering is $\mathcal{C}=\{\{T_1, T_2\}, \{T_4, T_5\}\}$, while the isolated clusters $\{T_3\}$ and $\{T_6\}$ are filtered out during the \textsf{Refine} step as anomalous or rare subsequences.
\end{example}

\noindent \textbf{Runtime Analysis of the Core \AHC.}
Unlike conventional hierarchical clustering, which considers all $O(N^2)$ cluster pairs, \AHC\ restricts comparisons to $k$-adjacent clusters, where $N$ is the number of extracted subsequences. As long as neighborhoods contain $O(k)$ clusters, each step of Algorithm~\ref{alg:ahc-rec-core} examines $O(kN)$ candidate pairs instead of $O(N^2)$, so the $N-1$ merges take $O(kN^2)$ time compared with $O(N^3)$ for the same procedure without adjacency constraints, and only $O(kN)$ linkages are maintained instead of $O(N^2)$. The parameter $k$ controls the tradeoff between efficiency and completeness: larger values consider more candidate merges and produce a hierarchy closer to standard hierarchical clustering, while smaller values improve efficiency by enforcing stronger locality constraints. Section~\ref{sec:ahc-rollback} introduces a rollback mechanism that mitigates the limitations of small $k$ values. As is standard in clustering analysis, distance and linkage computations are treated as unit-cost operations; the internal cost of a single distance computation is discussed in Section~\ref{sec:runtime-analysis}.

\subsection{Rollback Optimization}
\label{sec:ahc-rollback}

As discussed earlier in Section~\ref{sec:ahc}, restricting merges to $k$-adjacent clusters introduces a trade-off: very small $k$ improves efficiency but may temporarily block desirable merges between clusters that are close in time but not yet adjacent. The \emph{rollback optimization} addresses this limitation by allowing \AHC\ to operate with small $k$ while still recovering missed, near-temporal merges once they are detectable during recursion.

\begin{example} \label{ex:ahx-rec-justify}
From Example~\ref{ex:ahc} with $k{=}1$, $T_3$ and $T_6$ were left unmerged. The subsequences $T_1$, $T_2$, and $T_6$ all correspond to the same underlying pattern (A), and merging all three would form a larger, more coherent normal cluster. Without this merge, $T_6$ would incorrectly appear as an isolated anomaly. A larger $k$ (e.g., $k{=}4$) would place $C_6$ in the neighborhood of $C_2$ and merge $\{T_1,T_2,T_6\}$, preventing this false anomaly. However, such a large $k$ substantially increases comparisons. Rollback enables this merge without increasing $k$. \end{example}

Rollback alternates between a forward merge pass and a corrective step that undoes the merges that prevented a missed merge, when evidence of such a merge is discovered. The forward pass follows the core \AHC\ procedure, iteratively merging $k$-adjacent clusters using the current neighborhood structure. The corrective step uses both the value-based distance $d$ (and its induced linkage) and the temporal distance $\delta$ (Section~\ref{sec:preliminaries}). A temporal threshold $\delta_\textsf{max}$ limits how far apart two subsequences may be for a missed merge to be recovered; it is a parameter of \AHC\ (Table~\ref{tbl:param}).

\begin{algorithm}[t]
\SetKwInOut{Input}{Input}
\SetKwInOut{Output}{Output}
\caption{\textsf{RecClusterRollback}}
\label{alg:ahc-rec}
\SetAlgoLined
\Input{Clusters $\mathcal{C}$ and threshold $\ldist$ ($-\infty$ in the initial call)}
\Output{Clustering $\mathcal{C}$}
\lIf{$|\mathcal{C}| = 1$ \label{ln:rec-check}}{\KwRet $\mathcal{C}$}

$\ldist_\textsf{min} \leftarrow \infty;$\\
\ForEach{$C_i \in \mathcal{C}$ and $C_j \in C_i.\textsf{nb}$}{
    \If{$\ldist(C_i, C_j) < \ldist_\textsf{min}$}{
        $\ldist_\textsf{min} \leftarrow \ldist(C_i, C_j); C^1_\textsf{min} \leftarrow C_i; C^2_\textsf{min} \leftarrow C_j;$ \label{ln:rec-findmin}
    }
}
\If{$\ldist_\textsf{min} < \ldist$ \label{ln:d-cond}}{
    $(T^1_\textsf{min},T^2_\textsf{min}) \leftarrow \textsf{WitnessPair}(C^1_\textsf{min}, C^2_\textsf{min});$\\
    \If{$\delta(T^1_\textsf{min},T^2_\textsf{min}) \le \delta_\textsf{max}$ \textbf{\em and} $T^1_\textsf{min},T^2_\textsf{min}$ \textbf{\em not adjacent} \label{ln:delta-cond}}{
        \ForEach{$T \in \{T^1_\textsf{min},T^2_\textsf{min}\}$}{
            \While{$|\textsf{ClusterOf}(\mathcal{C},T).\textsf{sq}| > 1$ \label{ln:split}}{$\textsf{Split}(\mathcal{C},\textsf{ClusterOf}(\mathcal{C},T));$}
        }
        $C^1 \leftarrow \textsf{ClusterOf}(\mathcal{C},T^1_\textsf{min});$\\ $C^2 \leftarrow \textsf{ClusterOf}(\mathcal{C},T^2_\textsf{min});$\\
        $C^1.\textsf{nb}\leftarrow C^1.\textsf{nb}\cup\{C^2\};\ C^2.\textsf{nb}\leftarrow C^2.\textsf{nb}\cup\{C^1\};$ \label{ln:expand1}\\
        \KwRet $\textsf{RecClusterRollback}(\mathcal{C}, -\infty)$ \label{ln:infty}
    }
}
$C_\textsf{new} \leftarrow \textsf{Merge}(\mathcal{C}, C^1_\textsf{min}, C^2_\textsf{min});$ \label{ln:rec-merge}\\
$\mathcal{C} \leftarrow (\mathcal{C}\setminus\{C^1_\textsf{min},C^2_\textsf{min}\}) \cup \{C_\textsf{new}\};$ \label{ln:rec-update}\\
\KwRet $\textsf{RecClusterRollback}(\mathcal{C}, \ldist_\textsf{min})$ \label{ln:rec-call}
\end{algorithm}

\eat{
\begin{algorithm}[t]
\SetKwInOut{Input}{Input}
\caption{\textsf{Split}}
\label{alg:revise}
\SetAlgoLined
\Input{Clustering $\mathcal{C}$ and merged cluster $C$}
\ForEach{$C_i \in C.\textsf{mb}$ and $C' \in C_i.\textsf{nb}$}{
    $C'.\textsf{nb} \leftarrow (C'.\textsf{nb}\setminus\{C\}) \cup \{C_i\};$
}
$\mathcal{C} \leftarrow (\mathcal{C}\setminus\{C\}) \cup C.\textsf{mb};$ \label{ln:revise-final}
\end{algorithm}
}

Algorithm~\ref{alg:ahc-rec} extends the core \AHC\ with rollback detection. When rollback is enabled, \AHC\ calls $\textsf{RecClusterRollback}(\TClustering, -\infty)$ instead of $\textsf{RecCluster}(\TClustering)$ in Line~\ref{ln:ahc-recurse} of Algorithm~\ref{alg:ahc-core}. As in the core algorithm, the nearest neighboring pair is first identified (Lines~\ref{ln:rec-check}--\ref{ln:rec-findmin}). Rollback is triggered when the current minimum linkage $\ldist_{\textsf{min}}$ is smaller than the previously accepted linkage $\ldist$ (Line~\ref{ln:d-cond}). The algorithm then identifies a \emph{witness pair} $(T^1_{\textsf{min}},T^2_{\textsf{min}})$, i.e., the pair of subsequences, one from each cluster, with the smallest distance $d$, with ties broken by the smallest temporal distance $\delta$. For single linkage, this pair realizes $\ldist_{\textsf{min}}$; for Ward linkage, which is defined by cluster centroids rather than by a single pair, it identifies where the two clusters are closest. If the temporal distance of the witness pair is at most $\delta_{\max}$ and the two subsequences are not already adjacent, i.e., they are more than $k$ positions apart and not linked by an earlier rollback (Line~\ref{ln:delta-cond}), rollback is performed; otherwise clustering proceeds normally.

Rollback undoes exactly the merges that absorbed the witness subsequences. For each of them, $\textsf{Split}$ replaces the cluster containing it by its two children in $C.\textsf{mb}$ until the subsequence is a singleton again (Line~\ref{ln:split}); merges that contain neither witness are kept. $\textsf{Split}$ also restores neighborhoods, so that two clusters are neighbors iff they contain subsequences at most $k$ positions apart or linked by rollback. The two singletons are then linked as neighbors (Line~\ref{ln:expand1}), and clustering resumes with threshold $-\infty$ (Line~\ref{ln:infty}), since the merge that set the previous threshold may have been undone. Requiring that the witnesses are not already adjacent guarantees progress: otherwise, linking them would change nothing and the same rollback would repeat.
Rollback enables \AHC\ to operate with very small adjacency widths while still recovering meaningful near-temporal merges that would otherwise be missed. We evaluate the tradeoff between rollback frequency, runtime, and accuracy for varying $k$ in Section~\ref{sec:param}.

\begin{figure}
    \centering
    \includegraphics[width=0.9\linewidth]{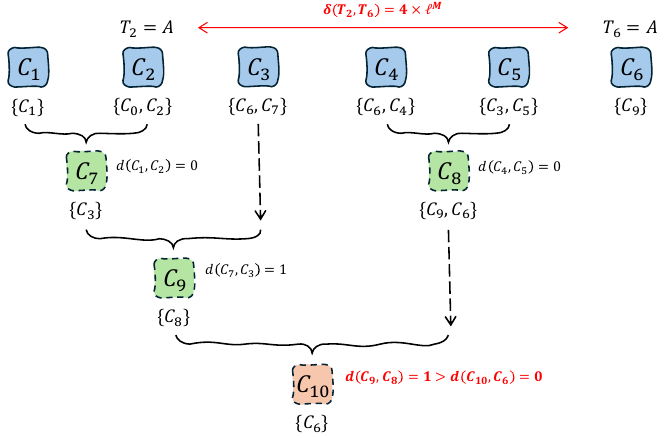}
    \caption{\AHC's first rollback.}
    \label{fig:rollback1}
    \vspace{3mm}
    \includegraphics[width=0.9\linewidth]{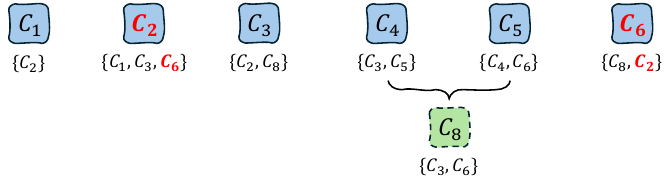}
    \caption{\AHC's second rollback.}
    \label{fig:rollback2}
    \vspace{3mm}
    \includegraphics[width=0.9\linewidth]{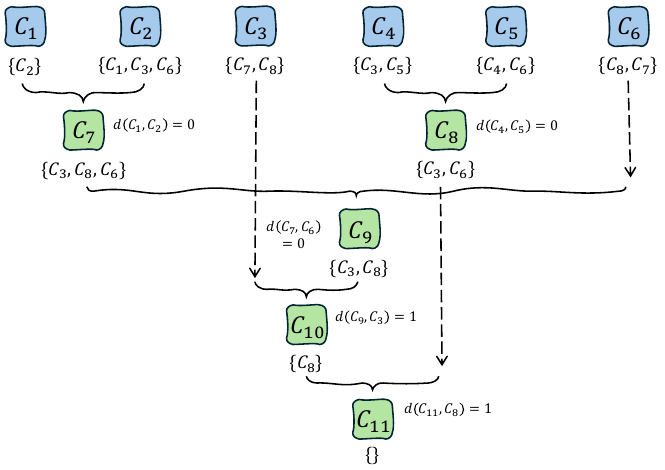}
    \caption{\AHC's second forward pass and final dendrogram.}
    \label{fig:rollback3}
\end{figure}

\begin{example} We revisit the setup of Example~\ref{ex:ahc}, now running \AHC\ with rollback enabled and $k{=}1$. We set $\delta_{\textsf{max}} = 4 \times \ell^M$, which is large enough for rollback to detect that the subsequences $T_2$ and $T_6$ (originating in clusters $C_2$ and $C_6$) are close in time and should be merged earlier. These are exactly the closest subsequences identified by \textsf{WitnessPair}; responsible for driving the rollback process.

Figure~\ref{fig:rollback1} shows the state immediately before rollback is triggered. The newly discovered linkage $\ldist(C_{10},C_6)=0$ is smaller than the previously accepted linkage $\ldist(C_9,C_8)=1$. This indicates that two highly similar clusters, $C_2$ (in $C_{10}$) and $C_6$, were not considered earlier because the forward pass was restricted to $k{=}1$ adjacency. The subsequences responsible for this small linkage—$T_2$ (from $C_2$) and $T_6$ (from $C_6$)—are found by \textsf{WitnessPair}, and their temporal distance $\delta(T_2, T_6) = 4 \times \ell^M \le \delta_{\textsf{max}}$ satisfies the rollback condition (Line~\ref{ln:delta-cond}). Thus, rollback is triggered at Line~\ref{ln:triggered}.


Figure~\ref{fig:rollback2} illustrates the rollback sequence. Recent merges are undone via \textsf{Split} (Line~\ref{ln:split}) until the algorithm reaches merges that are relevant to the missed subsequences $T_2$ and $T_6$. Clusters that do not contain either subsequence are skipped using the relevance test in Line~\ref{ln:irrelevant-test}, avoiding unnecessary splits. This backward traversal eventually reaches the two singleton clusters $C_2=\{T_2\}$ and $C_6=\{T_6\}$ whose merge should have occurred earlier.

Figure~\ref{fig:rollback3} shows the point where rollback \emph{terminates}. When both recovered clusters contain exactly one subsequence (Line~\ref{ln:rollback-stop}), rollback stops. At this moment, the algorithm expands their neighborhoods symmetrically (Lines~\ref{ln:expand1}--\ref{ln:expand2}) so that $C_2$ and $C_6$ now appear in each other's adjacency lists. The linkage threshold is reset to $\infty$ (Line~\ref{ln:infty}), and the next iteration of the \textsf{while} loop resumes the forward pass with this updated neighborhood structure.

This expanded adjacency allows the forward pass to merge $C_2$ and $C_6$ immediately, producing the correct higher-level cluster containing all three subsequences of the same pattern $\{T_1,\; T_2,\; T_6\}$. The final clustering (also shown in Figure~\ref{fig:rollback3}) contains this full normal-pattern cluster, and the cluster $\{T_4, T_5\}$, while avoiding the earlier false anomaly on $T_6$.\end{example}

\noindent \textbf{Runtime Analysis of \AHC\ with Rollback.}
Each rollback links two subsequences that are at most $\delta_\textsf{max}$ apart and were not adjacent before, so at most $N\lceil\delta_\textsf{max}/h\rceil$ rollbacks occur and \AHC\ always terminates. A rollback undoes only the merges that absorbed the two witnesses, at most $2(N-1)$, and redoing them costs the same as in the core algorithm, i.e., $O(kN^2)$ in the worst case. In practice, rollbacks are infrequent, and Section~\ref{sec:param} quantifies their overhead, which allows much smaller values of $k$.
\eat{
In terms of runtime, the rollback optimization adds only a bounded overhead to the core adjacency-constrained recursion.
While the base \AHC\ performs $O(kN)$ distance computations, rollbacks may temporarily undo and recompute merges.
In the worst case, a rollback can revisit a sequence of merges along a single path in the dendrogram, yielding an additional linear factor in the number of clusters.
Thus, the overall complexity remains $O(kN)$ on average, and $O(kN^2)$ in the pathological worst case.
In practice, however, rollbacks are infrequent and the observed cost remains close to the near-linear behavior of the core algorithm, while using much smaller $k$ values to achieve comparable clustering quality at significantly lower cost.
}

\section{Drift and Anomaly Co-Detection}\label{sec:normals}

This section presents details of how \soln\ manages normal patterns, and uses them to jointly detect drift and anomalies.  We begin with our approach for \emph{offline} subsequence anomaly detection as a simpler case than the online setting, and use it to introduce the notions of active/inactive normal patterns and anomaly scores, which are shared and reused in the online detection.
We first describe the offline setting (Section~\ref{sec:offline}), where the full \AHC\ segmentation is available. 
Here, overlapping cluster spans define the set of active patterns at each time, and each subsequence is either matched to one of these patterns or marked as anomalous. 
We then present the \emph{online} setting (Section~\ref{sec:online}), where pattern spans are unknown and must be inferred incrementally. 
In this setting, \soln\ maintains and updates the active set, assigns subsequences to their most plausible matching pattern, and triggers the creation of new normal patterns when none of the existing ones sufficiently explain recent behaviour. In Section~\ref{sec:online}, we describe how active patterns are maintained, how matching patterns are determined in the online setting, and how new patterns and anomalies are detected.



\begin{figure}
    \centering
    \includegraphics[width=0.8\linewidth]{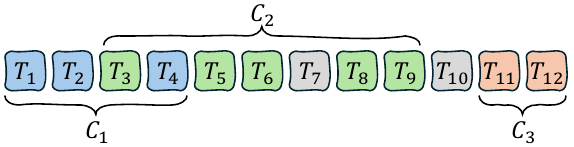}
    \caption{Offline anomaly and drift detection.}
    \label{fig:offline}
\end{figure}

\subsection{Offline Detection}\label{sec:offline}

In the offline setting, the entire time series $T$ is available, and we apply \AHC\ once over all subsequences of $T$. This produces a segmentation of $T$ into a set of \emph{clusters}, each representing a instance of a normal pattern occurring over a specific temporal region. 
Let $\TClustering = \{C_1, \ldots, C_{|\TClustering|}\}$ be the set of clusters returned by \AHC. Each cluster $C_r \in \TClustering$ contains a set of member subsequences $C_r.\text{sq} = \{T_{r_1}, \dots \}$. 
We define a temporal {\em scope} for $C_r$ by $\mathit{start}_r = \min\{ j : T_j \in C_r.\text{sq} \}$ and $ 
\mathit{end}_r = \max\{ j : T_j \in C_r.\text{sq} \}$. Thus, each $C_r$ describes and is an instance of a recurring pattern occurring between indices $\mathit{start}_r$ and $\mathit{end}_r$.

Different clusters may represent the same underlying pattern. To identify these, we group clusters whose average subsequences (centroids) are sufficiently similar. Given a set of clusters $\TClustering=\{C_1, \ldots, C_{|\TClustering|}\}$, we extract a normal model $M$ consisting of \emph{normal patterns}, with a one-to-many relationship between normal patterns and clusters: each cluster is an instance of exactly one normal pattern, and a normal pattern may be represented via multiple (cluster) instances.  For each normal pattern $\pattern^M_i$, let $\TClustering^i \subseteq \TClustering$ denote its associated set of clusters. The combined temporal footprint of $\pattern^M_i$ is the union of the spans of $\TClustering^i$.

We assume anomaly and drift detection is performed for each subsequence $T_j \in T, j \in [1,N]$. While deciding anomaly and drift for $T_j$, a normal pattern $\pattern^M_i$ is \emph{active} if there exists a cluster $C_r \in \TClustering^i$ such that $\mathit{start}_r \le j \le \mathit{end}_r$. If no such instance exists, $\pattern^M_i$ is \emph{inactive} while processing $T_j$. This definition captures the fact that different normal patterns may overlap in time, based solely on where their clusters occur in $T$. We use $\activeM^j \subseteq M$ to denote the set of normal patterns active for $T_j$.

Given the clustering $\TClustering$ of $T$, each subsequence $T_j$ in $T$ either belongs to exactly one cluster or to none. If $T_j \in C_r.\text{sq}$ for some cluster $C_r \in \mathcal{C}^i$, then $\pattern^M_i$ is the \emph{matching pattern} for $T_j$. If $T_j$ does not belong to any cluster, it has no matching pattern. Trivially, a matching pattern, if it exists, is one of the active patterns when processing $T_j$.

\begin{example} \label{ex:offline} Figure~\ref{fig:offline} illustrates a time series with subsequences
$T_1,\ldots,T_{12}$.
The AHC segmentation produces clusters $\mathcal{C}=\{C_1,C_2,C_3\}$ and a normal
model $M$ with three normal patterns
$\pattern^M_1,\pattern^M_2,\pattern^M_3$, corresponding to $C_1,C_2,C_3$,
respectively.
Subsequences belonging to each cluster (and its pattern) are shown in blue,
green, and orange, while gray subsequences do not belong to any cluster or
pattern. Pattern activity varies over time.
For $T_1$ and $T_2$, only $\pattern^M_1$ is active.
For $T_3$ and $T_4$, the active set is
$\{\pattern^M_1,\pattern^M_2\}$.
At $T_{10}$, the active set is empty. Matching patterns are indicated by color:
$T_1,T_2,$ and $T_4$ match $\pattern^M_1$;
$T_3,T_5,$ and $T_6$ match $\pattern^M_2$;
$T_{11}$ and $T_{12}$ match $\pattern^M_3$.
Subsequences $T_7$ and $T_{10}$ have no matching pattern.
\end{example}

\noindent \textbf{Anomaly Scores in the Offline Setting.} For subsequences with a matching pattern, we define their anomaly score using the distance to the matching pattern $s_j = \dist(T_j, \pattern^M_i)$ if $T_j$ matches $\pattern^M_i$. Note that $\pattern^M_i$ by definition is a subsequence itself. If $T_j$ does not belong to any cluster (does not have a matching pattern), but lies in a temporal scope of one or more pattern (one or more normal patterns are active), we assign the lowest possible distance among active patterns $s_j = \min_{\pattern^M_i \in \activeM^j} \; \dist(T_j, \pattern^M_i)$ where $\activeM^j$ denote the set of normal patterns active at index $j$. Finally, if $j$ lies outside the temporal footprint of all normal patterns (i.e., all patterns are inactive at $j$, i.e., $\activeM^j = \emptyset$ and trivually there wont be a matching pattern),  $s_j$ is assigned the maximum possible anomaly score signaling an anomaly. 

\begin{example} Continuing Example~\ref{ex:offline}, all subsequences except $T_7$ and $T_{10}$ are scored against their matching normal patterns. For $T_7$, which has no matching pattern but occurs while $\pattern^M_2$ is active, the score is computed as its distance to $\pattern^M_2$. For $T_{10}$, where no normal pattern is active, we assign the maximum possible
anomaly score. We also considered assigning the same maximum score to $T_7$ by ignoring the
active pattern; however, this uniformly harsher treatment led to lower detection accuracy in practice.\end{example}




\noindent \textbf{Offline Drift Detection.} Concept drift corresponds to changes in normal pattern  activity over time. \eat{Let $\activeM^j$ denote the set of normal patterns active at index $j$.} 
A drift event occurs whenever $\activeM^j \neq \activeM^{j-1}$. Thus, offline anomaly detection and drift detection both follow directly from the \AHC-derived segmentation: matching patterns provide pointwise explanations and anomaly scores, while changes in the active set characterize the evolution of normality over time.

\begin{example} Drifts occur when the set of active patterns changes. Specifically, a drift occurs from $T_2$ to $T_3$ when $\pattern^M_2$ becomes active. Further drifts occur from $T_4$ to $T_5$ and from $T_9$ to $T_{10}$, when $\pattern^M_1$ and $\pattern^M_2$ are deactivated, respectively. These transitions capture gradual shifts between the normal patterns $\pattern^M_1$ and $\pattern^M_2$. Finally, an abrupt drift occurs from $T_{10}$ to $T_{11}$, where $\pattern^M_3$ is activated following the deactivation of $\pattern^M_2$, marking a direct transition from $\pattern^M_2$ to $\pattern^M_3$. \end{example}

\setlength{\textfloatsep}{6pt}
\begin{algorithm}[t]
\SetKwInOut{Input}{Input}
\SetKwInOut{Output}{Output}
\caption{\detectP}
\label{alg:anomaly}
\SetAlgoLined

\Input{Time series $T$ and normal model $M$}
\Output{Anomaly scores $S=[s_1,\ldots,s_N]$}

\ForEach{$j \in [1,N]$}{
    $\mathcal{W} \gets [\max(1, j{-}W{+}1), j]$\; \label{ln:det-window}
    \lForEach{$(\pattern^M_i, \threshold^M_i, \freq^M_i) \in M$}{$m_{j,i} \gets \pmember(\seq,\pattern^M_i)$ \label{ln:det-membership}}
    $\activeM \gets \{\pattern^M_i \in M \mid \avgOf([m_{t,i}]_{t \in \mathcal{W}}) \ge \freq^M_i\}$\; \label{ln:det-active}
    \If{$\activeM = \emptyset$ \label{ln:det-trigger}}{
        $M' \gets \textsf{UpdateNormalModel}(\{T_t \mid t \in \mathcal{W}\}, M)$\; \label{ln:det-update}
        \ForEach{$(\pattern^M_i, \threshold^M_i, \freq^M_i) \in M' \setminus M$ \label{ln:det-backfill}}{
            \lForEach{$t \in \mathcal{W}$}{$m_{t,i} \gets \pmember(T_t,\pattern^M_i)$}
        }
        $M \gets M'$\;
        $\activeM \gets \{\pattern^M_i \in M \mid \avgOf([m_{t,i}]_{t \in \mathcal{W}}) \ge \freq^M_i\}$\; \label{ln:det-reactive}
    }
    $s_j \gets s_\textsf{max}$\; \label{ln:det-init-score}
    \lIf{$\activeM \neq \emptyset$}{$s_j \gets \min_{\pattern^M_i \in \activeM} \dist(\seq,\pattern^M_i)$ \label{ln:det-score}}
}
\KwRet $[s_1,\ldots,s_N]$\;
\end{algorithm}

\subsection{Online Detection}\label{sec:online}

In the online setting, subsequences $T_1,\ldots,T_N$ arrive sequentially as they are extracted from the stream $T$ using a sliding window. 
Unlike the offline case—where \AHC\ is applied once to the entire time series to directly produce clusters and normal patterns—the online setting must incrementally infer (i) which known patterns are currently active, (ii) which pattern best matches the current subsequence, (iii) whether the behaviour indicates drift, and (iv) whether a subsequence should be scored as anomalous, using only information available up to the current time step.

\soln\ operates online using two tightly coupled but conceptually distinct procedures:
(1) \emph{online detection}, which assigns anomaly scores based on a current normal model; and
(2) \emph{normal model update}, which augments this model when new normal behaviour is detected.
These procedures are formalized in Algorithms~\ref{alg:anomaly} and~\ref{alg:updateModel}, respectively.

Crucially, online detection assumes the availability of an \emph{initial normal model} $M$, extracted offline from a representative training time series.
This training data may correspond to a separate historical period or to an initial prefix of the same time series later processed online.
While $M$ serves as the starting point for detection, \soln\ does \emph{not} assume that this model remains fixed:
Algorithm~\ref{alg:updateModel} allows $M$ to be incrementally updated during online operation to account for drift and the emergence of new normal patterns.

We first describe the online detection procedure (Algorithm~\ref{alg:anomaly}), and then explain how and when the normal model is updated (Algorithm~\ref{alg:updateModel}).

\begin{figure}[t]
    \centering
    \includegraphics[width=0.99\linewidth]{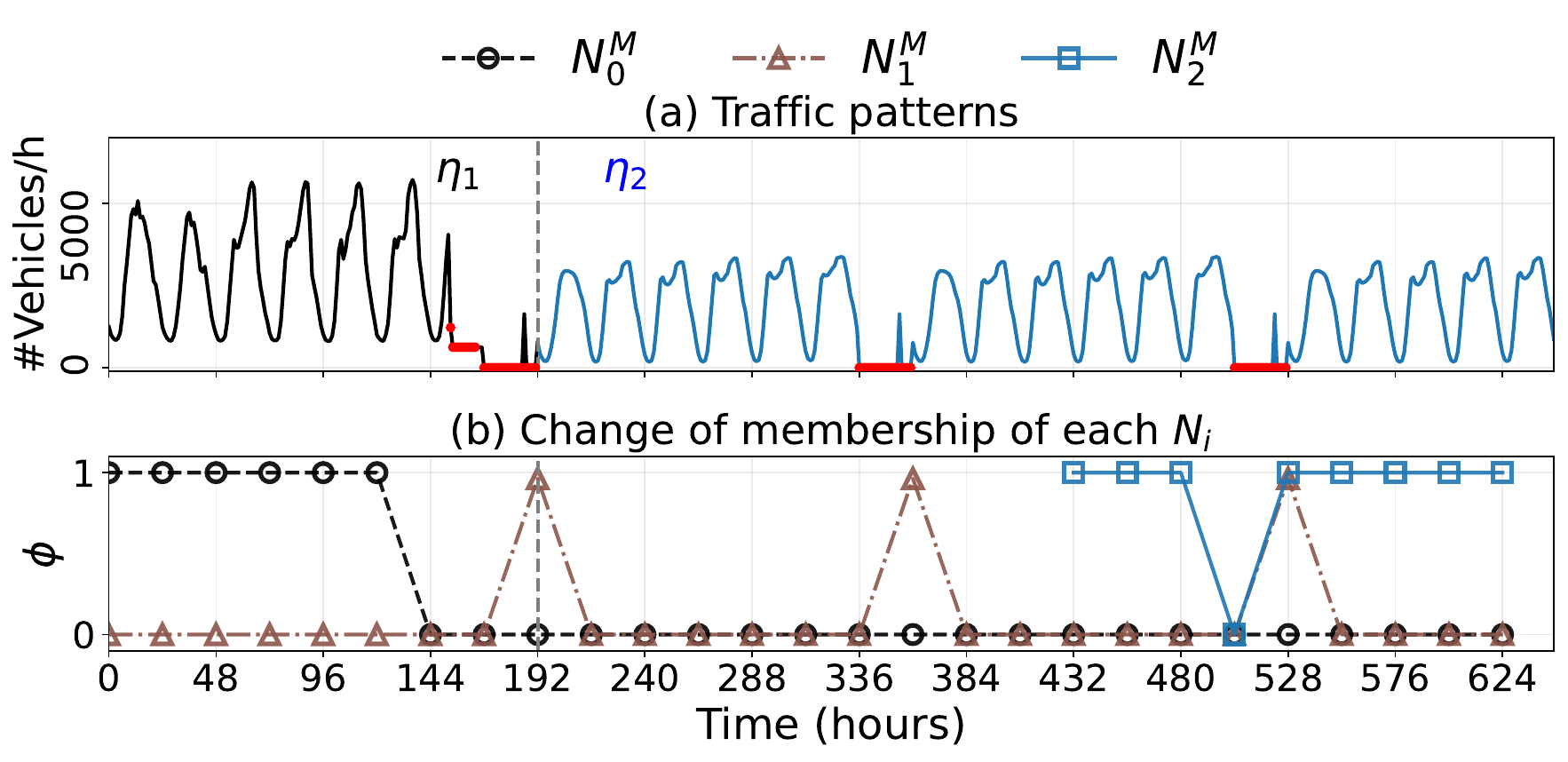}
    \caption{Active vs. inactive pattern activation. \eat{Keep as per Example 6 below?}} 
    \label{fig:new_drift}
\end{figure}

\paragraph{Maintaining Active Patterns}
Let $M=\{\pattern^M_1,\ldots,\pattern^M_m\}$ be the current normal model, where each pattern $\pattern^M_i$ is associated with a distance threshold $\threshold^M_i$ and a frequency threshold $\freq^M_i$. For each incoming subsequence $\seq$, Algorithm~\ref{alg:anomaly} computes its membership to every pattern, $m_{j,i}=\phi(\seq,\pattern^M_i)$, where $\phi$ is defined in Eq.~\ref{eq:membership} (Line~\ref{ln:det-membership}).
To capture recent behavior, \soln\ maintains a currency window $\mathcal{W}$ of the last $W$ subsequences (Line~\ref{ln:det-window}). A pattern $\pattern^M_i$ is considered active if its average membership over the window is at least $\freq^M_i$, and inactive otherwise. Alg.~\ref{alg:anomaly} recomputes the active set $\activeM$ at each step (Line~\ref{ln:det-active}), so a previously inactive pattern is reactivated as soon as its average membership again reaches $\freq^M_i$.

\eat{
\paragraph{Normal Model and Maintaining Active/Inactive Patterns} Let $M = \{\pattern^M_1,\ldots,\pattern^M_m\}$ denote the current normal model, where each pattern $\pattern^M_i$ is associated with a distance threshold $\threshold^M_i$ and a minimum frequency $\freq^M_i$. For each incoming subsequence $\seq$, Algorithm~\ref{alg:anomaly} computes its membership to every pattern in $M$ (Line~\ref{ln:det-membership}) $m_{j,i} = \phi(\seq,\pattern^M_i)$ where $\phi$ is the membership function defined in Eq.~\ref{eq:membership}. Higher membership values indicate stronger similarity between $\seq$ and $\pattern^M_i$.

To capture short-term temporal context, \soln\ maintains a recency window $W$ of the most recent subsequences. For each pattern $\pattern^M_i$, Algorithm~\ref{alg:anomaly} tracks the average membership of that pattern over the window ending at time $j$, denoted $\emph{\avgOf}([m_{k,i}]^{j}_{k=j-W})$. A pattern is considered \emph{active} and is added to the set of active patterns $\activeM$ if and only if $\emph{\avgOf}([m_{k,i}]^{j}_{k=j-W}) \ge \freq^M_i$.

This condition is checked incrementally for each pattern in Lines~\ref{ln:det-membership}–\ref{ln:det-active} of Algorithm~\ref{alg:anomaly}. Patterns whose average membership falls below their frequency threshold for varying $j$ are removed from the active set, while patterns that satisfy the condition are added. As a result, previously inactive patterns may be reactivated if their membership rises, naturally supporting recurring behaviour and gradual drift.}

\begin{example}
\eat{[MOVE to Section 5.2 as example?] After reading this, it is more of an example showing how the membership function works, rather than a comparative or scalability experiment that shows how many new or different normal patterns we identify.  Do we have an experiment that can show new or different normal patterns found over time for 2 or 3 datasets?  If not, I suggest moving this to the main body of the paper as an example when we introduce $\phi$.} \eat{\andri can find new patterns several times per data (depending on data), but we have no labels for the pattern chages. Therefore, even if we claim that \andri can find new patterns, we cannot put the results as \emph{accuracy} here (also, we don't have accuracy comparison, due to no baselines report it). }
Figure~\ref{fig:new_drift}(a) shows real traffic flow data from the California highway recordings where two patterns $\eta_1$ and $\eta_2$ occur before and after a drift~\cite{california-traffic}.  Figure~\ref{fig:new_drift}(b) shows the corresponding membership $\pmember$ scores that reflect activation of normal patterns $N^M_i$.   When $\eta_1$ occurs,  \eat{\fei{label the patterns $\eta_1, \eta_2$}} $\phi_0$ for $N^M_0$ (black) shows it's currently active. 
In contrast, $N^M_1$ is inactive (with low $\phi_1$ score), but shows intermittent, high $\phi_1$ scores reflecting similarity to anomalies at $t = \{192, 350, 528\}$, but continues to be inactive.  Following a window $W$ of continuous low $\pmember$, \andri triggers \AHC and identifies a new normal pattern $N^M_2$ (blue) at $t = 432$.  
\end{example}

\paragraph{Online Anomaly Scoring} After updating the active set $\activeM$ while processing the subsequence $T_j$, Algorithm~\ref{alg:anomaly} computes an anomaly score for $T_j$. If at least one pattern is active, the \emph{matching pattern} is defined as the active pattern with minimum distance to $\seq$:%
\vspace{2mm}
\begin{align}
\pattern^M_q
= \arg\min_{\pattern^M_i \in \activeM^j}
\dist(\seq,\pattern^M_i).
\end{align}
\vspace{-2mm}

The anomaly score is then set to $s_j = \dist(\seq,\pattern^M_q)$
(Line~\ref{ln:det-score}).This mirrors offline scoring, but only over currently active patterns. If $\activeM=\emptyset$, no existing pattern explains the recent behavior, and $\seq$ is assigned the maximum anomaly score $s_\textsf{max}$ (Line~\ref{ln:det-init-score}).



\paragraph{Normal Model Update and Drift Detection} Updating the normal model is handled by Algorithm~\ref{alg:updateModel}, which applies \AHC\ to a window $T^W$ of recent subsequences to extract candidate normal patterns. Separating model update from detection allows this procedure to be triggered under multiple conditions, rather than being hard-coded into anomaly scoring.

One natural trigger is the case where $\activeM = \emptyset$, indicating that none of the existing patterns explain the recent window. However, other triggers are also possible, such as: (i) sustained high anomaly scores over several time steps, (ii) a sharp increase in anomaly frequency, or (iii) periodic or scheduled model refresh. Any such policy may invoke Algorithm~\ref{alg:updateModel}.

Algorithm~\ref{alg:updateModel} first trains a local normal model $M^W$ on the window $T^W$
(Line~\ref{ln:update-train}). Each candidate pattern $\pattern^W_i$ is then compared against existing patterns in $M$ (Lines~\ref{ln:update-compare-start}–\ref{ln:update-compare-end}). A candidate is accepted only if it is sufficiently distinct from all existing patterns and occurs frequently enough. Accepted patterns are added to $M$ (Line~\ref{ln:update-add}) and may immediately become active in subsequent detection steps.

\begin{figure}
    \centering
    \includegraphics[width=0.85\linewidth]{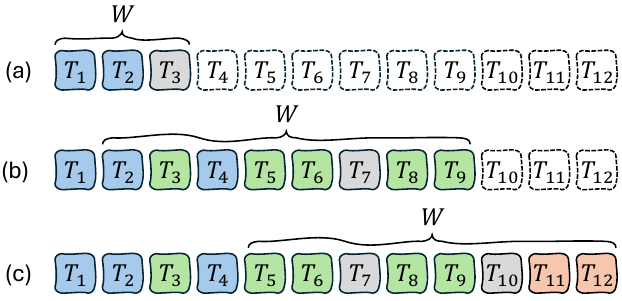}
    \caption{Online with a recency window $W$.}
    \label{fig:online}
\end{figure}

\begin{example}
Figure~\ref{fig:online} illustrates \soln\ in the online setting across three phases of the stream.
We assume that the initial normal model $M$ contains two patterns, $\pattern^M_1$ and $\pattern^M_2$
(shown in blue and green, respectively, as in Example~\ref{ex:offline}).
A third pattern, $\pattern^M_3$ (orange), is not present in the initial normal model—for example,
because this behaviour was absent in the training data.

In (a), after receiving $T_1,T_2,$ and $T_3$, the recency window activates only $\pattern^M_1$,
while $\pattern^M_2$ remains inactive.
In (b), after observing $T_1,\ldots,T_9$, both $\pattern^M_1$ and $\pattern^M_2$ become active due to
sustained membership in the window.
In (c), multiple recent anomalies ($T_7$ and $T_{10}$) trigger a model update.
Applying \AHC\ produces two clusters: one corresponding to the existing pattern $\pattern^M_1$,
and another forming a new pattern $\pattern^M_3$, which is added to $M$ and subsequently activated.

Due to the dynamics of pattern activation in the online setting, subsequences
$T_3, T_7, T_{10},$ and $T_{11}$ are reported with high anomaly scores.
In particular, $T_3$ and $T_{11}$ receive high scores because normal patterns require time to become active.
For instance, $\pattern^M_2$ is not yet active when processing $T_3$ and only becomes active around $T_5$.
As a result, $T_3$ is compared solely against the active pattern $\pattern^M_1$, which is dissimilar,
leading to a high anomaly score.
Similarly, the newly discovered pattern $\pattern^M_3$ becomes active only after observing both
$T_{11}$ and $T_{12}$.
Consequently, $T_{11}$ is scored against an incomplete active set, resulting in a high anomaly score.

These effects are inherent to real-time processing.
We discuss next how allowing a small reporting delay mitigates such transient false positives.
\end{example}

By decoupling online detection from normal model update, \soln\ supports flexible and extensible drift-handling strategies. Algorithm~\ref{alg:anomaly} focuses on incremental scoring using the current model, while Algorithm~\ref{alg:updateModel} governs how and when new normal patterns are incorporated. This design enables unified online anomaly detection and drift adaptation while remaining consistent with the offline formulation.

\eat{
\setlength{\textfloatsep}{6pt}
\begin{algorithm}[t]
\SetKwInOut{Input}{Input}
\SetKwInOut{Output}{Output}
\caption{\detectP}
\label{alg:anomaly}
\SetAlgoLined

\Input{Time series $T$ and normal model $M$}
\Output{Anomaly scores $S=[s_1,\ldots,s_N]$}

\ForEach{$j \in [1,N]$}{
    $\activeM \gets \emptyset$\; \label{ln:det-init}
    \ForEach{$(\pattern^M_i, \threshold^M_i, \freq^M_i) \in M$}{
        $m_{j,i} \gets \pmember(\seq,\pattern^M_i)$\; \label{ln:det-membership}

        \If{$\emph{\avgOf}([m_{k,i}]^{j}_{k=j-W}) < \freq^M_i$}{
            $\activeM \gets \activeM \setminus \{\pattern^M_i\}$\; \label{ln:det-active-check}
        }
        \Else{
            $\activeM \gets \activeM \cup \{\pattern^M_i\}$\; \label{ln:det-active-update}
        }
    }
    $s_j \gets \infty$\; \label{ln:det-init-score}
    \If{$\activeM \neq \emptyset$}{ \label{ln:det-has-active}
        $\pattern^M_q \gets 
        \operatorname*{arg\,min}_{\pattern^M_q \in \activeM} \dist(\seq,\pattern^M_q)$\; \label{ln:det-best-pattern}
        $s_j \gets \dist(\seq,\pattern^M_q)$\; \label{ln:det-score}
    }
}
\KwRet $[s_1,\ldots,s_N]$\;
\end{algorithm}
}

\setlength{\textfloatsep}{6pt}
\begin{algorithm}[t]
\SetKwInOut{Input}{Input}
\SetKwInOut{Output}{Output}
\caption{\textsf{UpdateNormalModel}}
\label{alg:updateModel}
\SetAlgoLined

\Input{Time window $T^W$ and normal model $M$}
\Output{Updated normal model $M$}

$M^W \gets \trainP(T^W)$\; \label{ln:update-train}
\ForEach{$(\pattern^W_i,\threshold^W_i,\freq^W_i) \in M^W$}{
    $w \gets \textsf{false}$; $j \gets 1$\; \label{ln:update-compare-start}
    \While{$\neg w \wedge j \le |M|$}{
        \lIf{$d(\pattern^W_i,\pattern^M_j) \leq \threshold^M_j$}{
            $w \gets \textsf{true}$
        }
        $j \gets j+1$\;
    } \label{ln:update-compare-end}
    \lIf{$\neg w$}{
        $M \gets M \cup \{(\pattern^W_i,\threshold^W_i,\freq^W_i)\}$ \label{ln:update-add}
    }
}
\KwRet $M$\;
\end{algorithm}

In the strictly online (real-time) setting, which is what described above and in Algorithm~\ref{alg:anomaly}, the score $s_j$ is computed and reported as soon as all points of $\seq$ are observed. In settings that allow delayed detection (see Section~\ref{sec:preliminaries}), the system may wait up to $\latency$ additional time points before reporting $s_j$, enabling confirmation of whether the current region reflects a concept drift (leading to the creation of a new normal pattern) or an isolated anomaly. Accordingly, \soln\ supports both real-time scoring and delayed (window-confirmed) scoring depending on latency requirements.

Importantly, incorporating latency does not require any fundamental changes to the detection algorithm. Given a delay parameter $\latency$, the anomaly score $s_j$ is computed when the window $W$ includes additional subsequences as determined by $\latency$. Active pattern discovery proceeds as before, except that it may consider a limited number of subsequences occurring after $T_j$, rather than relying solely on past subsequences. Access to these future subsequences can improve pattern selection and, in some cases, lead to better detection performance than strictly real-time execution. Pattern matching and anomaly score computation remain unchanged, operating on the resulting set of active patterns. Consequently, the core scoring logic, pattern matching procedure, and model update rules are preserved. This design treats latency as an output-level mechanism that enhances robustness to noise or transient deviations without increasing algorithmic complexity or compromising the algorithm’s online nature.

\begin{example} Consider the same stream as in the previous example, but assume that \soln\ allows a reporting delay of $\ell$ subsequences, meaning that a subsequence is reported as anomalous only after observing $\ell$ subsequent subsequences. This delay enables the algorithm to look one step ahead before finalizing anomaly decisions. With $\ell=1$, subsequences such as $T_3$ and $T_{11}$ are no longer reported as anomalies. In the case of $T_3$, the activation of $\pattern^M_2$ shortly afterward provides a better matching normal pattern, reducing its anomaly score. Similarly, for $T_{11}$, the emergence and activation of $\pattern^M_3$ after observing $T_{12}$ explains the behaviour, eliminating the false positive. More generally, changes in the active set of normal patterns—i.e., concept drift—occur between
$T_4$ and $T_5$, and again between $T_{11}$ and $T_{12}$. In the online setting, such changes are detected with an inherent delay compared to the offline case, as expected. Allowing a bounded latency $\latency$ smooths this transition, improving both anomaly precision and drift detection stability without sacrificing real-time operation.\end{example}

\subsection{Runtime Analysis}\label{sec:runtime-analysis}

Let $N$ denote the number of scored subsequences extracted from $T$, and let $|M|$ be the number of normal patterns maintained in the model. We write $c_d$ for the cost of a single distance (or membership) computation between a subsequence and a pattern (e.g., $c_d=O(\ell)$ for direct $\ell$-length distances).

\emph{Offline detection} learns the normal model once (dominated by the cost of \AHC\ on the training series, analyzed above) and then scores all $N$ subsequences against the learned model. In a straightforward implementation, scoring requires finding the closest matching pattern per subsequence, which costs $O(N\cdot |M|\cdot c_d)$ in total.

\emph{Online detection} processes the subsequence stream incrementally. For each incoming subsequence, updating memberships and selecting the closest active pattern costs $O(|M|\cdot c_d)$ per step. The frequency-based active-set maintenance can be implemented in $O(1)$ amortized time per pattern by maintaining rolling counts over the last $W$ subsequences. When the active set becomes empty, the algorithm may invoke $\trainP$ on a recent window of size $W$, incurring an additional cost $c_W$; since such retraining is triggered only when normal patterns become inactive, the overall online cost remains linear in $N$ with occasional retraining overhead. The memory footprint is $O(|M|)$ plus the bookkeeping required for the $W$-length window (or $O(|M|)$ with rolling counts).

\vspace{-1ex}
\section{Evaluation}
\vspace{-1ex}

\subsection{Experimental Setup}
\label{sec:expsetup}
We implement our algorithms using  Python 3.11 with sklearn, scipy, stumpy~\cite{law2019stumpy} libraries, on an Ubuntu server with an Intel(R) Xeon(R) w5-2455x CPU and an NVIDIA GeForce RTX 4090 GPU.  For baselines, we use the TSB-UAD repository~\cite{TSB-UAD-2022}, and code from the original authors' repository.

\begin{table} [t]
\begin{small}
\begin{center}
\renewcommand{\arraystretch}{1.1}
\caption{Parameters (defaults in bold). \label{tbl:param}}
\begin{tabular}{  l | l  | l}
      \hline
      \textbf{Parm.} & \textbf{Values} & \textbf{Description}  \\
      \hline \hline
      $\cwindow $ & 10, \textbf{20}, 30, 40, 50 ($\times \ell$) & currency window size\\
      \hline
      $\ell^M$ & \textbf{2}, 3, 4 ($\times \ell$) & normal pattern length \\
      \hline
      \sizeC & 0.5, \textbf{1}, 2, 4, 6, 8, 10 (\%$|T|$) & min cluster size \\
      \hline
      $k$ & {1}, \textbf{2}, 3, 5, 7 & \AHC $k$-neighbors  \\
      \hline
      $\delta_{\textsf{max}}$ & 10, \textbf{20}, 30, 40, 50 ($\times \ell$) & max. temporal distance\\ 
      \hline
\end{tabular}
\end{center}
\end{small}
\end{table}

\stitle{Datasets.} We use six real datasets, also used in existing work~\cite{INN-2020, OmniAnomaly-2019}.  \eat{Table~\ref{tab:datasets} summarizes their characteristics.} All our data and source code are publicly available~\cite{datasite}.

\noindent (1) \uline{\climate}~\cite{canada-climate}: Canadian climate data recording hourly precipitation levels from weather base stations.  The data contains 44k points with approximately 2.5\% representing point and sequential anomalies, and seasonal drift naturally occurring in the data.  Anomalies are labelled based on 24-hour precipitation levels exceeding regional, and seasonal daily mean and standard deviations from the last 10 to 30 years~\cite{canada-climate-normal}.  

\noindent (2) \uline{\traffic}~\cite{california-traffic}: California highway recordings of total traffic flow, average occupancy, and average speed (miles/hr) at hourly rates. The data contains 63k recordings with a 2\% anomaly rate.  Anomaly and drift labels are computed according to deviations from normal traffic patterns that include variations between weekdays vs. weekends, seasonality, and holidays.   Further details can be found in the appendix~\ref{sec:appendix}.  

\noindent (3) \uline{\nab}~\cite{NAB-2017, TSB-autoad-2025}: the \uline{N}umenta \uline{A}nomaly \uline{B}enchmark contains real data with labelled anomaly periods. We select the AWS server, and traffic streams, totalling 6k with 10\% anomaly rate.  

\noindent (4) \uline{\ucr}~\cite{UCRArchive2018}: environmental monitoring data from the UCR time series classification archive, measuring air temperature recordings with injected anomalies, and varying normal baseline fluctuations.  The data contains 12k points with 1\% anomalies. 

\noindent (5) \uline{\smd}~\cite{OmniAnomaly-2019}: the \uline{S}erver \uline{M}achine \uline{D}ataset is widely used in multivariate time series anomaly detection, containing 28 entities (machines) with 38 metrics per entity from a large Internet company with concept drift and anomalies due to workload variations and system failures.  The data size is 24k with 2\% anomaly rate. \eat{We evaluate 28-machines, each of data originally contained originally 38-dimensions.} 

\noindent (6) \uline{\cat}~\cite{catsv2, TSB-autoad-2025}: the multivariate \uline{C}ontrolled \uline{A}nomalies \uline{T}ime \uline{S}eries (v2) data records commands, external and environmental stimuli, and telemetry readings of simulated complex dynamic systems.  The data contains 240k points with 3.7\% anomaly rate on average. \\

\begin{table*}[t]
\centering
\caption{Comparative accuracy across datasets. }
\label{tab:comparison}
\scriptsize
\renewcommand{\arraystretch}{1.3}
\setlength{\tabcolsep}{3.5pt}
\begin{tabular}{lc||c|c|c|c|c|c|c|c|c|c|c|c}
\hline\hline
& \multirow{2}{*}{} 
& \multicolumn{2}{c|}{\climate} 
& \multicolumn{2}{c|}{\traffic} 
& \multicolumn{2}{c|}{\nab}
& \multicolumn{2}{c|}{\ucr}
& \multicolumn{2}{c|}{\smd} 
& \multicolumn{2}{c}{\cat} \\
\cline{3-14}
\multicolumn{2}{c||}{Method}
& \multicolumn{1}{c|}{AUC}
& \multicolumn{1}{c|}{peak-F1}
& \multicolumn{1}{c|}{AUC}
& \multicolumn{1}{c|}{peak-F1}
& \multicolumn{1}{c|}{AUC}
& \multicolumn{1}{c|}{peak-F1}
& \multicolumn{1}{c|}{AUC}
& \multicolumn{1}{c|}{peak-F1}
& \multicolumn{1}{c|}{AUC}
& \multicolumn{1}{c|}{peak-F1} 
& \multicolumn{1}{c|}{AUC}
& \multicolumn{1}{c}{peak-F1}\\
\hline\hline
\multirow{3}{*}{\rotatebox[origin=c]{90}{offline}}
& \andrioff & \textbf{0.90$\pm$0.02} & \textbf{0.29$\pm$0.05}  & \textbf{0.86$\pm$0.05} & \textbf{0.45$\pm$0.05} & \textbf{0.81$\pm$0.16}  & \textbf{0.58$\pm$0.26} & \textbf{0.93$\pm$0.06} & 0.20$\pm$0.15 & \textbf{0.89$\pm$0.11} & \textbf{0.35$\pm$0.17}  & \textbf{0.67$\pm$0.10} & 0.18$\pm$0.06\\ 
& \cabd        & 0.68$\pm$0.02 & 0.22$\pm$0.04  & 0.43$\pm$0.05 & 0.10$\pm$0.04 & 0.51$\pm$0.06 & 0.20$\pm$0.04 & 0.51$\pm$0.24 & {0.07$\pm$0.09} & 0.70$\pm$0.10 & 0.20$\pm$0.12 & 0.52$\pm$0.05 & 0.09$\pm$0.04\\ 
& \norma       & {0.86$\pm$0.02} & {0.28$\pm$0.05}  & {0.70$\pm$0.10} & {0.27$\pm$0.09} & 0.76$\pm$0.19 & {0.53$\pm$0.26} & {0.91$\pm$0.06} & \textbf{0.25$\pm$0.19} & {0.84$\pm$0.13} & {0.28$\pm$0.17}   & 0.61$\pm$0.12 & \textbf{0.19$\pm$0.07} \\
\hline
\multirow{5}{*}{\rotatebox[origin=c]{90}{online}}
& \soln  & \textbf{0.88$\pm$0.02} & \textbf{0.29$\pm$0.05} \ & \textbf{0.82$\pm$0.07} & \textbf{0.38$\pm$0.10} & \textbf{0.77$\pm$0.16} & \textbf{0.52$\pm$0.23} & \textbf{0.96$\pm$0.04} & \textbf{0.27$\pm$0.15} &  \uline{0.89$\pm$0.10} & 0.33$\pm$0.14  & \textbf{0.69$\pm$0.13} & \uline{0.24$\pm$0.12}\\
& \sand        & \uline{0.84$\pm$0.03} & 0.29$\pm$0.04 & \uline{0.75$\pm$0.06} & \uline{0.31$\pm$0.06} & 0.56$\pm$0.24 & \uline{0.34$\pm$0.16} & 0.63$\pm$0.26 & 0.05$\pm$0.07 &  0.68$\pm$0.28 & 0.25$\pm$0.13  & 0.56$\pm$0.09 & 0.10$\pm$0.04\\

& \tranad      & 0.68$\pm$0.06 & 0.27$\pm$0.05 \ & 0.31$\pm$0.06 &  0.06$\pm$0.03 & 0.55$\pm$0.10 & 0.24$\pm$0.10 & 0.66$\pm$0.11 & 0.04$\pm$0.03 & 0.84$\pm$0.15 & \textbf{0.43$\pm$0.26}  & 0.58$\pm$0.15 & 0.20$\pm$0.11 \\
& \arcus       & 0.78$\pm$0.06 & 0.23$\pm$0.05 & 0.54$\pm$0.05 & 0.06$\pm$0.02 & \uline{0.64$\pm$0.07} & 0.27$\pm$0.05 & \uline{0.75$\pm$0.08} & 0.05$\pm$0.03  &  0.69$\pm$0.08 & 0.08$\pm$0.04  & 0.52$\pm$0.01 & 0.09$\pm$0.03\\
& \divad       & 0.80$\pm$0.06 & 0.24$\pm$0.08 & 0.39$\pm$0.08 & 0.07$\pm$0.06 & 0.60$\pm$0.12 & 0.31$\pm$0.15 & 0.65$\pm$0.15 & 0.04$\pm$0.03  & 0.80$\pm$0.16 & {0.42$\pm$0.28}   & 0.60$\pm$0.09 & 0.17$\pm$0.10\\
& \meter       & 0.64$\pm$0.04 & 0.21$\pm$0.04  & 0.68$\pm$0.08 & 0.17$\pm$0.10 & 0.54$\pm$0.11 & 0.24$\pm$0.11 & 0.54$\pm$0.16 & 0.03$\pm$0.03 & 0.85$\pm$0.12 & 0.36$\pm$0.24   & 0.60$\pm$0.14 & 0.17$\pm$0.08\\
& \candi       & 0.68$\pm$0.04 & 0.14$\pm$0.03 & 0.64$\pm$0.09 & 0.23$\pm$0.13 & 0.59$\pm$0.13 & 0.32$\pm$0.16 & 0.73$\pm$0.15 & \uline{0.15$\pm$0.19}  & \textbf{0.91$\pm$0.10} & \uline{0.43$\pm$0.22}  & 0.69$\pm$0.04 & \textbf{0.34$\pm$0.14}\\
\hline\hline
\end{tabular}
\end{table*}

\stitle{Comparative Baselines.} We evaluate \soln\ against eight baselines in offline and online settings.  We compare \andrioff against the first two  baselines below, and \soln\ against the latter six baselines in online settings.

\sstab 
(1) \uline{\norma}\cite{Norma-2021}: identifies normal patterns offline using hierarchical clustering. A subsequence anomaly score is computed as the weighted sum of its distances to the normal-model centroids.  We compare against \andrioff. 

\noindent (2) \uline{\cabd}\cite{INN-2020}: an anomaly and change point detection algorithm using inverse nearest neighbour (INN) clustering to differentiate these points. We compare against \andrioff.

\noindent (3) \uline{\sand}\cite{boniol2021sand}: a $k$-shape, clustering-based method that processes $T$ in batches~\cite{k-shape-2015}, and adaptively detects anomalies in online settings. We set the batch size 5000, similar to the original method.  

\noindent (4) \uline{\tranad}\cite{TranAD-2022}: a transformer-based, deep learning model  with adversarial training to amplify small deviations, for anomaly detection over multivariate data. 

\noindent (5) \uline{\arcus}\cite{arcus-22}: online, deep anomaly detection with  adaptive model pooling to detect anomalies with a combination of models over multivariate time series.

\noindent (6) \uline{\divad}\cite{divad-2025}: multi-encoder, multivariate, anomaly detection model that learns domain invariant and domain-specific representations.  Anomalies are identified as deviations from domain-invariant encodings.

\noindent(7) \uline{\meter}\cite{METER-23}: by training a base detection model to capture central concepts,  anomalies are identified via parameter shifts of the base model. \eat{concept adaptation online anomaly detection model that dynamically generates and shifts the base model's parameters.} 

\noindent (8) \uline{\candi}\cite{CANDI-26}: using a trained, anomaly detector base model, deviations are modelled via temporal convolutions and attention  that selectively identify and adapt to potential false positives. \\

We reduce the dimensionality of the \smd\ and \cat\ datasets to minimize redundancy.  We remove constant and co-linear dimensions using a variance inflation factor (VIF)~\cite{montgomery2020introduction, vif-2022}.    We apply \soln\ (and other baseline univariate methods) to each resulting dimension, and merge anomaly scores using $l_2$-norm. We apply multivariate methods (\tranad, \arcus, \divad, \meter, and \candi) to the dimension-reduced datasets.

\stitle{Metrics and Parameters.} We measure accuracy using Area Under the Receiver Operating Characteristic Curve (AUC-ROC), and peak-F1.  Table~\ref{tbl:param} summarizes the model parameters.  
\begin{align}
\label{eq:metric}
\mathrm{AUC\text{-}ROC} &= -\int_0^1 \mathrm{TPR}(t)\,d\mathrm{FPR}(t),\\
\mathrm{Peak\text{-}F1} &= \max_{t\in[0,1]} \frac{2\,\mathrm{Precision}(t)\,\mathrm{Recall}(t)}{\mathrm{Precision}(t)+\mathrm{Recall}(t)},\\
\text{where}\; \;
\mathrm{TPR}(t)&= \frac{TP(t)}{TP(t)+FN(t)}, \notag\\
\mathrm{FPR}(t) &= \frac{FP(t)}{FP(t)+TN(t)} \notag
\end{align}



\begin{itemize}[nolistsep,leftmargin=*]
\item \uline{$W$}: defines the horizon for recurrence. We set $W=20\times \ell$ by default: small $W$ may miss drifts, while large $W$ increases cost and may reduce sensitivity. 

\item \uline{$\ell$, $\ell^M$}: $\ell$ is chosen to capture a complete pattern via sensitivity testing.  We set $\ell^M=2\times \ell$, consistent with \norma, and use $2\ell$ when longer motifs are expected.

\item \uline{\sizeC}: we set 1\% of data as the default: small \sizeC\ set   anomalies towards normal patterns, and larger \sizeC\ lead to shorter normal patterns being considered as anomalies.


\item \uline{$\delta_{\textsf{max}}$}: defines the proximity between the \textsf{WitnessPair}, to decide whether to merge non-consecutive clusters. We set $\delta_{\textsf{max}} = 20\times\ell$, similar to $W$ in \AHC.
\end{itemize}

\begin{figure*}[t]
\centering
\vspace{-2mm}
\begin{subfigure}[t]{\linewidth}
    \vspace{0pt}
    \centering
    \includegraphics[width=\linewidth]{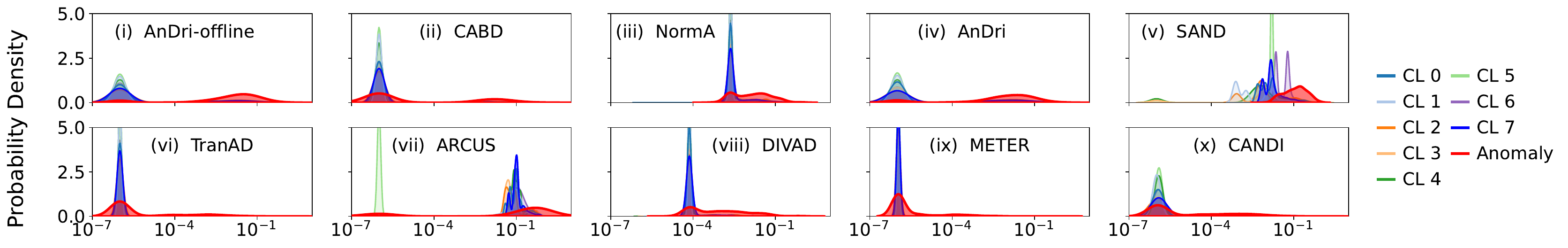}
    \vspace{-5mm}
    \caption{\climate}
    \label{fig:kde_climate}
\end{subfigure}
\begin{subfigure}[t]{\linewidth}
    \centering
    \includegraphics[width=\linewidth]{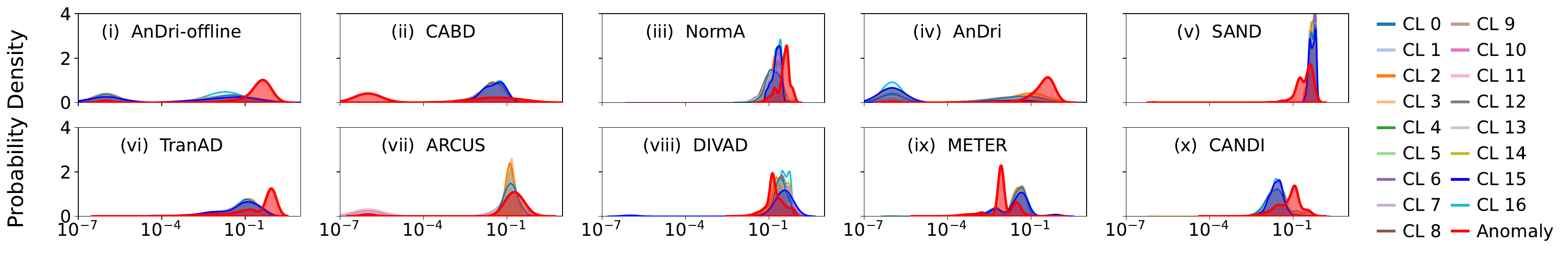}
    \vspace{-5mm}
    \caption{\traffic}
    \label{fig:kde_traffic}
\end{subfigure}
\begin{subfigure}[t]{\linewidth}
    \centering
    \includegraphics[width=\linewidth]{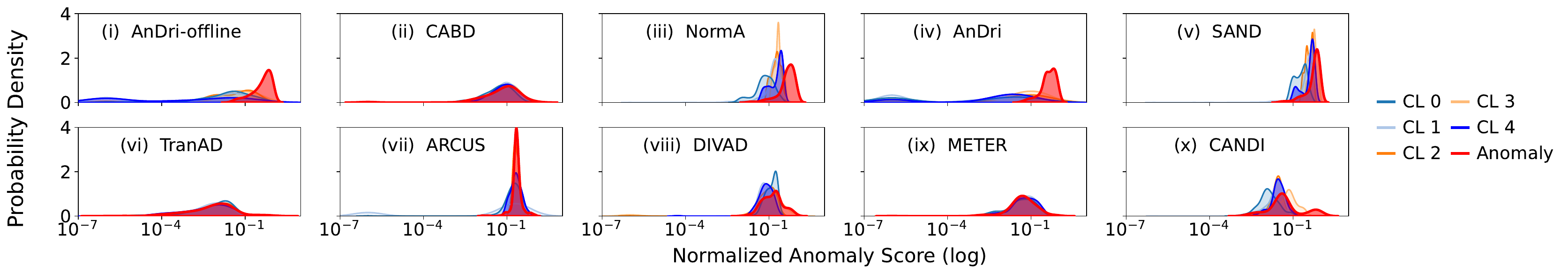}
    \vspace{-5mm}
    \caption{\nab}
    \label{fig:kde_nab}
\end{subfigure}
\caption{Comparative KDE curves of normalized anomaly scores for \AHC\ clusters vs.  anomalies.}
\label{fig:kde_score}
\vspace{-2mm}
\end{figure*}

\begin{figure}[t]
\centering
\begin{subfigure}[t]{\linewidth}
    \vspace{5pt}
    \centering
    \includegraphics[width=\linewidth]{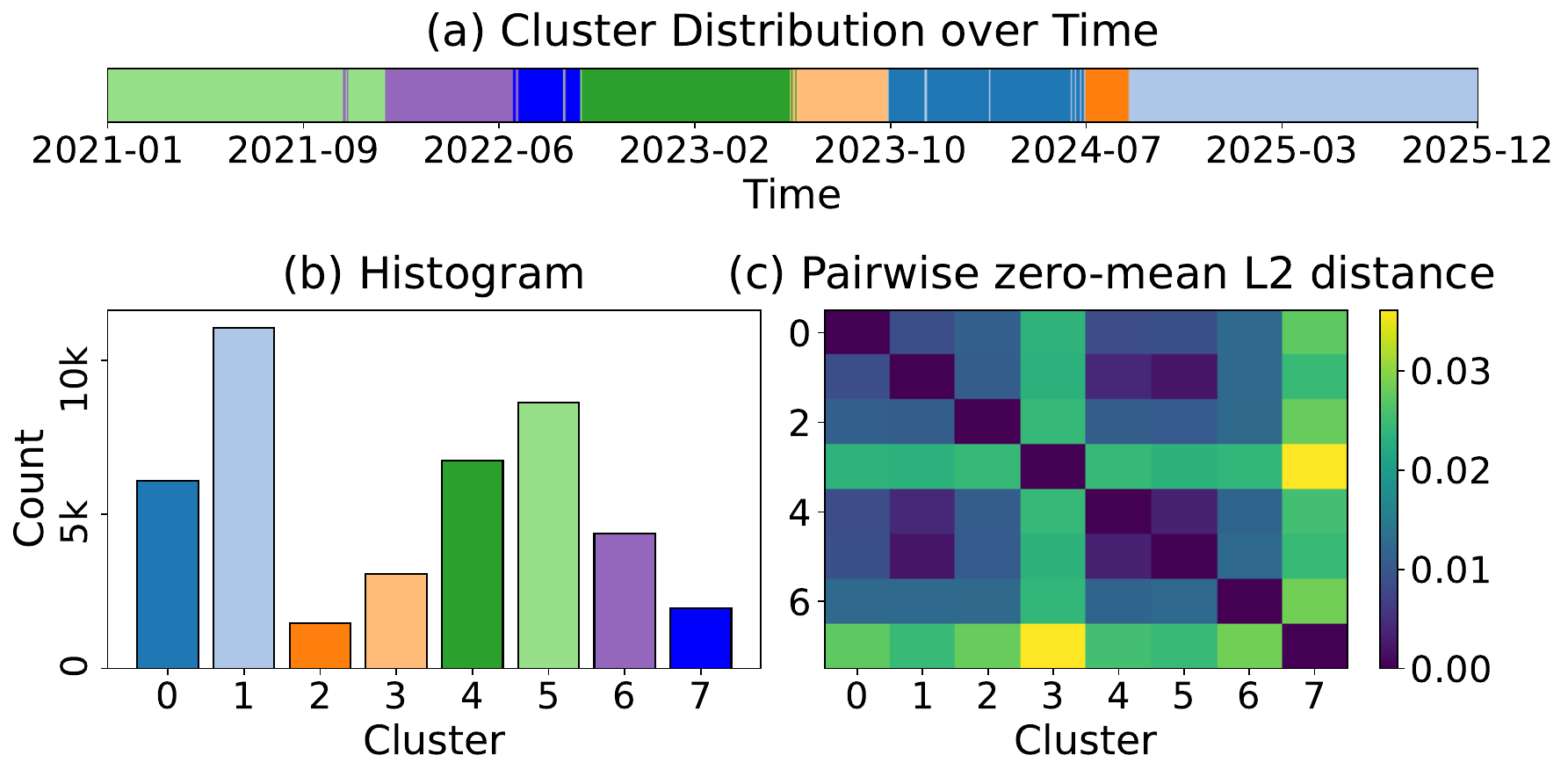}
    \vspace{-5mm}
    \caption{\climate}
    \label{fig:cls_climate}
\end{subfigure}
\begin{subfigure}[t]{\linewidth}
    \vspace{5pt}
    \centering
    \includegraphics[width=\linewidth]{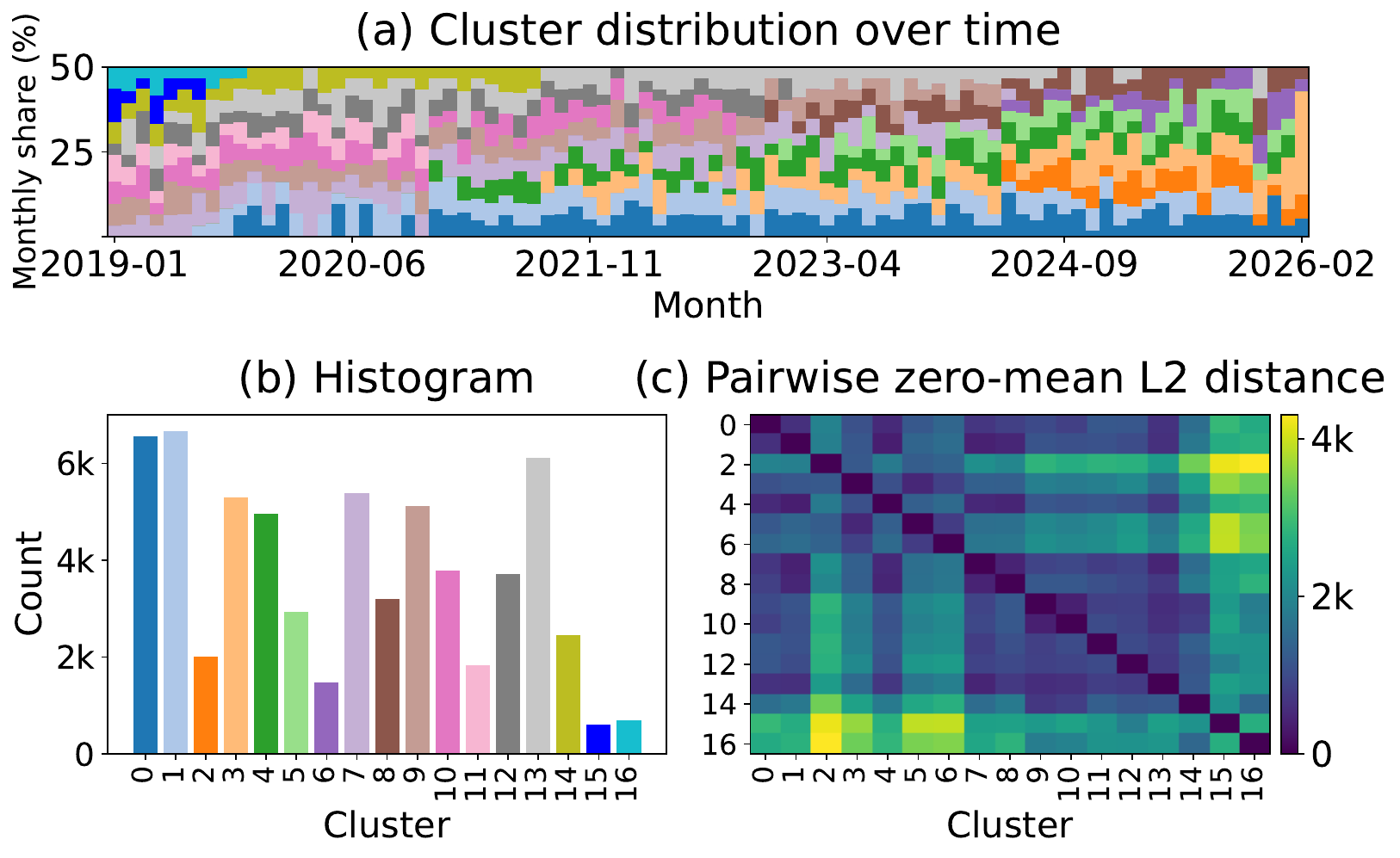}
    \vspace{-5mm}
    \caption{\traffic}
    \label{fig:cls_traffic}
\end{subfigure}
\begin{subfigure}[t]{\linewidth}
    \vspace{5pt}
    \centering
    \includegraphics[width=\linewidth]{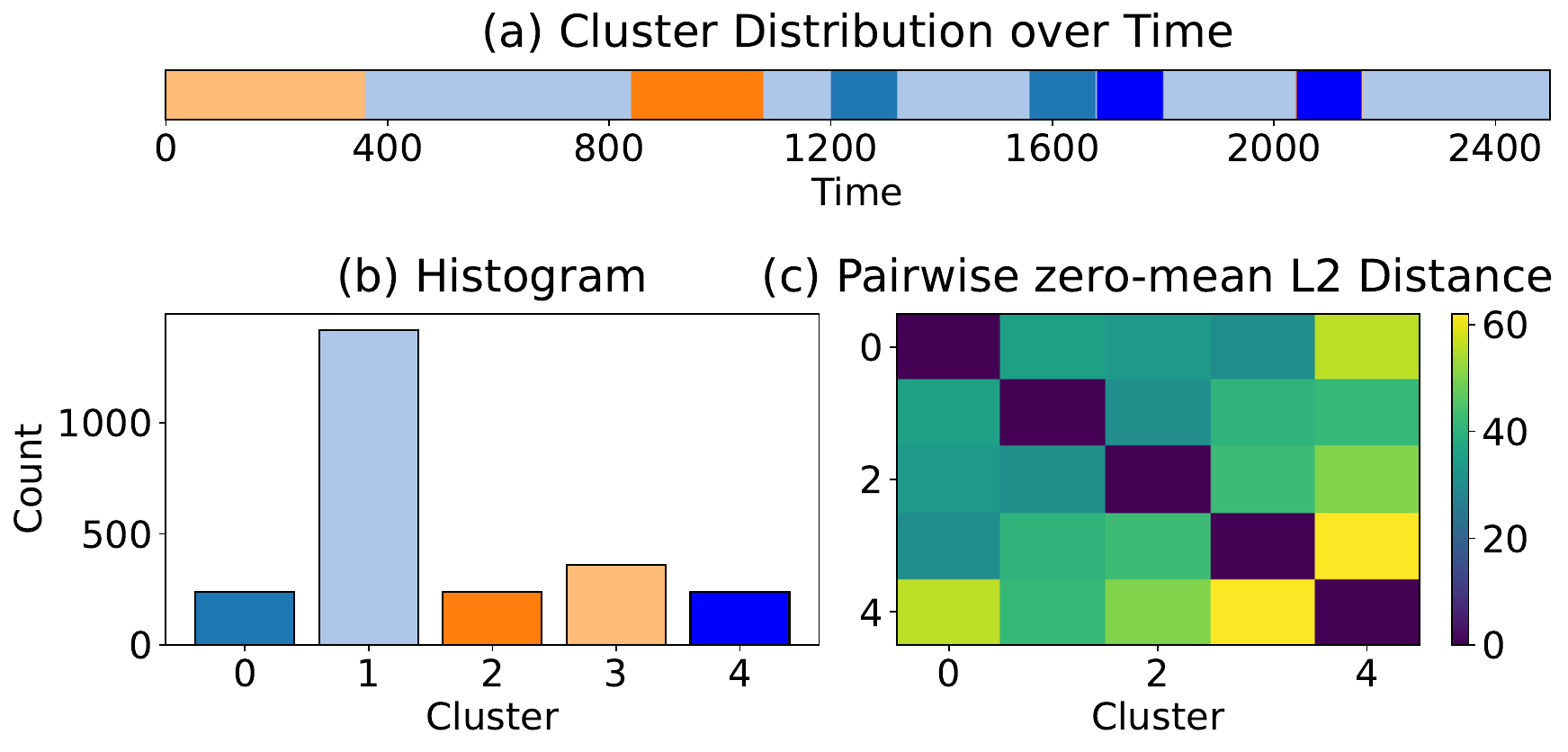}
    \vspace{-5mm}
    \caption{\nab}
    \label{fig:cls_nab}
\end{subfigure}
\captionof{figure}{Cluster distributions over time, frequency, similarity.}
\label{fig:cls_dist}
\end{figure}

\subsection{Overall Comparative Accuracy}

\noindent \textbf{Exp-1: Comparative baseline accuracy.}  Table~\ref{tab:comparison} shows the comparative AUC and peak-F1, where the best and second-best scores are shown in bold, and underline, respectively, for each class of offline vs. online methods.  \andrioff\ outperforms all baselines across all datasets between 2-55\% AUC.  Unlike existing methods that apply conventional hierarchical clustering, and merge subsequences on similarity alone~\cite{Norma-2021}, \andrioff's \AHC considers temporal proximity to prefer similar and temporally close subsequences as normal patterns.  This enables \andrioff to recognize a broader scope of normal baselines than \cabd and \norma, e.g., frequent rainfall in summer (\climate), holidays similar to weekend commuting patterns (\traffic), and seasonal changes (\nab, \ucr).  \andrioff\ outperforms in peak-F1 in the majority of datasets up to 18\%, and is comparable (within 5\%) in the \ucr and \cat dataset.

In online settings, \andri\ outperforms all baselines in both AUC (by 23\%) and peak-F1 (by 47\%) across all four univariate datasets. Given a broader set of baseline normal models (learned offline from \AHC), \andri's ability to adaptively add and refresh (active) normal patterns, allows for more fine-grained anomaly detection, and improved accuracy.  For the multivariate datasets \cat\ and \smd, \andri achieves top or extremely competitive AUC accuracy (within 2\%).  When strong cross-correlations exist between a subset of dimensions, \andri achieves lower peak-F1 scores.  We are working on developing native multivariate methods as our next step.

\noindent \textbf{Exp-2: Differentiating Anomalies from Drift.}  
Figure~\ref{fig:kde_climate} - Figure~\ref{fig:kde_nab} show the Kernel Density Estimate (KDE) plots of anomaly-score distributions for: (i) clusters identified by \AHC, and (ii) labelled anomalies, over the \climate, \traffic, and \nab\ datasets across all baselines.  Since each cluster represents a distinct normal pattern, the overlap between the anomaly distribution (shown in red) and the cluster distributions (non-red colours) indicate a baseline method's ability to differentiate anomalies from drift. A method that assigns consistently higher anomaly scores to true anomalies than to all normal clusters will exhibit well-separated KDE curves, thereby achieving high detection accuracy using a single threshold. 

One of \andri's key differentiators is its ability to identify normal patterns that occur with high frequency and similarity with high temporal locality (and not necessarily frequent over the entire time series), which are often ignored by existing baselines.  Figure~\ref{fig:cls_climate}(a) shows the temporal distribution of cluster occurrence, and their frequency histogram (Figure~\ref{fig:cls_climate}(b)), including such an exemplar normal pattern (cluster 7, dark blue) in the \climate dataset.  Cluster 7 is infrequent over the entire time series, but occurs with high temporal locality (around 2022-06), and is dissimilar from other normal patterns (as shown via the similarity matrix Figure~\ref{fig:cls_climate}(c)).  Similar normal patterns are found in the \traffic and \nab datasets, shown as Cluster 15 and Cluster 4 in blue, respectively, in Figure~\ref{fig:kde_traffic} and Figure~\ref{fig:kde_nab}.  
Similar to \climate, Figure~\ref{fig:cls_traffic} and Figure~\ref{fig:cls_nab} show the temporal distribution of cluster occurrence, their frequency histogram, and the similarity matrix of normal patterns. We highlighted Cluster 15 for \traffic\ and Cluster 4 for \nab datasets, which is temporally frequent but globally rare, and dissimilar from the other clusters, in blue. Due to the complex and frequent normality changes from weekdays, weekends, and holidays, we selected $k=5$ for \traffic, leading temporal occurrence of normal at a time can be multiple. Therefore, Figure~\ref{fig:cls_traffic}(a) shows monthly cluster occupation. 
\eat{(Temporal distribution graphs, and histograms for these datasets can be found in ~\cite{techreport}).  }
Figure~\ref{fig:kde_climate}, Figure~\ref{fig:kde_traffic}, Figure~\ref{fig:kde_nab} show that \andrioff and \andri consistently achieve the most distinctive and largest separation between the anomaly and cluster distributions; reflected in higher accuracy scores (as per Table~\ref{tab:comparison}).  The small size and distinctiveness of these clusters make them particularly challenging to differentiate from anomalies, as shown by the overlap between the blue and red distributions across all baseline KDE curves. However, \andrioff\ and \andri\ effectively separate these challenging normal clusters from anomalies.

\begin{figure}[t]
    \centering
    \includegraphics[width=1.05\linewidth]{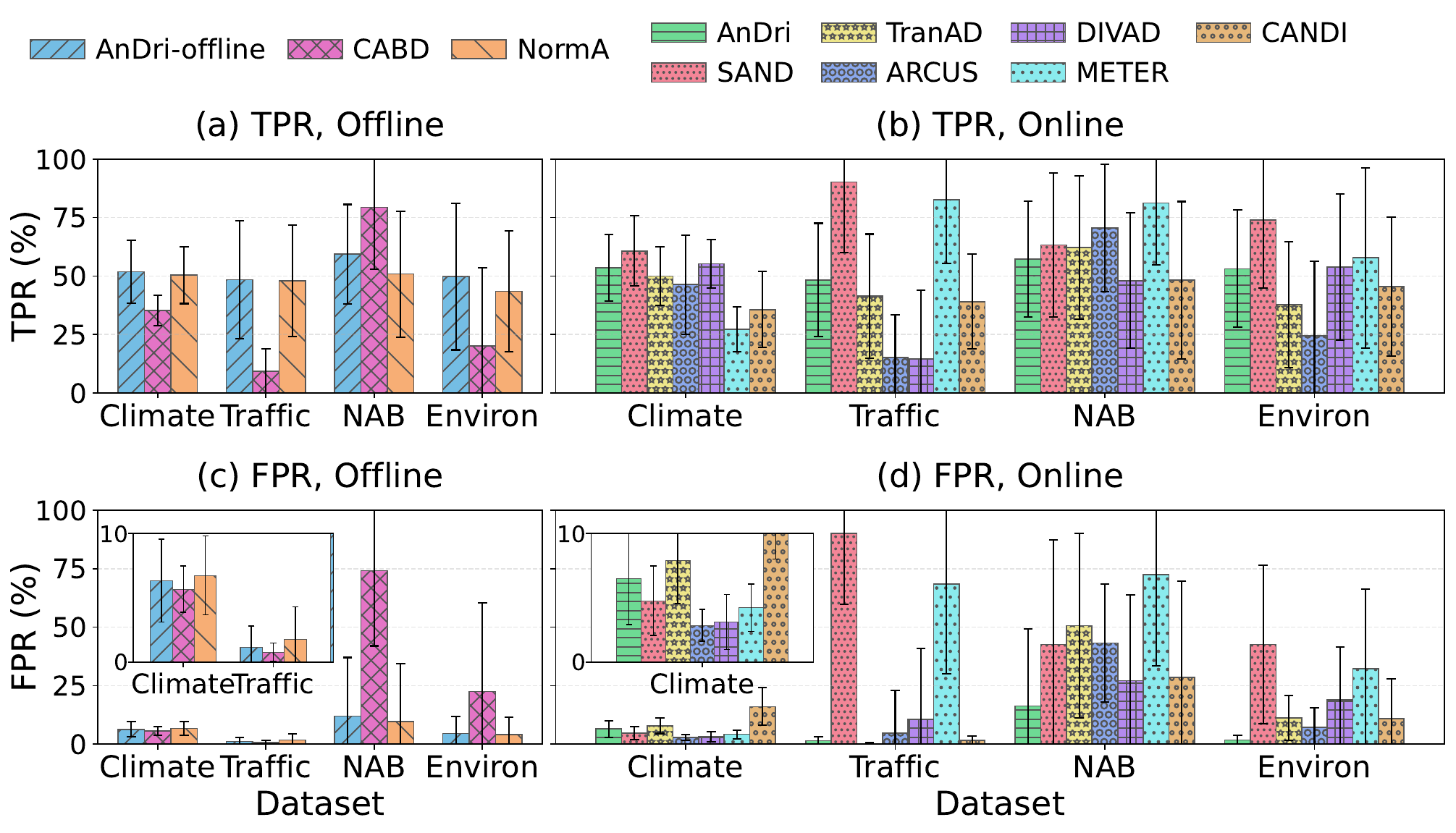} 
    \caption{\andri effectiveness during drift ($\uparrow$ TPR, $\downarrow$ FPR). \eat{\fei{The climate and traffic FPR offline comparisons are not viewable (c).  Please put x-labels on (a) and (b), you are assuming reviewers will match and align the labels with the baseline groups per dataset.}}}
    \label{fig:tpr_fpr}
\end{figure}

\subsection{Concept Drift Adaptation}

\noindent \textbf{Exp-3: Drift adaptation.}  We evaluate \andri's ability to identify anomalies during periods of drift, i.e.,  transitions from one normal pattern to another pattern.  Existing baselines often confuse such transitions as anomalous behaviour, therefore leading to an increased false positive rate (FPR).  We compute the true positive rate (TPR), and FPR over the four univariate datasets (multivariate datasets exhibit varying drift points across dimensions, and is out of scope for \andri).  Figure~\ref{fig:tpr_fpr}(a) \eat{\fei{Fill in figure numbers.}} shows that \andrioff achieves competitive TPR (7\% higher than \norma on average), and Figure~\ref{fig:tpr_fpr}(c) shows reduced FPR (by an average -33\%, and -6.5\%  than \norma), due to: (i) \AHC clustering that identifies more distinctive clusters and normal patterns based on temporal locality; (ii) computing anomaly scores against baseline normal patterns that are \emph{most similar} rather than a weighted average across all normal patterns~\cite{Norma-2021}.  In online settings, while baselines such as \divad and \candi achieve comparable TPR rates in univariate settings, they incur higher FPR rates due to missed inter-variable correlations.   \andri achieves the best balance towards high TPR and low FPR rates against all baselines and across all datasets.  \andri\ achieves comparable TPR rates to SAND, but SAND suffers from delayed response times due to weight adjustments between data chunks.  This consequently leads to mis-classification of drifts as anomalies, and higher FPR rates.  \andri balances this tradeoff with its adaptive membership function ($\phi$) to recognize active vs. inactive patterns in a more timely manner, leading to 76\% lower FPR rates on average.

\begin{figure}[t]
    \centering
    \includegraphics[width=0.99\linewidth]{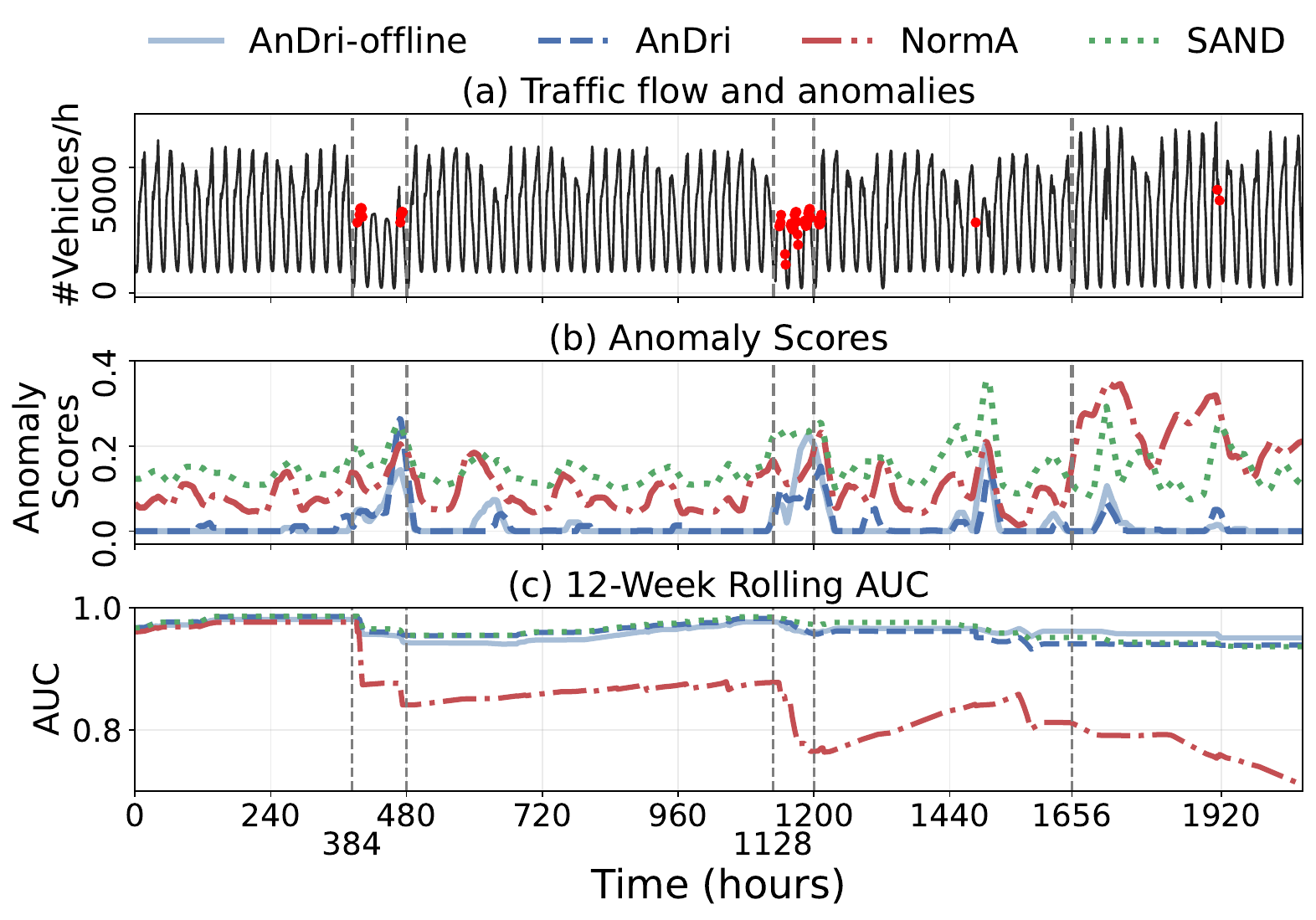}
    \caption{Anomaly scores and 12-Week rolling AUCs for a station in \traffic\ data under drifts. \eat{Suggest changing x-axis, as in Fig 14, and need to add text of $t$ in hours starting from a specific date.  Do this once starting with this figure.}}
    \label{fig:drift_traffic}
\end{figure}

Figure~\ref{fig:drift_traffic}(a) shows an example of traffic flow patterns, anomalies (shown in red), and changes in traffic flows (representing drift, bounded by dotted lines), where the x-axis represents the cumulative hours starting at $t=0$ (May 3, 2023).   Figure~\ref{fig:drift_traffic}(b) and Figure~\ref{fig:drift_traffic}(c) show the corresponding  anomaly scores, and AUC, respectively.  All methods (except \norma) adapt during the drift periods ($t \in [384, 480], [1128, 1200]$), with continually high AUC following the drift.  \norma shows increased anomaly scores during the drift, and consequently, lower AUC thereafter; likely due to its sensitivity of computing anomaly scores as a weighted sum across all normal patterns.  \sand also exhibits high anomaly scores leading to increased likelihood of false positives.

\eat{
\begin{figure}[t]
    \centering
    \includegraphics[width=0.99\linewidth]{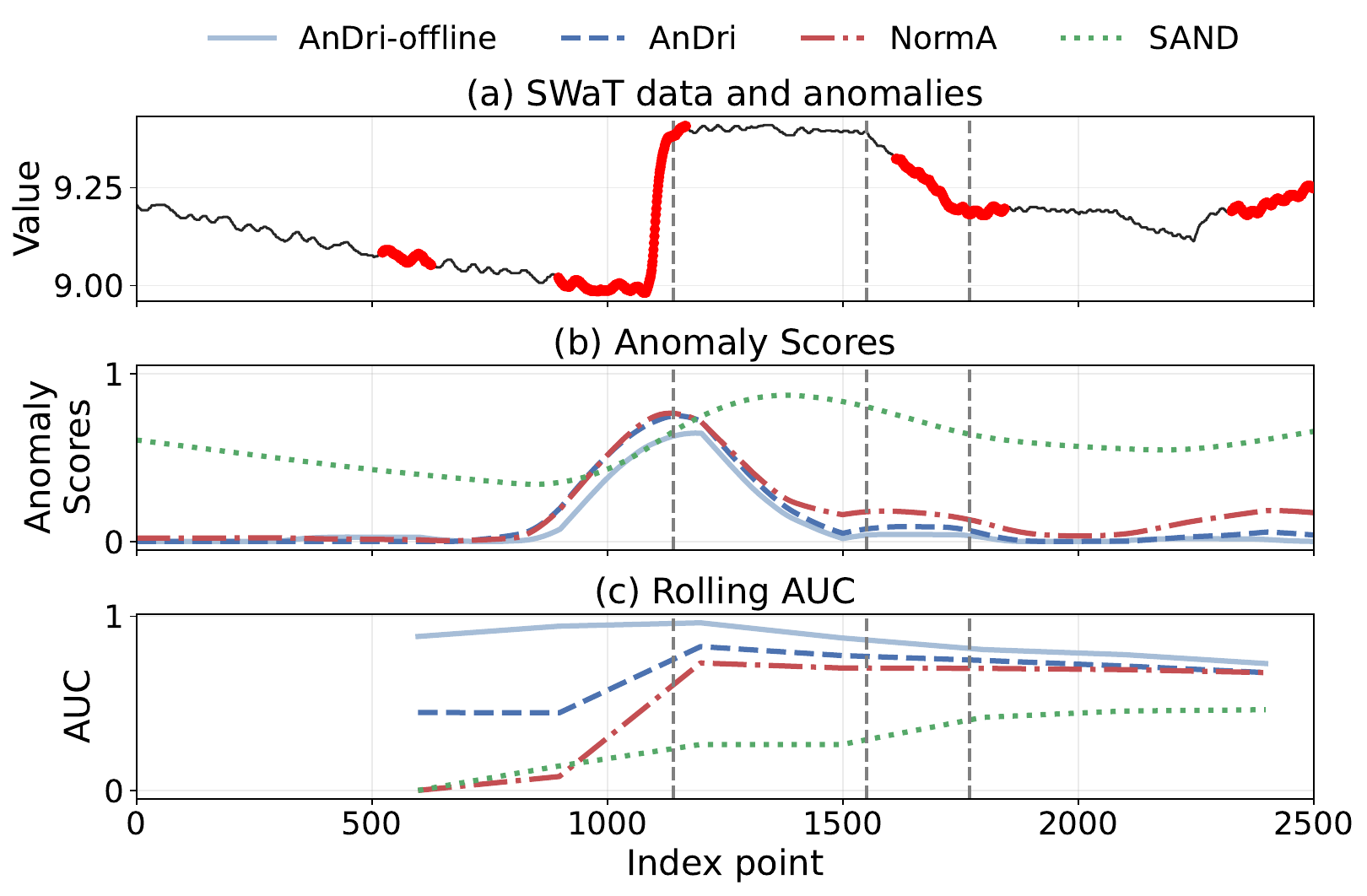}
    \caption{Anomaly scores and 2000-points rolling AUCs for a sensor in \swat\ data under drifts.}
    \label{fig:drift_swat}
\end{figure}
}

\begin{figure}[t]
    \centering
    \includegraphics[width=0.99\linewidth]{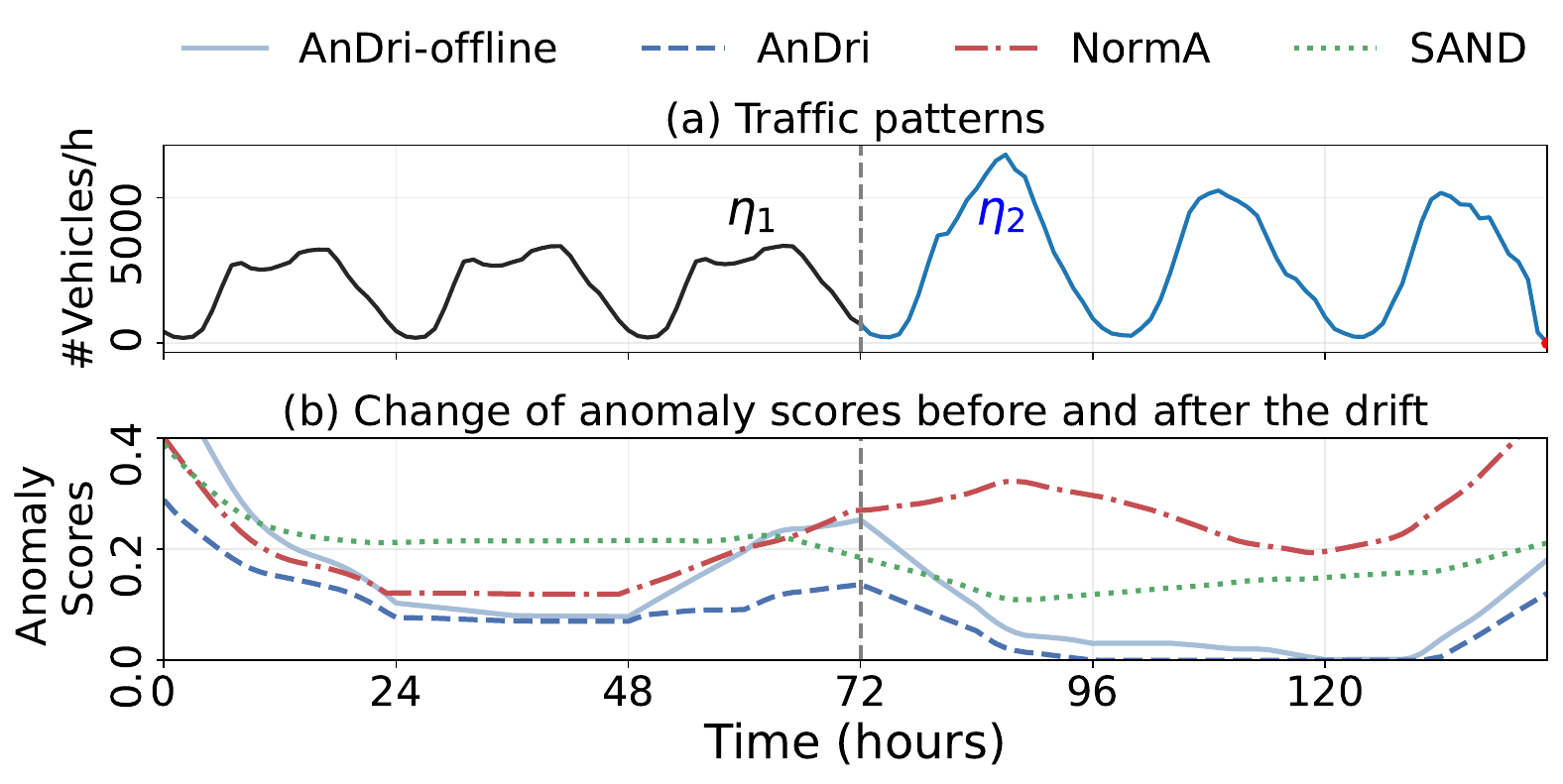}
    \caption{Comparative anomaly scores during drift (dotted line).  \eat{\fei{Change to "hours" (plural), as "hour" can mean exact time.  Do for all graphs.}   \fei{My earlier comment about the legend was only about the drift point, not about the methods, include legend for the curves for (b).}
    \fei{(i) This is not daily pattern if the dates in (b) are over a week; (ii) remove the labels before/after drift, and rename to normal patterns $\eta_1, \eta_2$; (iii) where are the red anomalies?  Otherwise, remove the entire legend; (iv) remove "in 2022" irrelevant.}} 
    }
    \label{fig:short_drift}
\end{figure}

\begin{figure}[t]
    \centering
    \includegraphics[width=0.99\linewidth]{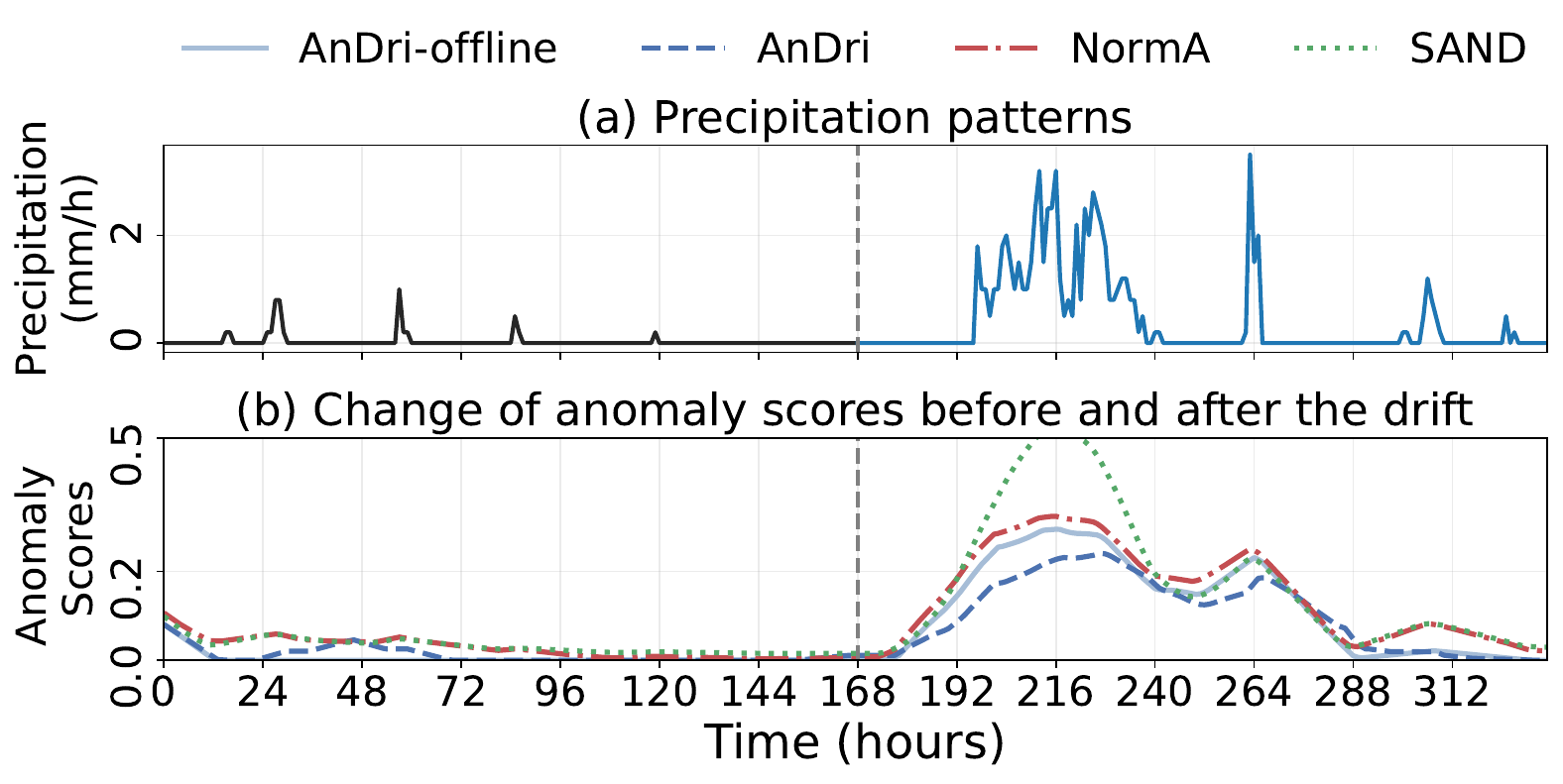}
    \caption{Comparative anomaly scores during drift (dotted line). 
    }
    \label{fig:short_drift_climate}
\end{figure}

\begin{figure}[t]
    \centering
    \includegraphics[width=0.99\linewidth]{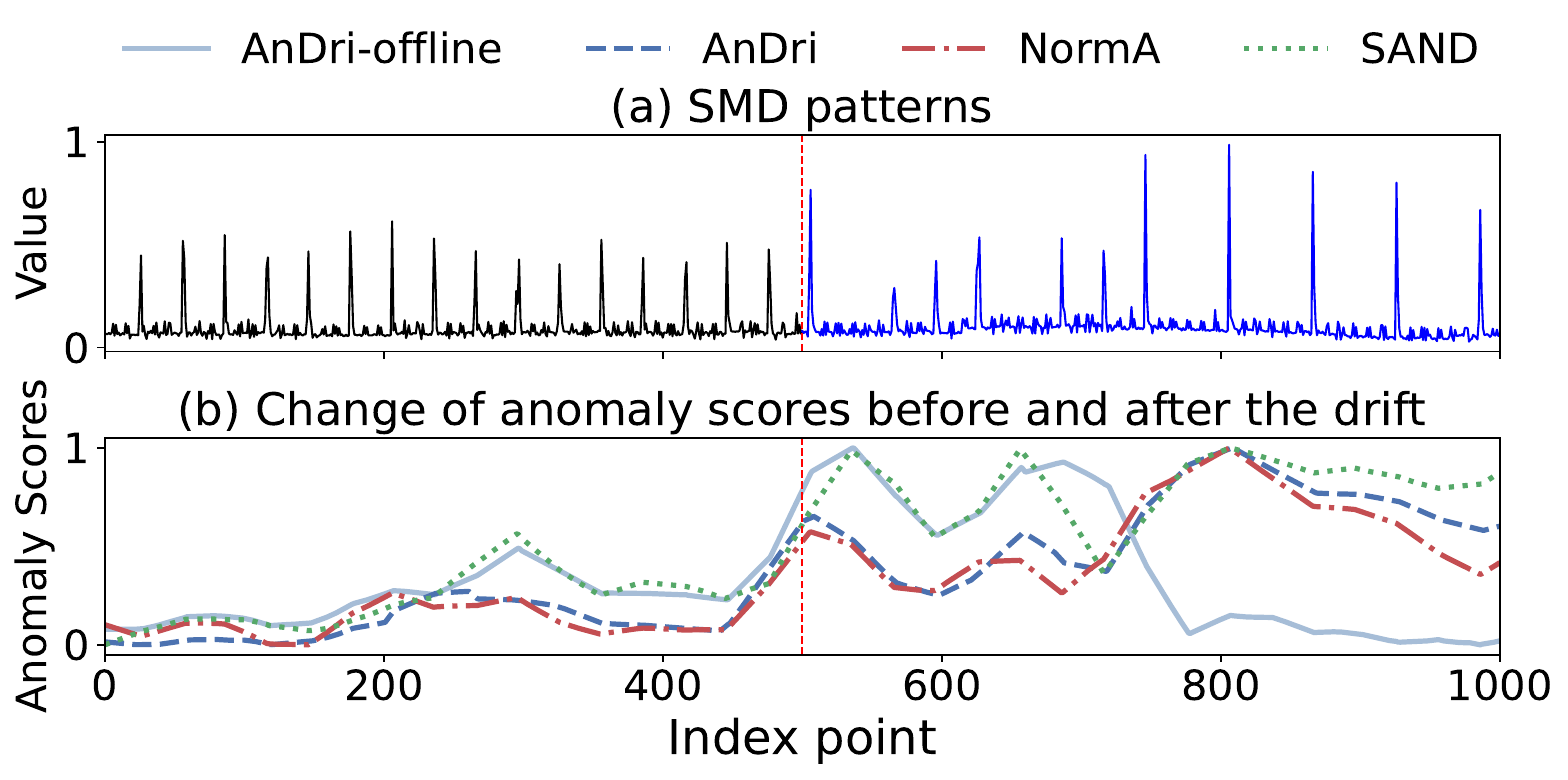}
    \caption{Comparative anomaly scores during drift (dotted line). 
    }
    \label{fig:short_drift_smd}
\end{figure}

\noindent \textbf{Exp-4:  Localized normality.} 
Figure~\ref{fig:short_drift}(a) shows two traffic flow patterns  $\eta_1, \eta_2$ (measured in \#vehicles/hr) occurring before and after a drift $t=72$, We note that $\eta_2$ is localized to a small window of one week, and seen previously.   The corresponding anomaly scores are shown in Figure~\ref{fig:short_drift}(b), where \andri scores are consistently lower than baseline methods \norma and \sand.  Notably, the scores for \andri decrease following the drift, signifying its recognition of $\eta_2$, and its (high) similarity to previously seen patterns.  In contrast,  \sand adapts slowly showing gradual decrease, whereas  \norma scores increase; signifying its inability to recognize $\eta_2$. 
Figure~\ref{fig:short_drift_climate} and Figure~\ref{fig:short_drift_smd} show changes in precipitation patterns in Wabush and in the records of an \smd\ machine (machine-3-1), respectively, along with their corresponding anomaly scores by the univariate methods. In Figure~\ref{fig:short_drift_climate}(a), \soln\ adapts to the change in the precipitation pattern during $t=[192, 240]$, resulting in lower anomaly scores than the other methods. Similarly, although the frequency and peak values in Figure~\ref{fig:short_drift_smd}(a) change during $t=[500, 700]$ but \soln-offline identifies this shift in normal patterns and produces lower anomaly scores.

\begin{figure}[t]
    \centering
    \includegraphics[width=0.99\linewidth]{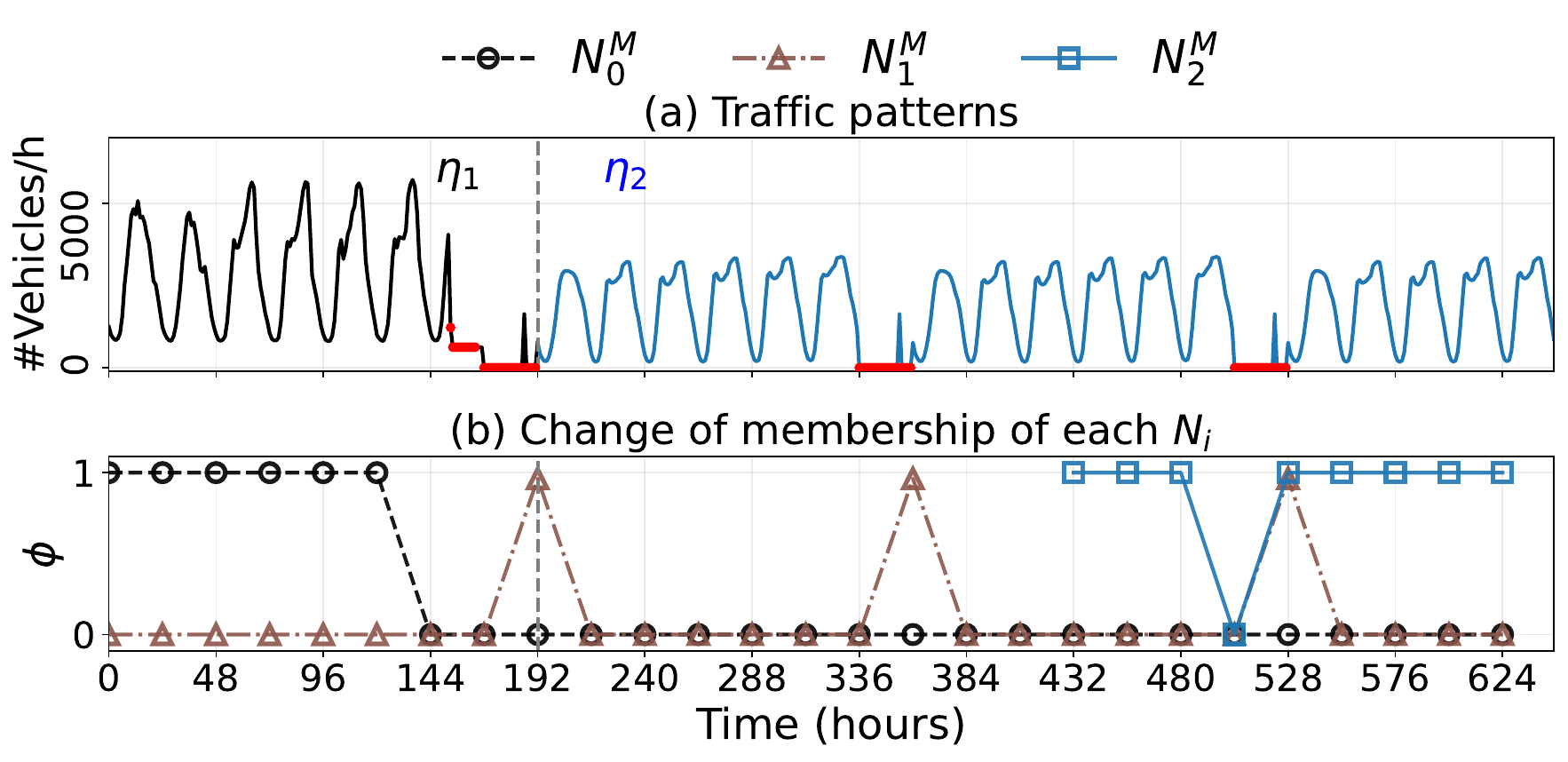}
    \caption{Active vs. inactive pattern activation (\traffic). } 
    \label{fig:new_drift}
\end{figure}

\begin{figure}[t]
    \centering
    \begin{subfigure}[t]{\linewidth}
        \centering
        \includegraphics[width=\linewidth]{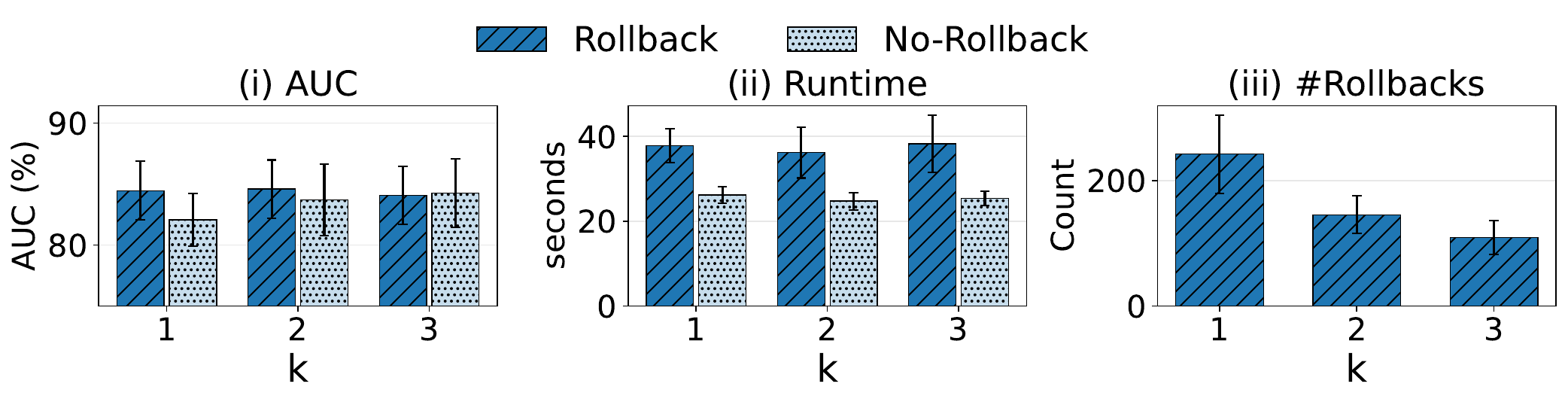}
        \vspace{-6mm}
        \caption{\climate}
        \label{fig:rollback_climate}
    \end{subfigure}
    \begin{subfigure}[t]{\linewidth}
        \centering
        \includegraphics[width=\linewidth]{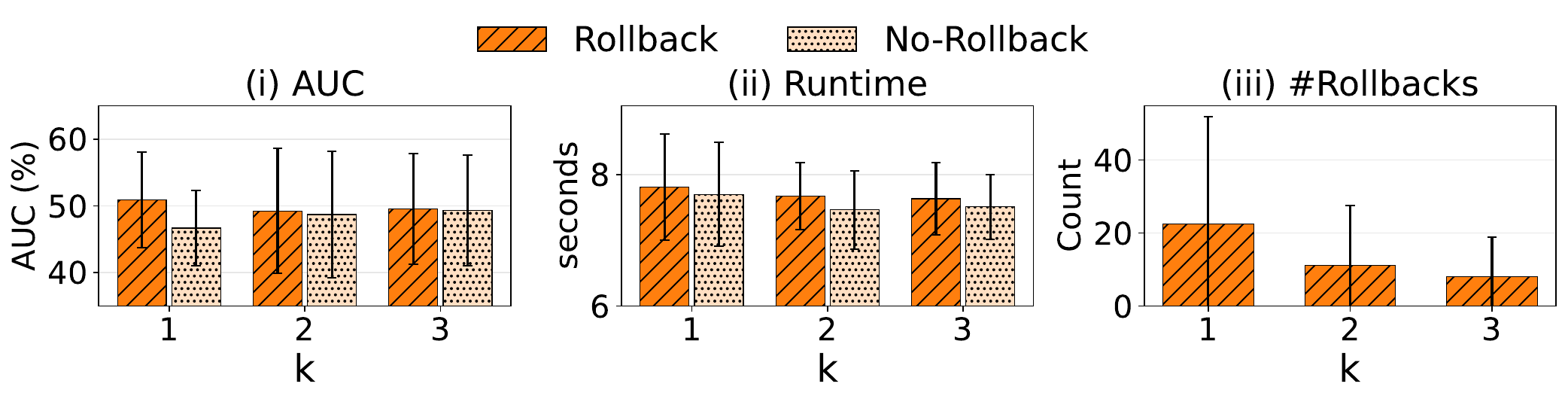}
        \vspace{-6mm}
        \caption{\nab}
        \label{fig:rollback_nab}
    \end{subfigure}
    \begin{subfigure}[t]{\linewidth}
        \centering
        \includegraphics[width=\linewidth]{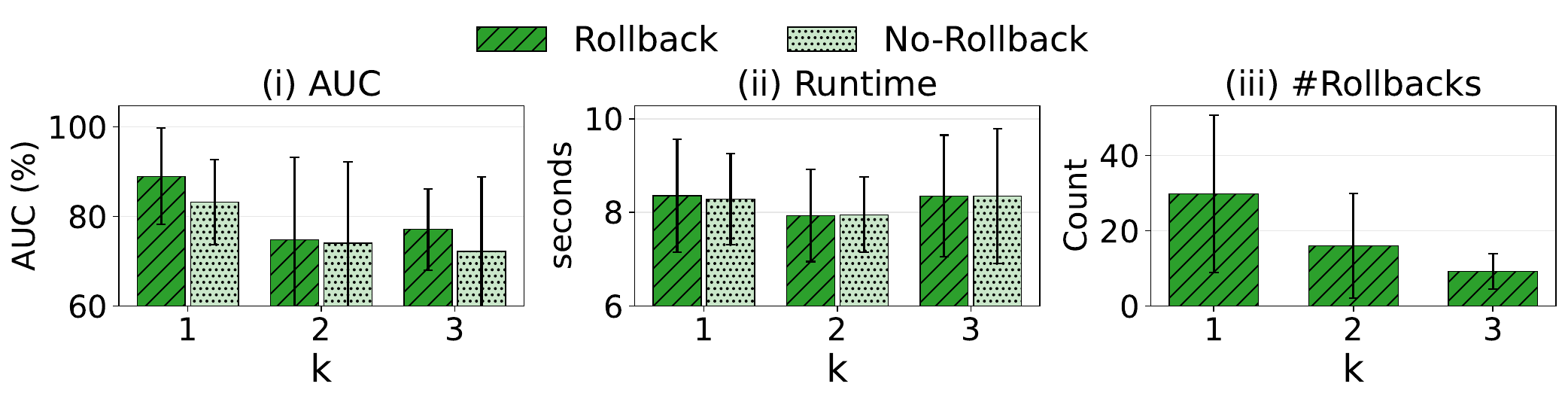}
        \vspace{-6mm}
        \caption{\ucr}
        \label{fig:rollback_env}
    \end{subfigure}
    \caption{Rollback performance for varying $k$.}
    \label{fig:rollback_varying_k}
\end{figure}

\begin{figure}[t]
    \centering
    \includegraphics[width=0.99\linewidth]{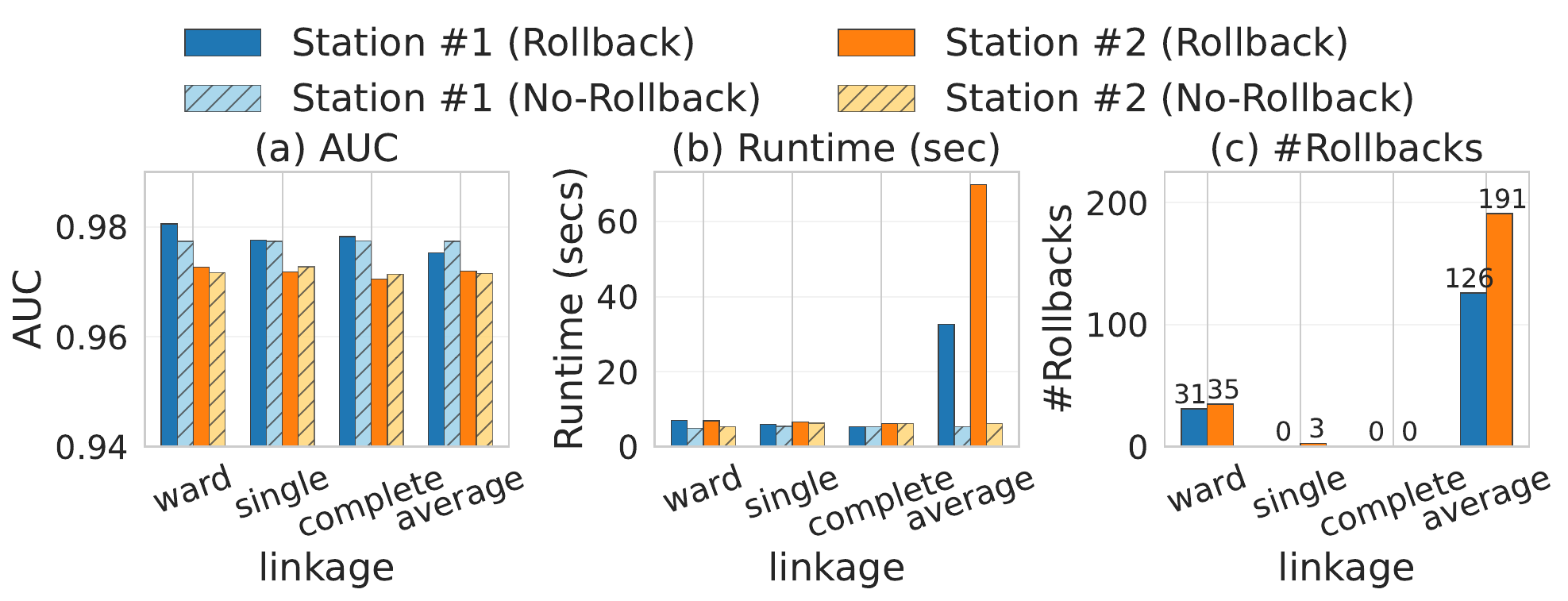} \\
    \caption{Rollback performance for varying linkage (\climate). }
    \label{fig:climate_varying_linkage}
\end{figure}

\begin{figure}[t]
    \centering
    \includegraphics[width=0.99\linewidth]{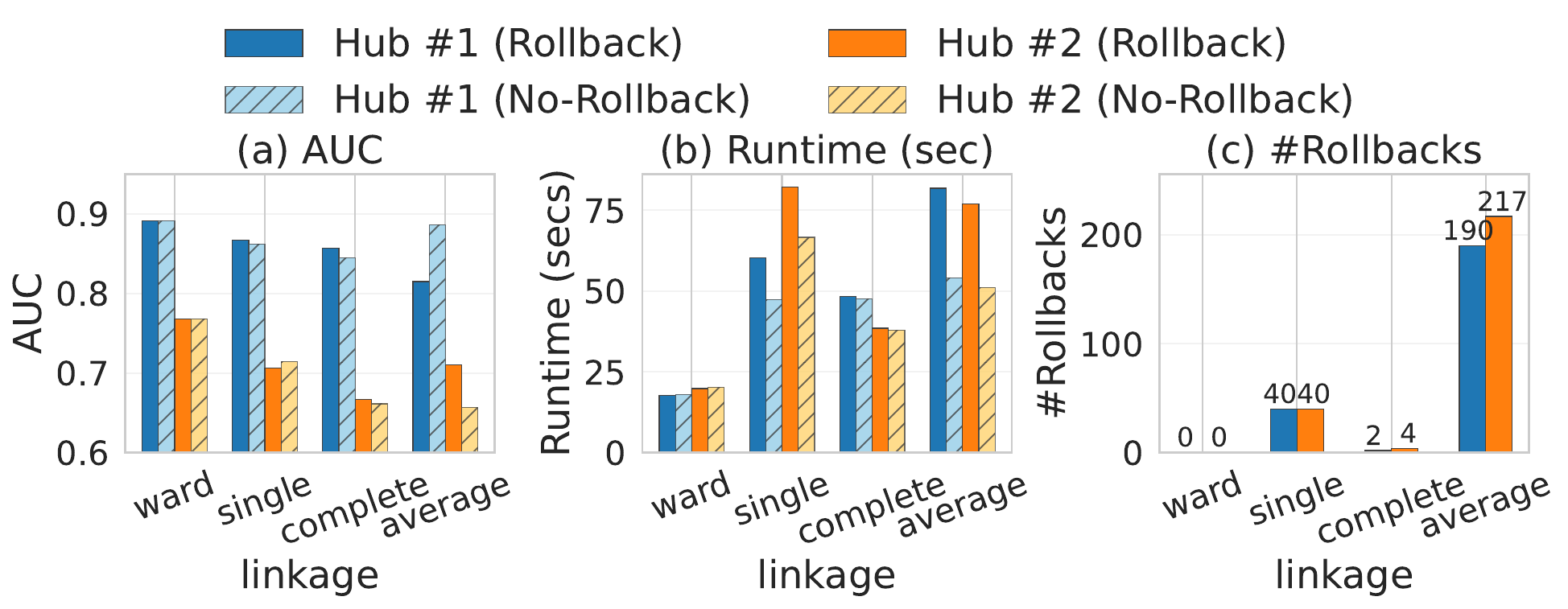} \\
    \caption{Rollback performance for varying linkage (\traffic).}
    \label{fig:traffic_varying_linkage}
\end{figure}

\subsection{Ablation Studies}\label{sec:param}


\noindent \textbf{Exp-5: Rollback effectiveness with varying-$k$.}
We study the effectiveness of our rollback optimization for small $k$ by measuring the accuracy (AUC) vs. runtime tradeoff, and rollback occurrence, across the \climate, \nab, \ucr datasets, as shown in Figure~\ref{fig:rollback_varying_k}. Recall that the rollback mechanism allows \AHC to operate with small $k$ (to improve runtime efficiency) while recovering missed merges from non-adjacent (nearby) clusters.  Figure~\ref{fig:rollback_varying_k} shows that \andri with rollback optimization achieves higher accuracy than no rollback, with +4.0\%, +1\%, +2\% AUC at $k = 1, 2, 3$, respectively, on average across all datasets. As expected, we observe the largest accuracy gains for smaller $k$.  The rollback optimization provides flexibility to select small $k$ while maintaining high accuracy, e.g., Figure~\ref{fig:rollback_climate} shows that \andri achieves 84.5\% AUC at $k = 1$ (with rollback) vs. 84.2\% at $k=3$ (without rollback).  As merges are undone in the corrective step, this induces a runtime overhead, \eat{between 0.2-51.2\%,} i.e., +16\% at $k = 1$, +16.5\% at $k = 2$, and +17\% at $k = 3$.  \eat{I do not understand these numbers, as the runtime overhead does not look like 50\% but also not as small as 1\%} This is also proportional to the decreasing number of rollbacks that are incurred as $k$ increases, i.e., -41\% less rollbacks from $k = 1$ to $k = 2$, and -27\% from $k=2$ to $k=3$.

\begin{figure}[t]
    \centering
    \includegraphics[width=0.99\linewidth]{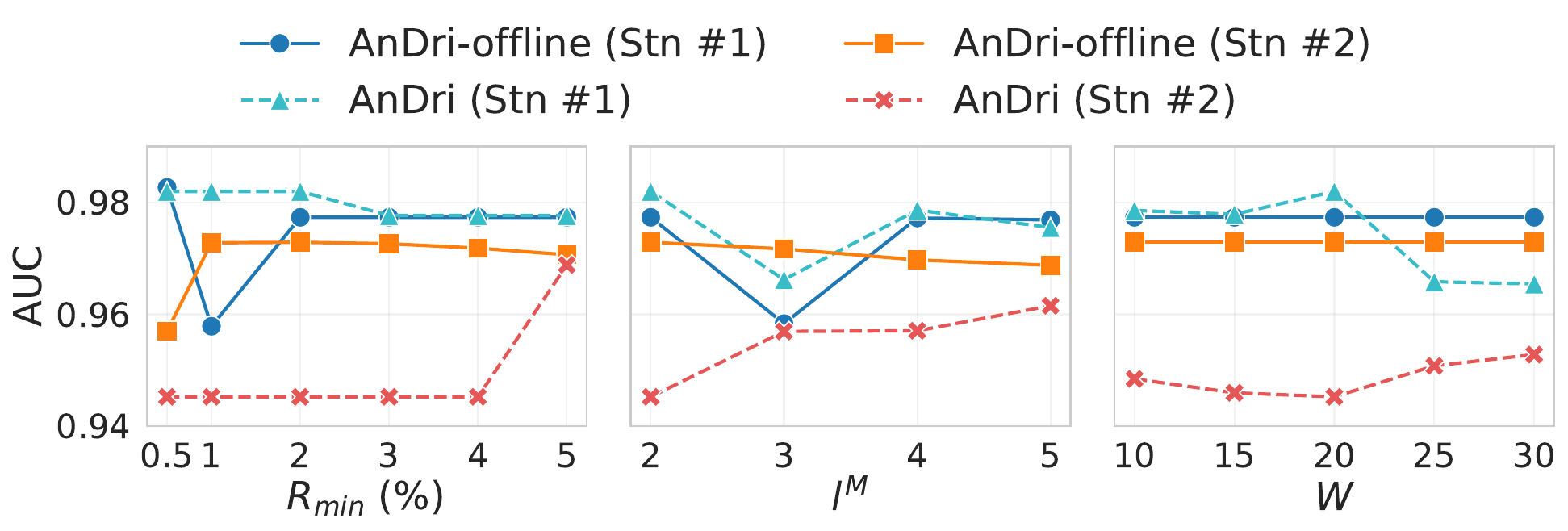} \\
    \caption{Comparative AUC for varying \sizeC, $\ell$, \cwindow (\climate). }
    \label{fig:climate_varying_param}
\end{figure}

\begin{figure}[t]
    \centering
    \includegraphics[width=0.99\linewidth]{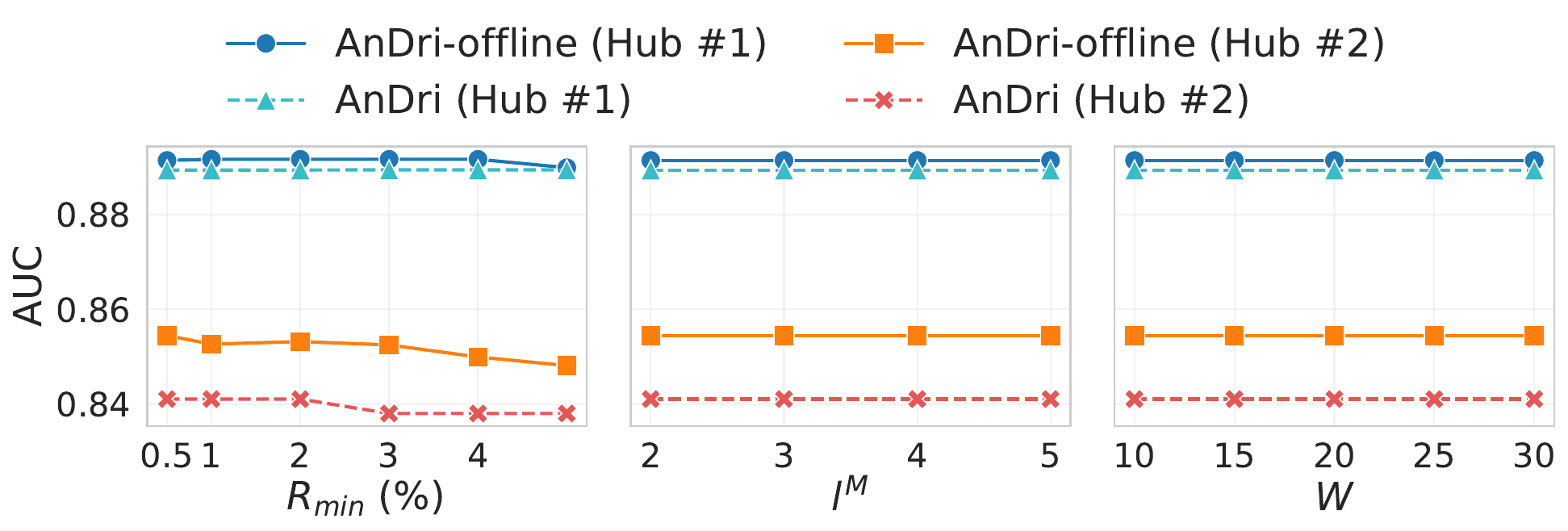} \\
    \caption{Comparative AUC for varying \sizeC, $\ell$, \cwindow (\traffic). }
    \label{fig:traffic_varying_param}
\end{figure}

\noindent \textbf{Exp-6: \AHC effectiveness.}  We evaluate the effectiveness of our \AHC clustering algorithm, by replacing it with two existing clustering methods, namely agglomerative hierarchical clustering (HC)~\cite{hierarchical-2016}, \eat{fei{Please include a citation to HC alg description},} and $K$-shape clustering~\cite{k-shape-2015}, used in \norma~\cite{Norma-2021} and SAND~\cite{boniol2021sand}, respectively. $K$-shape partitions $T$ into $K$ clusters, sharing the same shape using shift-invariant distances that are robust to time shifts and amplitude scaling.  Table~\ref{tab:ablation_feature} shows the comparative AUC and F1 for each clustering method, across the \climate, \traffic, and \nab datasets.  \AHC outperforms HC and $K$-shape by an average, +21\% AUC, and between +12-18\% F1. 
We observe larger accuracy gains over the \traffic and \nab datasets (up to 33\% AUC, 28\% F1) due to similarity among (i) weekdays; (ii) holidays and weekend, traffic flow patterns; and (iii) similar network traffic and CPU load due to spatio-temporal locality references. \eat{\fei{Can you be more specific what this means in the NAB dataset?  What type of system behaviour?} \fei{fill in reasons for temporally close patterns in nab datasets}}  \AHC clusters these similar subsequences with high temporal locality.  

\noindent \textbf{Exp-7: Varying linkage metrics.}  We evaluate the impact of varying linkage metrics: single, complete, average and Ward over the \climate and \traffic datasets, in Figure~\ref{fig:climate_varying_linkage} and Figure~\ref{fig:traffic_varying_linkage}, respectively. All metrics achieve comparatively similar AUC in the \climate dataset (Figure~\ref{fig:climate_varying_linkage}(a)), and the average metric shows 6x slower runtime (Figure~\ref{fig:climate_varying_linkage}(b)).  In contrast, Ward achieves an AUC gain between +3\% -- +7\% in the \traffic dataset (Figure~\ref{fig:traffic_varying_linkage}(a)), and a 55--83\% speedup in runtime compared to other metrics at $k = 5$ (Figure~\ref{fig:traffic_varying_linkage}(b)).  Figure~\ref{fig:climate_varying_linkage}(c) and Figure~\ref{fig:traffic_varying_linkage}(c) show that single and complete linkage metrics produce fewer rollbacks than average and Ward. Rollbacks with single linkage are mostly limited during the early clustering stages (e.g., singleton merges), and become rare as clusters grow in size.  The complete linkage metric rarely triggers rollbacks except some exceptional cases, such as merging relatively large clusters that contain non-adjacent subsequences. In contrast, average linkage triggers frequent rollbacks as it is more sensitive to small differences within and between clusters.  Using Ward linkage also triggers rollbacks during the initial merging stage when $k=1$, but no rollbacks are observed in our experiments with $k=5$, since a larger $k$ helps to cluster similar but non-adjacent subsequences. From our evaluation, Ward linkage provides the best accuracy-to-runtime tradeoff, particularly for $k > 1$.

\eat{
We observe that all linkage metrics performed well on the \climate data (in Figure~\ref{fig:climate_varying_linkage}), with Ward outperforming all others by 1.5\% AUC and F1, but at a cost of 10.8\% slower than complete and single, and average taking 6x longer in runtime.  For the \traffic\ data, Ward achieves +3--7\% and +4--13\% in AUC and peak-F1, respectively,  and a 55--83\% speedup over the other metrics (at $k=5$, in Figure~\ref{fig:traffic_varying_linkage}).  From our evaluation, Ward linkage provides the best accuracy-to-runtime tradeoff, particularly for $k > 1$.}
\eat{We observe that all linkage metrics performed well on the \climate\ data in terms of AUC, and Ward, single, and complete shows relatively similar cost, compared to the average, which takes 6x longer in runtime as shown in Figure~\ref{fig:climate_varying_linkage}. In Figure~\ref{fig:traffic_varying_linkage} for the \traffic\ data, Ward achieves +3--7\% \eat{and +4--13\%} in AUC \eat{and peak-F1, respectively,}  and a 55--83\% speedup over the other metrics (with $k=5$).  From our evaluation, Ward linkage provides the best accuracy-to-runtime tradeoff, particularly for $k > 1$.}

Figure~\ref{fig:climate_varying_param} and Figure~\ref{fig:traffic_varying_param} show AUC performance for varying parameters: \sizeC, $\ell^M$, and \cwindow, over two selected \climate stations, and two \traffic\ hubs, respectively, for \andri and \andrioff.  Figure~\ref{fig:climate_varying_param}(a) shows that varying  \sizeC\ has little effect on AUC,  especially when \sizeC $\geq 0.02$. For smaller values of \sizeC, station \#2 (which has more frequent and heavier rainfall) shows lower AUC due to the mis-classification short abnormal rainfalls as normal.  In contrast, the \traffic\ time-series shows the AUC insensitivity to varying \sizeC. Larger $k$ values also lead to better differentiation between short normal patterns and anomalies across different \sizeC values.   There are no significant changes when varying either $\ell^M$ or \cwindow\ on both datasets, as shown in Figure~\ref{fig:climate_varying_param}(b) and Figure~\ref{fig:climate_varying_param}(c) and Figure~\ref{fig:traffic_varying_param}(b), Figure~\ref{fig:traffic_varying_param}(c). 
\subsection{Efficiency and Scalability.}

\begin{figure}[t]
    \centering
    \includegraphics[width=1.05\linewidth]{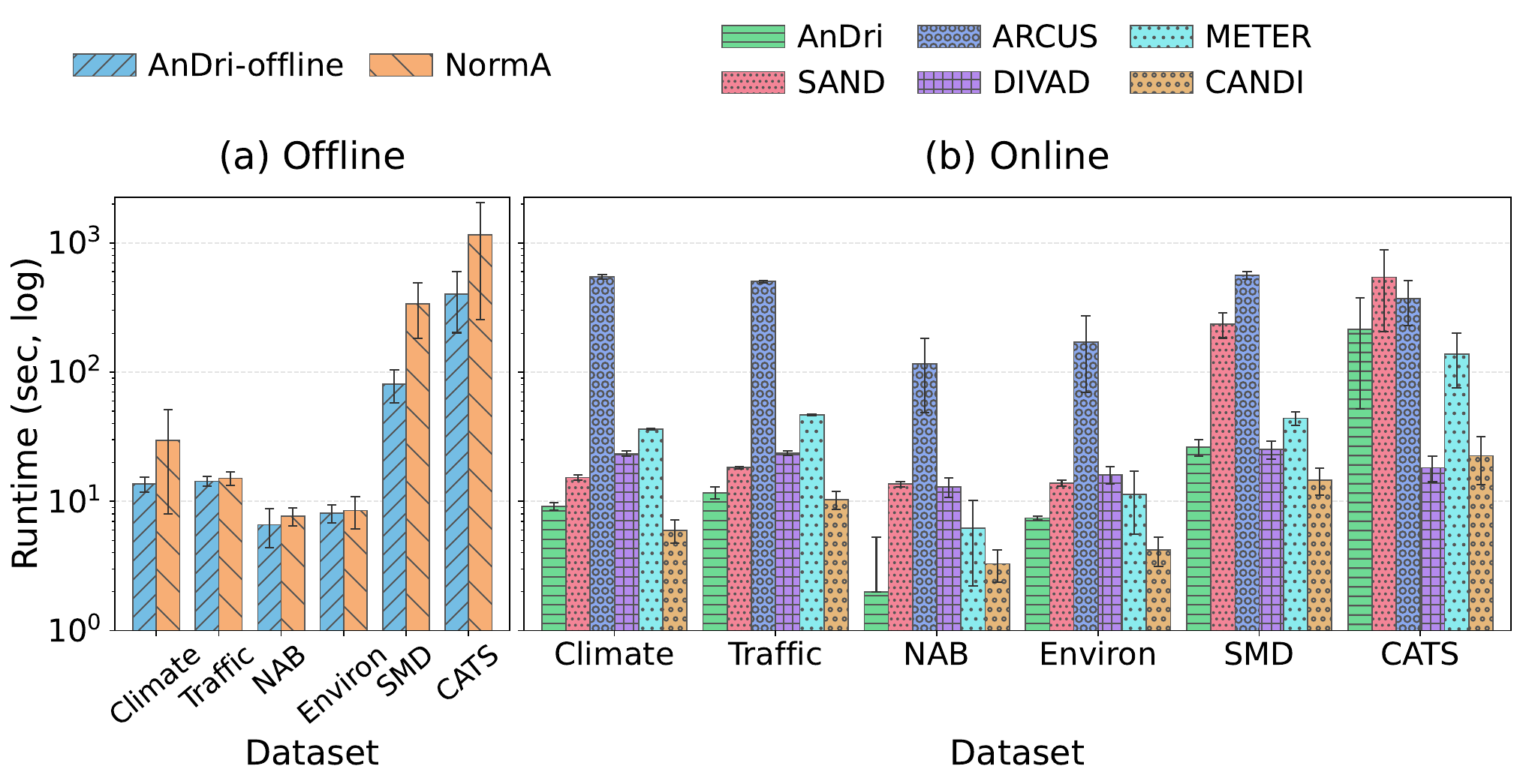}
    \vspace{-3mm}
    \caption{Comparative baseline runtime. \eat{\fei{Remove tranAD.  figure is a bit too small to read the online method comparisons so trying to remove one method to help readability. Enlarge sub-figure labels, too hard to find, spacing between labels and legend and figure.  } }}
    \label{fig:runtime_analy}
\end{figure}

\noindent \textbf{Exp-8: Comparative runtimes.} We evaluate \andri runtime performance against the top baseline methods with the closest accuracy (from Table~\ref{tab:comparison}). Figure~\ref{fig:runtime_analy} reports the comparative runtimes (in log-scale) for offline and online settings.  Figure~\ref{fig:runtime_analy}(a) shows \andrioff is up to 1.4x faster than \norma\ due to it's adjacent clustering of the $k$-neighbouring subsequences.  Figure~\ref{fig:runtime_analy}(b) shows that \andri is 1.5-7x faster than 
\sand\ as re-clustering is only done when the membership function decreases (indicating dissimilarity from $M_A$) versus re-clustering for each data chunk in \sand.  \divad and \candi\ process multivariate dimensions (\smd, \cat) in a single pass, leading to faster runtimes, while \andri processes each dimension separately. As mentioned, a native multivariate detection method is our next step.  In contrast, \andri achieves comparable \eat{(within 2.6 sec)} or superior runtimes (by 64\%) across the \climate, \traffic, \nab, and \ucr datasets. \andri keeps track of inactive patterns, where in cases of recurring drift these inactive patterns are activated from memory without the need for re-clustering.

\begin{figure}[t]
    \centering
    \includegraphics[width=1.05\linewidth]{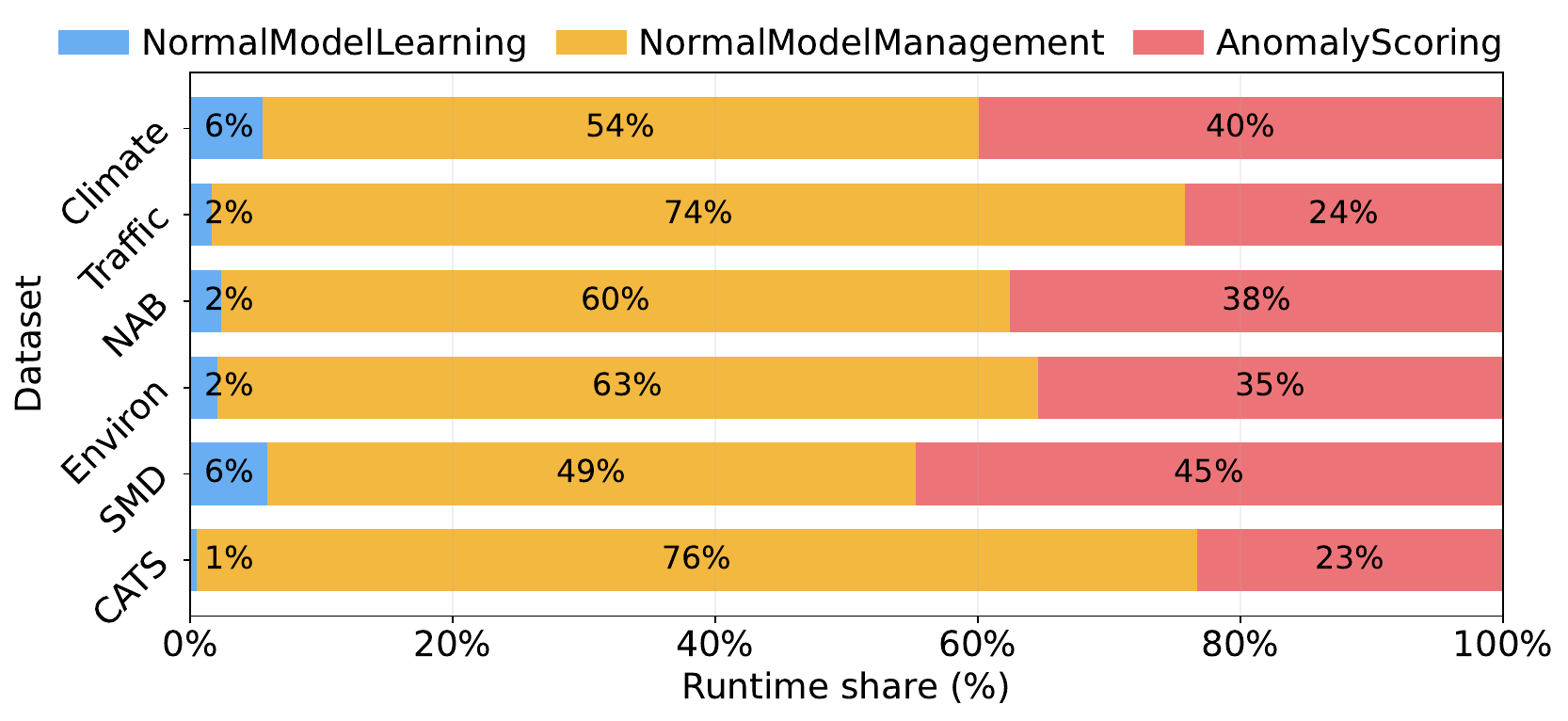}
    \vspace{-3mm}
    \caption{Runtime breakdown.}
    \label{fig:runtime_break}
\end{figure}

\noindent \textbf{Exp-9: Runtime breakdown for each datasets.}  Figure~\ref{fig:runtime_break} shows the runtime breakdown (by component as per Figure~\ref{fig:overview}) for each dataset.  \kw{Normal Model Learning} accounts for up to {6\%} of the total time, where \AHC\ and rollback operations dominate this proportion, particularly for the \climate\ dataset due to precipitation changes and frequent seasonal drifts.  \eat{\fei{if there are more, fill in reasons why learning is higher for some datasets.}}  \kw{Normal Model Management} incurs {49\% - 76\%} of the total time mostly due to activation and de-activation of patterns, affecting datasets with increased drift

\begin{table}[t]
\centering

\caption{\AHC clustering effectiveness. \eat{\fei{Make the font size the same as Table 2.}}}
\vspace{-2mm}
\label{tab:ablation_feature}
\small
\setlength{\tabcolsep}{3pt}
\renewcommand{\arraystretch}{0.95}
\begin{tabular*}{\columnwidth}{@{\extracolsep{\fill}}llcc@{}}
\toprule
Dataset & Clustering & AUC & F1 \\
\midrule
\climate & \textbf{AHC}
  & $\mathbf{0.90 \pm 0.02}$ & $\mathbf{0.29 \pm 0.05}$ \\
& HC
  & $0.78 \pm 0.03$ & $0.25 \pm 0.05$ \\
& $K$-shape
  & $0.79 \pm 0.03$ & $0.26 \pm 0.05$ \\
\midrule
\traffic & \textbf{AHC}
  & $\mathbf{0.86 \pm 0.05}$ & $\mathbf{0.45 \pm 0.05}$ \\
& HC
  & $0.53 \pm 0.07$ & $0.17 \pm 0.05$ \\
& $K$-shape
  & $0.59 \pm 0.17$ & $0.18 \pm 0.12$ \\
\midrule
\nab & \textbf{AHC}
  & $\mathbf{0.81 \pm 0.16}$ & $\mathbf{0.58 \pm 0.26}$ \\
& HC
  & $0.69 \pm 0.17$ & $0.47 \pm 0.23$ \\
& $K$-shape
  & $0.57 \pm 0.22$ & $0.34 \pm 0.18$ \\
\bottomrule
\end{tabular*}
\end{table}

\section{Related Work}

\label{sec:rw}


\noindent \textbf{Time series anomaly detection.}
\eat{Recent anomaly detection techniques have focused on identifying sequential anomalies, which is more difficult than point anomalies. Sequential anomalies occur in a time series when a series of abnormal data points occur when compared with its neighbors~\cite{systematic_review-21, Eval-2022, Review-anomaly-2022, liu-overview-24}.}
Existing solutions search for repeated pattern motifs~\cite{MatrixProfile1-2016, lu2022matrix}, or learn normal behaviour patterns from historical data to identify discords, which are patterns that largely deviate from the normality \eat{thus many of previous methods are based on finding motif (repeated pattern of subsequence) or discord (the most different subsequence in the time series)}~\cite{Norma-2021, boniol2021sand}. \eat{to understand the pattern of normal and abnormal subsequences.} This includes methods such as NormA~\cite{Norma-2021} and SAND~\cite{boniol2021sand} that cluster a given time series to identify clusters of sequences, representing normal models, which are assumed to be frequent, periodic, and span the majority of the dataset.  Patterns that are periodic and occur with a sufficient frequency throughout the dataset are  identified as normal.  However, deviations from these assumptions, e.g., normal behaviours that are frequent over short time intervals, variations in the period, or having varying baseline normal behaviours, create ambiguity to differentiate normal vs. abnormal patterns in time series data. 

\noindent \uline{DNN-based methods.} DNN-based methods identify anomalies when the observed data significantly deviates from the prediction. Such approaches typically use convolutional neural networks (CNNs)~\cite{DeepAnt-2019}, long-short term memory (LSTM)~\cite{LSTM-AD-2015}, or transformer~\cite{TranAD-2022} structures.  Recent work has combined these with generative models such as variational autoencoders (VAEs)~\cite{LSTM-VAE-2017, OmniAnomaly-2019} or generative adversarial networks (GANs)~\cite{MAD-GAN-2019, TranAD-2022}.  Despite these advances, these models often require extensive tuning of hyper-parameters, and typically suffer from slow adaptation rates to concept drift leading to misclassifications. 

\noindent \uline{Drift-aware multivariate methods.} Recent work extend DNN-based multivariate methods towards drift-aware anomaly detection or drift-adaptation, to improve robustness under changes in \emph{domains}~\cite{divad-2025, vanda-24}, \emph{data distributions}~\cite{arcus-22, adtcd-23, METER-23}, \emph{anomaly patterns}~\cite{cutaddpaste-2024}, or \emph{test-time adaptation (TTA)~\cite{M2N2-24, CANDI-26}}. DIVAD~\cite{divad-2025} focuses on learning domain-invariant features for high-dimensional data, while VANDA~\cite{vanda-24} transfers and adapts models across domains. ARCUS~\cite{arcus-22} trains multiple models and switches between models at a given time, whereas ADTCD~\cite{adtcd-23} incrementally updates the model when a drift is detected. METER~\cite{METER-23} avoids full retraining of the model, but shifting parameters with hyper-network with excluding anomalies. On the contrary, CutAddPaste~\cite{cutaddpaste-2024} generalizes the model through self-supervised training with synthetically generated anomalies. M2N2~\cite{M2N2-24} and CANDI~\cite{CANDI-26} introduced TTA that adjust a pre-trained model during test with curated small samples, while keeping both fixed pre=trained model and adaptive detector.  However, these methods primarily  focus on extracting inter-variable relationships for feature extraction or model adaptation, rather than managing recurring normal patterns, and distinguishing concept drifts from anomalies based on temporal locality.

\noindent \uline{LLM-based methods. } Recent studies explore the use of LLMs for time series anomaly detection. These methods convert numerical values into textual representations and leverage the reasoning power of pretrained LLMs~\cite{llmtsad-24, llmad25, delving25}. While these methods can provide interpretable anomaly reasoning, these studies report limited effectiveness on long, multivariate, and non-stationary time series compared to specialized time series anomaly detection methods~\cite{llmtsad-24, llmad25}.

\noindent \uline{Anomaly Model Selection.} Recent benchmark studies have shown that no single anomaly detection method can effectively handle various anomalies, due to the diverse characteristics of time series and dynamic anomalous behaviours~\cite{TSB-UAD-2022, Eval-2022}. To address this challenge, anomaly model selection methods select the most appropriate detector for a given time series~\cite{chooseWisely23, unsupervisedselection23, mospat22} or ensemble of selected detectors based on time-series characteristics~\cite{autotsad24, msad25, ramses26}. For example, AutoTSAD~\cite{autotsad24} evaluates candidate detectors using synthetically injected anomalies, while Goswami et al.~\cite{unsupervisedselection23} rank detectors in a supervised manner. 
The primary challenge is how to reliably identify the best detector while minimizing computational overhead. Since detector ranking requires pre-running multiple models, these methods are limited to offline selection and difficult to apply in online settings to switch detectors,  or adaptive parameter updates under drift.

\noindent \textbf{Concept-drift detection.} 
Concept drift detection has been widely studied in streaming and time-series settings. Most drift detectors monitor changes in global statistical properties, such as mean, variance, or distributional divergence. Classical approaches include error-rate–based detectors (e.g., DDM and EDDM) that track increases in prediction error~\cite{DDM-2004,EDDM-2006}, window-based methods (e.g., ADWIN) that compare statistics across sliding windows~\cite{ADWIN-07}, and distribution-based tests that detect shifts using statistical distances~\cite{Kifer-2004,PageHinkley-1954}. Adaptive ensemble methods, such as Adaptive Random Forest~\cite{ARF-17}, extend these ideas by updating or replacing model components when drift is detected. While effective at identifying stream-level distribution shifts, these methods operate on global summaries and do not explicitly model subsequence structure or localized temporal patterns. CABD studies anomaly detection with change points over groups of distributionally similar sequential data points~\cite{INN-2020}, but considers a limited notion of change. Seasonal trend decomposition techniques have also been adapted for application-specific settings involving events~\cite{BQCD-2021}, fraud~\cite{ARCPD-2021}, and evolving patterns~\cite{pca-15,alippi2016hierarchical}. Unfortunately, anomaly detection and concept drift detection have largely been studied in isolation~\cite{ARF-17,DeepGBM-2019,DDG-DA}.

\noindent \textbf{Dynamic Clustering over Streams.}  Dynamic clustering methods such as CluStream~\cite{clustream}, stream-based $k$-means~\cite{CMMGM02}, SAND~\cite{boniol2021sand} incrementally update cluster structure to changes in the data.  Stream-based $k$-means update cluster centres,  or merge clusters as necessary for incoming data, but remain highly sensitive to the order of arrival of the data points. While CluStream shares similar objectives to \soln, the interaction of their online vs. offline components, to compute micro vs. macro clusters, respectively, differs from AnDri in two ways.  First, if an incoming data point is dissimilar to existing clusters, the creation of singleton clusters requires that existing clusters be removed or merged.  This consequently removes the representative normal patterns,  and merged clusters cannot be undone, eliminating the matching of historical normal patterns.  Second, the continuous, re-clustering necessarily invokes greater overhead, whereas \AHC is invoked only when active patterns fail to explain recent data. Between invocations, \soln\ only updates activation status, avoiding unnecessary re-clustering, allowing re-activation of previously-learned patterns rather than having to re-learn them.

\section{Conclusion}
\label{sec:conclusion}

We propose \soln, a framework for anomaly detection in the presence of concept drift.  We introduce the notion of a \emph{dynamic normal model} where normal patterns are not fixed, but can be activated, deactivated or added over time.  This enables \soln\ to adapt to concept drift and anomalies over time.  \soln\ computes anomaly scores by identifying the most similar active pattern, and avoids the need to aggregate over all patterns, improving precision when multiple normal patterns co-exist.  Lastly, we introduce \AHC, a new clustering method for learning normal patterns that respect their temporal locality; critical for detecting short-lived, but recurring patterns that are overlooked by existing methods.  We present a rollback optimization for \AHC that allows smaller $k$ values to be used for greater efficiency but without compromising accuracy.  As next steps, we intend to explore techniques to dynamically adjust the currency window size, and native extensions to multivariate time-series.






\bibliographystyle{IEEEtran}
\bibliography{concept-drift}

\appendices
\section{Augmenting labels for the \traffic\ data}
\label{sec:appendix}

The \traffic\ dataset from Caltrans Performance Measurement System (PeMS) is organized into nine districts. District 4 (Bay Area) contains 3,855 stations across 29 freeways, each consisting of bi-directional lanes (e.g., eastbound/westbound or northbound/southbound). Along U.S. Route 101 North, there are 563 stations, most of which began data collection in either 2009 or 2016. We first selected 70 stations with the fewest missing records from 2009 to February 2026, and retained 22 stations for our experiments. 

Since the original dataset does not provide labels for anomalies or drifts, we augmented both types of labels. The test time series spans from January 2019 to February 2026. Drift labels are assigned to the official COVID-19 pandemic intervention period, from March 19, 2020 to June 15, 2021~\cite{CA_N33_20_2020, CA_Reopening_2021}. 
During this period, stay-at-home policies significantly reduced traffic volume due to decreased commuting activities such as travel to workplaces and schools. In addition, we added drift labels for summer break, from July 1 to August 31, winter break, from December 3rd week to January 1st week, and the week of Thanksgiving, November 4th week. 
While changes in traffic volume during this period are expected, the traffic patterns evolve gradually and maintain temporal consistency within the period. Therefore, we consider this period as a drift, representing a shift in normal patterns, rather than as anomalies.

The anomaly labels were augmented using the weighted mean and standard deviation of historical weekly data. To do this, we first collected data from the past and next $r_w =4$ weeks based on the current week, as well as data from the same week in the same month of previous years. Then, we computed a weighted sum and standard deviation using an exponentially weighted moving average (EWMA). The weights were set to 0.5 for the weekly component $\alpha_w$ and 0.5 for the yearly component $\alpha_y$, both relative to the current time. For example, the same period last year is assigned a weight of $0.5$, the period two years ago has a weight for $0.5\times 0.5$, and similarly, last week has a weight of 0.5, while the corresponding week from the last year has a weight of $0.5\times 0.5$. The equation form of above are:

\begin{align}
w(i, y) &= \alpha_w^{|i|} \cdot \alpha_y^{y} \\
\mu_t &= \frac{\sum_{y \in year} \sum_{i \in week} w(i,y)\, x_{t+i}^{(y)}}
{\sum_{y \in year} \sum_{i \in week} w(i,y)} \\
\sigma_t^2 &= \frac{\sum_{y \in year} \sum_{i \in week} w(i, y)\, \left(x_{t+i}^{(y)} - \mu_t\right)^2}
{\sum_{y \in year} \sum_{i \in week} w(i,y)} \\
\end{align}

\noindent where $week$ is a set of neighboring weeks (e.g., $t\pm2, ..., t\pm4)$ and $year$ is a set of corresponding weeks in previous years. 

For this computation, we set holiday flags to both fixed-date holidays (e.g., New Year, Christmas, and Veteran day) and week-based holidays (e.g., Martin Luther King Jr. Day (3rd Monday of January), Presidents' Day (3rd Monday of February), Memorial Day (Last Monday of May), and etc.). When collecting weekly data, observations whose day-level flags do not match with the target weeks are excluded. It prevents holidays from influencing non-holiday traffic patterns, and vice versa. For instance, the normal traffic range for Presidents' Day estimated using only past Presidents' Days or holiday Mondays within a $\pm r_w$-week window. This ensures that even holidays show different traffic patterns from regular weekdays, they are not augmented as anomalies as long as they are consistent with patterns observed in past holidays. 

Additionally, for fixed-date holidays such as Veterans Day, only data from the same holiday in previous years are used, including observed shifts when the holiday falls on a weekend. To align the comparable periods across the year, we adopt the ISO week standard~\cite{iso8601} and additionally match week-based holidays. 

Finally, we define the normal range as $\mu_t \pm 3\times \sigma_t$ and assign an anomaly label of $1$ to any observation that falls outside this range.  Also, to validate the sequence of anomalies, we assign the \emph{gap} as 6-hours ($g=6$), where any interval between two consecutive anomalies shorter than the \emph{gap} is also labeled as anomalous, thus treating the entire range as a sequential anomaly.  


%
% ...

\end{document}